\documentclass[a4paper,fleqn,usenatbib]{mnras}

\usepackage{newtxtext,newtxmath}

\usepackage[T1]{fontenc}
\usepackage{ae,aecompl}

\usepackage{graphicx} 
\usepackage{amsmath} 
\usepackage{amssymb} 

\usepackage{amssymb}
\usepackage{amsmath}
\usepackage{lscape}
\usepackage{morefloats}
\usepackage{bigints}
\usepackage[font=small,labelfont=bf]{caption}
\usepackage{graphicx}
\usepackage{ulem}
\usepackage{mathtools}
\usepackage[titletoc,toc,title]{appendix}

\usepackage{float}
\usepackage{wrapfig}
\usepackage{lscape}
\usepackage{rotating}
\usepackage{breqn}
\usepackage{epstopdf}
\usepackage[bottom]{footmisc}

\title[Neutrinos, Baryons and Intrinsic Alignments]{Towards determining the neutrino mass hierarchy: weak lensing and galaxy clustering forecasts with baryons and intrinsic alignments}

\author[D. Copeland, A.N. Taylor, Alex Hall]{
David Copeland,$^{1}$\thanks{E-mail: dcope@roe.ac.uk}
Andy Taylor,$^{1}$
Alex Hall, $^{1}$
\\
$^{1}$Institute for Astronomy, University of Edinburgh, Royal Observatory, Blackford Hill, Edinburgh EH9 3HJ
}

\date{Accepted XXX. Received YYY; in original form ZZZ}

\pubyear{2019}

\bibpunct[, ]{(}{)}{;}{a}{}{,}
\AtBeginShipout{%
  \ifnum\value{page}>1 %
    \typeout{* Additional boxing of page `\thepage'}%
    \setbox\AtBeginShipoutBox=\hbox{\copy\AtBeginShipoutBox}%
  \fi
}

\begin{document}
\label{firstpage}
\pagerange{\pageref{firstpage}--\pageref{lastpage}}
\maketitle

\defcitealias{Mead15}{M15}
\defcitealias{Mead16}{M16}

\begin{abstract}
The capacity of Stage IV lensing surveys to measure the neutrino mass sum and differentiate between the normal and inverted mass hierarchies depends on the impact of nuisance parameters describing small-scale baryonic astrophysics and intrinsic alignments. For a Euclid-like survey, we perform the first combined weak lensing and galaxy clustering Fisher analysis with baryons, intrinsic alignments, and massive neutrinos for both hierarchies. We use a matter power spectrum generated from a halo model that captures the impact of baryonic feedback and adiabatic contraction. For weak lensing, we find that baryons cause severe degradation to forecasts of the neutrino mass sum, $\Sigma$, approximately doubling $\sigma_{\Sigma}$. We show that including galaxy clustering constraints from Euclid and BOSS, and cosmic microwave background (CMB) \textit{Planck} priors, can reduce this degradation to $\sigma_{\Sigma}$ to 9\% and 16\% for the normal and inverted hierarchies respectively. The combined forecasts, $\sigma_{\Sigma_{\rm{NH}}}=0.034\, \rm{eV}$ and $\sigma_{\Sigma_{\rm{IH}}}=0.034\, \rm{eV}$, preclude a meaningful distinction of the hierarchies but could be improved upon with future CMB priors on $n_s$ and information from neutrinoless double beta decay to achieve a 2$\sigma$ distinction. The effect of intrinsic alignments on forecasts is shown to be minimal, with $\sigma_{\Sigma}$ even experiencing mild improvements due to information from the intrinsic alignment signal. We find that while adiabatic contraction and intrinsic alignments will require careful calibration to prevent significant biasing of $\Sigma$, there is less risk presented by feedback from energetic events like AGN and supernovae.
\end{abstract}


\vspace*{-20mm}
\section{Introduction}
Neutrino cosmology is an intersection of particle physics and astrophysics that has the potential to greatly enrich both fields. Flavour oscillations measured in solar, atmospheric, reactor and accelerator experiments have constrained the mass-squared differences, $\Delta m_{12}^2$ and $|\Delta m_{23}^2|$, for the three neutrino mass eigenstates but cannot determine the absolute masses \citep[e.g.,][]{Maltoni04, Fogli06}. The sign of the largest splitting defines two possible orderings, known as the normal (NH, $\Delta m_{23}^2 > 0$) and inverted (IH, $\Delta m_{23}^2 < 0$) hierarchies. Lower bounds on the sum of the masses for each hierarchy, $\Sigma_{\rm{NH}} \gtrsim 0.06\,\rm{eV}$ and $\Sigma_{\rm{IH}} \gtrsim 0.1\,\rm{eV}$, can be arrived at numerically from the oscillation constraints \citep{Lesgourges06}. Cosmology has been providing increasingly tight upper bounds. Data from the Lyman-$\alpha$ forest in combination with \citet{Planck15} cosmic microwave background (CMB) constraints \citep{Palanque-Delabrouille15, Yeche17} and the more recent works by \citet{Vagnozzi17} and \citet{Planck18} analysing CMB and baryon acoustic oscillation (BAO) data all find $\Sigma \lesssim 0.12\,\rm{eV}$ (95\% confidence level). Though this starts to confront the limits of the IH, it remains an open question as to whether future cosmological data will be able to distinguish between the hierarchies by constraining the mass sum. The degeneracy between hierarchies is partially broken by different sets of free-streaming scales associated with the mass eigenstates, but these differences are sufficiently small for a fixed mass sum that there is no realistic prospect of achieving a detection in the near future \citep{Hall&Challinor12}. 

Forthcoming Stage IV surveys, such as Euclid \citep{Laurejis11} and LSST \citep{LSST09}, are aiming to achieve  $\Sigma$ errors small enough to determine the hierarchy by using large-scale structure (LSS) probes like tomographic weak gravitational lensing \citep{Bartelmann&Schneider01} and spectroscopic galaxy clustering. The former uses multiple photometric redshift bins to extract the influence of cosmological quantities on the expansion history and the growth of the dark matter distribution \citep{Albrecht06, Peacock06}. Understanding how galaxies trace the underlying matter distribution through clustering measurements provides an additional repository of information on the properties of the Universe \citep{Hauser73, Davis83}. This is subject to a range of systematics and statistical effects including redshift uncertainty \citep[see e.g.,][]{Wang13}, galaxy bias \citep[e.g.,][]{Desjacques18, Vagnozzi18} and anisotropies due to the Alcock-Paczynski (AP) effect \citep{Alcock&Paczynski79, Ballinger96, Ruggeri17} and redshift space distortions \citep{Kaiser87, Hamilton98}. 

These probes are expected to observe the imprint of neutrinos through their suppression of matter clustering. Neutrinos are very light and so have large free-streaming lengths. For scales below these they do not cluster. For a fixed total matter density, this results in matter perturbations being slightly smoothed. However, the cumulative effect on the matter power spectrum, $P\left(k\right)$, by $z=0$ is a decrease on linear scales by a factor, $\sim 1-8f_{\nu}$, in linear theory, where $f_{\nu} \equiv\Omega_{\nu}/\Omega_m$ is the ratio of the neutrino and total matter density parameters \citep{Hu98}. On non-linear scales, neutrinos are modelled through perturbative methods \citep[e.g.,][]{Pietroni08, Lesgourgues09, Levi16} or accounted for in simulations. Generally, the most accurate approaches for the latter have implemented neutrinos as separate low-mass collisionless particle species alongside cold dark matter (CDM) to fully capture their non-linear evolution \citep[e.g.][]{Bird12, Massara14, Liu18}. These indicate that the effect of neutrino suppression on haloes depends on their mass, leading to a greater power reduction than linear theory on intermediate scales of $k\sim1 h\, \mathrm{Mpc}^{-1}$ where larger haloes are significant. On deeper non-linear scales where smaller haloes predominate, the power, while suppressed overall, is boosted relative to the linear prediction.  

Depending on the choice of probes, methodology and underlying model, 1-$\sigma$ errors on the neutrino mass have been forecast broadly within the range $0.02-0.04\, \rm{eV}$ for Euclid-like surveys \citep[e.g.,][]{Carbone11, Audren13, Boyle18a}. A major systematic neglected in these forecasts is the redistributing of matter on halo scales by baryonic astrophysics. The mechanisms governing adiabatic contraction via radiative cooling or localized high-energy releases from active galactic nuclei (AGN) and supernovae \citep[see e.g.,][]{Rudd08, vanDaalen14} are not well-understood. AGN feedback can decrease the fraction of baryons in inner halo regions by several factors and bloat haloes outwards as heated gas expands through the matter distribution \citep[e.g.,][]{Duffy10}. This has been well-demonstrated by high-resolution hydrodynamic simulations such as the OverWhelmingly Large Simulations \citep[OWLS;][]{Schaye10}. A degree of counterbalance is provided on non-linear scales by the infall of dark matter into gravitational potential wells that have been deepened by the small-scale clustering of cooling gas \citep{Gnedin04, Jing06, Rudd08, Duffy10}. Overall, $P\left(k\right)$ experiences a $\sim$30\% suppression in the mildly non-linear regime, which gives way to a net increase of power through adiabatic contraction on deeply non-linear scales of $k\sim10 h\, \mathrm{Mpc}^{-1}$ \citep[e.g.,][]{Semboloni11}. As neutrinos and baryons both suppress matter power on weakly non-linear scales it is possible that a partial degeneracy exists between the two effects. It is therefore necessary to simultaneously constrain them both. Such degeneracies could potentially be broken by the scale-dependence of the baryonic effect and the redshift-dependence of neutrino suppression. 

Marginalizing over the uncertainty in baryon processes can significantly impact parameter forecasts. For example, \citet{Copeland18} predict 40\% degradations to constraints of the dark energy parameter space and up to 80\% for other cosmological parameters for a Euclid-like weak lensing survey \citep[see also][]{Semboloni11, Zentner12, Mohammed14}. Significant biases in neutrino mass measurements were found by \citet{Natarajan14} when failing to account for baryons, using simple arguments to model both components in $P\left(k\right)$, although the impact of marginalizing over baryon parameters was not explored. More recently, \citet{Parimbelli18} found that the neutrino mass is unlikely to be significantly biased by degeneracies with baryons, based on a Markov Chain Monte Carlo (MCMC) method using weak lensing and galaxy clustering that is limited by excluding freedom in the cosmological parameters. This paper explores similar territory but addresses different key questions. We are primarily interested in the degradations to error forecasts from marginalizing over baryonic effects, although the separate issue of bias in the neutrino mass itself is examined in the context of miscalibrations in baryon modelling. The specific focus of this paper is how degradation impacts the capacity to measure $\Sigma_{\rm{NH}}$ and $\Sigma_{\rm{IH}}$, and whether it will be feasible to distinguish between the mass hierarchies. To do this we perform a Fisher analysis of the full range of cosmological parameters probed by Euclid-like weak lensing and galaxy clustering surveys\footnote{It is important to emphasize that while this work uses Euclid survey parameters and specifications from e.g., \citet{Laurejis11} and \citet{Amendola18}, this paper is not the product of a Euclid Consortium collaboration, and the results presented here should not be taken as official Euclid forecasts.}. 

We employ analytic modifications of the halo model to capture the effects of neutrinos and baryons on $P\left(k\right)$ and the weak lensing convergence power spectrum, $C_{\ell}$. \citet{Mead15} (hereafter \citetalias{Mead15}) provide an empirically driven baryon prescription by calibrating internal halo structure relations to match several implementations of multiple sources of baryonic physics in OWLS. We use this halo model variant, HMCODE, in our work. Improving upon the accuracy of HALOFIT \citep{Smith03, Takahashi12} by up to a factor of 2 in the non-linear regime, it achieves $\simeq 5$ percent accuracy in $P\left(k\right)$ for scales $k \leq 10\,h\,\mathrm{Mpc}^{-1}$ and redshifts $z \leq 2$. For a wider discussion of the value of this approach compared to treatments that fit directly the stellar and gas physics within the halo \citep[e.g.,][]{Semboloni11, Mohammed14, Fedeli14}, we refer the reader to \citet{Copeland18}. \citet{Mead16} (hereafter \citetalias{Mead16}) extend their model to account for non-linear neutrino influences by modifying the parameters governing spherical collapse, as this will be directly impacted by the reduction in matter clustering. Fitting to the simulations of \citet{Massara14} reproduces $P\left(k\right)$ to within a few percent up to $k=10\,h\,\mathrm{Mpc}^{-1}$ for multiple redshifts. This is a slightly better performance than the \citet{Bird12} fitting formula. 

In our weak lensing analysis, we address one further source of uncertainty. On large scales, gravitational fields cause tidal distortions in galaxies, which contribute to correlations between the intrinsic ellipticities and gravitational shear of galaxies. Failing to account for these intrinsic alignments (IA) can significantly bias lensing parameter estimates \citep{Joachimi11, Troxel15}. 

Our work is the first to combine baryons, IA and massive neutrinos in a self-consistent Fisher framework in the context of distinguishing the NH and IH in Stage IV surveys. By applying this framework to weak lensing, spectroscopic galaxy clustering and CMB probes, we present a uniquely comprehensive analysis of the current prospects for determining the hierarchy. This paper proceeds as follows. In \S~\ref{sec:nubhm} we outline the neutrino and baryon modifications to the halo model. We present our results in \S~\ref{sec:res} for the mass constraints on the normal and inverted hierarchies, analyze the degradation due to baryons and intrinsic alignments, and explore the information gains available from combining weak lensing and galaxy clustering for a Euclid-like survey with priors from BOSS \citep{BOSS17} and \textit{Planck}. The significance of model bias is assessed in \S~\ref{sec:modbias}, before we conclude in \S~\ref{sec:conc}. The LSS probes and Fisher formalism we use to make forecasts are explained in Appendix~\ref{sec:probes}. 

\vspace*{-5mm}
\section{Neutrinos and Baryons In The Halo Model}
\label{sec:nubhm}
For a review of the halo model and how the matter power spectrum before including neutrinos and baryons is modelled, we refer the reader to Appendix~\ref{sec:halo_model}.
\begin{figure}
\includegraphics[width=\columnwidth]{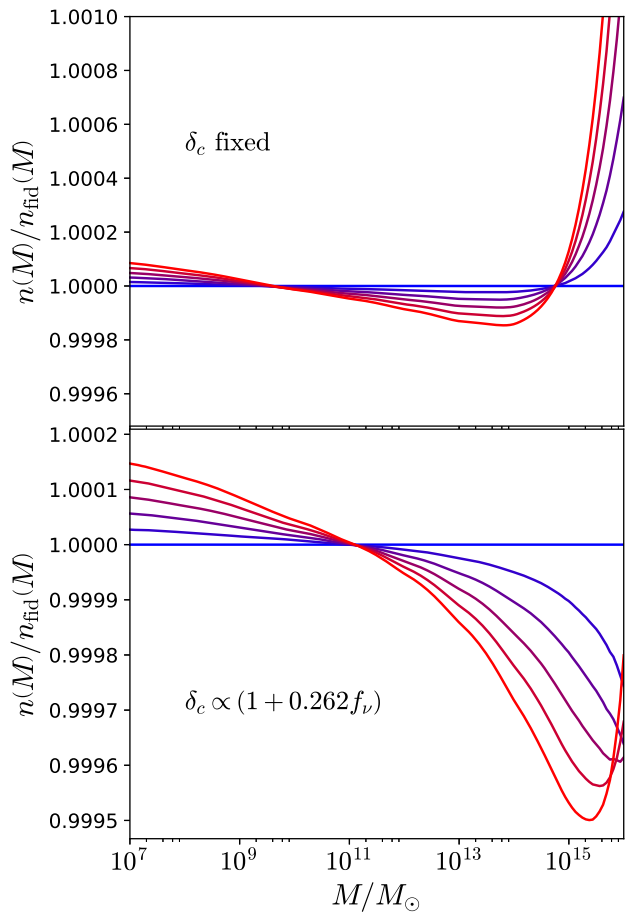}
\caption{Response of the halo mass function at $z=0$ to the neutrino mass sum for the normal hierarchy, with the horizontal blue line corresponding to the fiducial case using the minimal NH neutrino mass sum, $\Sigma_{\rm{NH,min}}=0.06\, \rm{eV}$ and increasingly red curves corresponding to $\Sigma$ in the range $\Sigma_{\rm{NH,min}} < \Sigma_{\rm{NH}} \leq 1.1\,\Sigma_{\rm{NH,min}}$. Top panel: the mass function is sensitive to the mass sum through $\sigma\left(M\right)$, which in turn depends on $\Sigma$ through the linear matter power spectrum. There is no dependence of $\delta_c$ on $\Sigma$ impacting the response. Bottom panel: the mass function depends on $\Sigma$ through $\sigma\left(M\right)$ and $\delta_c$, the latter being sensitive to $f_{\nu}$ via equation~\eqref{eq:dc}.}
\label{fig:massfunc} 
\end{figure}
\subsection{Modelling neutrino effects}
\label{subsec:modellingneutrinos}
We use the CAMB Boltzmann code \citep{Lewis00} to generate linear $P\left(k\right)$ for the NH and IH. To capture the effect of neutrinos on non-linear scales we use the modifications to the halo model prescribed by \citetalias{Mead16}. It should be noted that our version of HMCODE has been slightly altered from that of \citetalias{Mead15} and \citetalias{Mead16}, partly to correct for small instabilities that we originally encountered. Massive neutrinos impact the matter distribution primarily by damping matter perturbations as they free-stream through structure. For fixed $\Omega_m$\footnote{We define the total matter density parameter, $\Omega_m=\Omega_c+\Omega_b+\Omega_{\nu}$, as the sum of contributions from cold dark matter, baryons and neutrinos.}, the suppression of matter clustering is well-approximated as depending only on $f_{\nu}$. This affects the spherical collapse overdensity, $\delta_c$, and the virial density, $\Delta_v$, which are characteristic descriptors of halo structure. The former affects the shape of the mass function and determines the halo concentration by defining the formation redshift at which a certain fraction of mass has undergone spherical collapse. The latter defines limits on the density profile. By introducing a linear correction in terms of $f_{\nu}$, such that
\begin{equation}
\delta_c \propto 1+0.262f_{\nu}
\label{eq:dc} 
\end{equation} 
and
\begin{equation}
\Delta_v \propto 1+0.916f_{\nu},
\label{eq:dv}
\end{equation}
\citetalias{Mead16} achieve a percent-level fit to N-body massive neutrino and CDM simulations by \citet{Massara14} up to $k=10\,h\,\mathrm{Mpc}^{-1}$. These relations can be interpreted as the damping of matter fluctuations due to neutrino free-streaming resulting in a smaller fraction of perturbations collapsing by a given redshift, equivalent to raising the density threshold for collapse. Potential improvements could be attained by implementing more sophisticated methods by \citet{Ichiki12} or \citet{LoVerde14} in which, for example, non-linear neutrino clustering effects are considered. However, the \citetalias{Mead16} fit is sufficiently accurate for our purposes and permits us to use $f_{\nu}$ as a free parameter in fast power spectrum calculations for forecasts. In practice, we treat the mass sum, $\Sigma$, as the actual free parameter, via the relation,
\begin{equation}
f_{\nu} = \frac{\Omega_{\nu}}{\Omega_m} \simeq \frac{\Sigma}{94.1\, \Omega_m h^2\, \rm{eV}}.
\end{equation}
The \citet{Massara14} simulations are designed to test the halo model with neutrinos, specifically by providing accurate halo power spectrum responses to different neutrino masses. This makes the \citetalias{Mead16} fits to these simulations particularly relevant for our purposes, in which an accurate power spectrum is less important than the sensitivity of the power spectrum to changes in cosmology for making forecasts. Other implementations of neutrinos in the halo model \citep[e.g.,][]{Takahashi12, Bird12} generate slightly different non-linear responses in the power spectrum. It should be noted that none of these treatments is entirely robust, being subject to the limited accuracy of the underlying halo model fit on non-linear scales and the challenge of properly accounting for neutrino non-linear clustering effects. There may be future scope to improve on these approaches by extending the methodology of \citet{Mead17} and \citet{Cataneo18}, who demonstrate the increased accuracy on non-linear scales available to the halo model by incorporating spherical collapse and growth results into power spectrum responses. 

In Figure~\ref{fig:massfunc} we show the impact of varying $\Sigma$ on the halo mass function with and without the modification to $\delta_c$ in equation~\eqref{eq:dc}. Neutrinos have a very small impact on the matter distribution on non-linear scales so we have illustrated it through the ratios of the mass function to the case with the minimal NH neutrino mass sum, $\Sigma_{\rm{NH,min}}=0.06\, \rm{eV}$. 

Considering first the case with $\delta_c$ unmodified, we find that for fixed $\Omega_m$ and $\sigma_8$, increasing the neutrino mass increases the mass function for high halo masses in the top panel of Figure~\ref{fig:massfunc}. This reflects the fact that the suppression of power on small scales from neutrino free streaming requires that linear power is boosted on large scales to keep $\sigma_8$ fixed. As a result, the variance of the matter distribution is increased for the corresponding large mass scales. This is the regime where $\sigma\left(M\right)$ is smallest, and hence where $\nu$ is largest. At these scales the mass function is dominated by an exponential tail, as shown in equation~\eqref{eq:massfunc}. This leads to small changes in $\nu$, from varying $\Sigma$, generating large changes in $n\left(M\right)$.

The non-linear regime begins on smaller mass scales where the clustering is damped by higher $\Sigma$, reducing the probability of matter fluctuations being dense enough to collapse and contribute to $n\left(M\right)$. Equation~\eqref{eq:massfunc} is normalized to the mean density, so in order to satisfy the condition that integrating the mass function over all scales returns $\bar{\rho}$, massive neutrinos amplify $n\left(M\right)$ for very small $M$. However, these are highly non-linear scales that are inaccessible to Stage IV surveys and so of limited interest. The higher threshold for collapse in equation~\eqref{eq:dc} suppresses the number density of high-mass haloes. This will propagate into a reduction of matter power on the weakly non-linear scales corresponding to such objects. The suppression of clustering due to $\Sigma$ increases by a factor of $\sim 5$ with this change to $\delta_c$, although it should be emphasized that the overall effect of massive neutrinos on the mass function remains very small.

The impact of massive neutrinos on the density profile has a less significant (but still non-trivial) contribution to the matter power spectrum compared to changes in the mass function. The collapse overdensity partially controls the formation redshift determining the concentration relation in equation~\eqref{eq:conc}. At late times, regions are more likely to collapse due to the amplitude of matter fluctuations being greater. Increasing $\delta_c$ makes halo formation rarer, favouring later times over earlier times, and therefore lowering $z_f$. The resulting reduction in halo concentration translates to an increased scale radius for fixed $r_v$. Consequently, for a fixed halo mass, the density is suppressed on small scales, with a milder boost on large scales. However, when the reduction of $r_v$ through increasing $\Delta_v$ via $\Sigma$ is also accounted for, this picture is complicated. If we consider this effect in isolation, for fixed mass and concentration, the density on small scales must increase to reflect the fact that haloes are smaller when meeting the increased density threshold for virialization. Interestingly, when the sensitivity to both $\delta_c$ and $\Delta_v$ is included, the two contributions mostly cancel, leaving the halo density profile with minimal sensitivity to the neutrino mass on small scales. The relative insensitivity of the density profile, and more importantly the mass function, highlights, at the outset, the challenges of using the non-linear regime to constrain $\Sigma$.      

\subsection{Baryonic contributions}
\label{subsec:baryonparams}
We adopt the approach of \citetalias{Mead15} to capture the effects of baryonic astrophysics on haloes\footnote{This subsection is a compressed version of a more comprehensive discussion originally presented in \citet{Copeland18}, to which we refer the interested reader. An important addition is the real-space halo density profile with a baryonic feedback modification in equation~\eqref{eq:real_profile_eta0}. This was illustrated in Figure 2 of \citet{Copeland18} but we present the expression for the first time here.}. Internal halo structure relations are modified to include two parameters, $A_B$ (referred to as $A$ in \citetalias{Mead15}) and $\eta_0$, that mimic adiabatic contraction via radiative cooling, and baryonic feedback (from e.g., AGN and supernova sources) respectively. These are fit to a range of simulations with different implementations of baryons (and a dark matter only case) provided by OWLS. Fiducial values, $A_{B,\rm{fid}}=3.13$ and $\eta_{0,\rm{fid}}=0.603$, are determined by fitting to power spectra produced by the emulator code, COSMIC EMU, designed for the high resolution N-body simulations from the Coyote Universe project \citep{Heitmann09, Heitmann10, Lawrence10, Heitmann14}. For this fit, HMCODE achieves $\simeq 5$ percent accuracy for the redshifts and scales of interest to Stage IV surveys. 

Other methods for modelling baryons introduce additional density profiles describing specific physical contributions. An example of an alternative method for parameterizing baryons is that of \citet{Schneider&Teyssier15}, whose `baryon correction model' incorporates star and gas components to their density profiles to reproduce specific physical contributions. This model is used by  \citet{Parimbelli18} in their analysis of neutrinos and baryons. We advocate instead the \citetalias{Mead15} philosophy of making empirical corrections to halo model relations because these are primarily motivated by generating the most accurate power spectrum, which is the underlying probe used for our forecasts. For our purposes we require an accurate understanding of the response of halo structures to baryons rather than a detailed framework for the underlying physics. Similar reasoning has previously been applied by surveys such as KiDS \citep{Hildebrandt17} for using the \citetalias{Mead15} model to account for baryons in their data analyses.

In \citetalias{Mead15}, adiabatic contraction is captured by varying the amplitude, $A_B$, of the concentration factor in equation~\eqref{eq:conc}. Radiative cooling of baryonic matter causes it to cluster, thereby deepening gravitational wells into which dark matter falls, and so leading to the adiabatic contraction of the total matter distribution. On non-linear scales, simulations show that the effect has an impact of several percent \citep{Duffy10, Gnedin11}. This manifests as a reduced scale radius in the halo profile, suppressing the density on large scales and increasing it at smaller scales for a fixed virial radius and halo mass. This is demonstrated in Figure 1 of \citet{Copeland18}, which shows density profiles for a range of $A_B$ values.

Baryonic feedback from sources like AGN and supernovae is far less straightforward to model. Significant energy exchanges with their environments involve isotropic radiative transfer or highly directional jets in the case of the former \citep{Schaye10, vandaalen11, Martizzi14} and multiple sources heating surrounding gas in the latter \citep{Pontzen&Governato12, Lagos13}. In both cases envelopes of gas can expand on scales comparable to the virial radius, so a significant range of scales from subparsecs to megaparsecs are affected. This makes modelling feedback analytically difficult, a challenge that is exacerbated by the effect on haloes depending on mass as well as scale. As simulations have demonstrated \citep[e.g.,][]{Pontzen&Governato12, Teyssier13, Martizzi13}, similar AGN mechanisms for the expulsion of gas from central regions for both lower and higher mass haloes can remove substantial baryonic matter from the former while merely bloating the latter outwards in a less catastrophic process. 

This range of effects is captured in \citetalias{Mead15} by transforming the scale of the window function,
\begin{equation}
u\left(k\mathopen{|}\mathclose M\right)\longrightarrow u\left(\nu^{\eta}k\mathopen{|}\mathclose M\right),
\end{equation}
by the mass-dependent factor, $\nu=\frac{\delta_c}{\sigma\left(M\right)}$. We find that in real space the modified density profile takes the form
\begin{align}
\rho\left(r,M\right) = \frac{\rho_s}{\nu^{3\eta}}\frac{1}{\left(\frac{r}{\nu^{\eta}r_s}\right)\left[1+\left(\frac{r}{\nu^{\eta}r_s}\right)\right]^2}, && r \leq \nu^{\eta}r_v.
\label{eq:real_profile_eta0}
\end{align}
We refer the reader to Figure 2 in \citet{Copeland18}, which shows that for more positive values of $\eta$ higher mass haloes (characterized by $\nu> 1$) are increasingly bloated while lower mass haloes ($\nu < 1$) are left relatively diminished by the outright removal of gas. \citetalias{Mead15} achieve their best power spectra fits, including for dark-matter-only simulations, when using a non-zero value of $\eta$ with a redshift dependence. The parameter was decomposed such that
\begin{equation}
\eta=\eta_0-0.3\,\sigma_8\left(z\right),
\end{equation}
with the constant, $\eta_0$, controlling the strength of the feedback impact on halo structure.

\citet{Copeland18} extend HMCODE by introducing an inner halo core that could parameterize the effects of baryons or, potentially, other physics on the smallest scales of structure. However, we do not include this degree of freedom in this work, and simply use the NFW profile, with $A_B$ and $\eta_0$ as our baryon parameters.  

\begin{figure*}
\centering
\includegraphics[width=\textwidth]{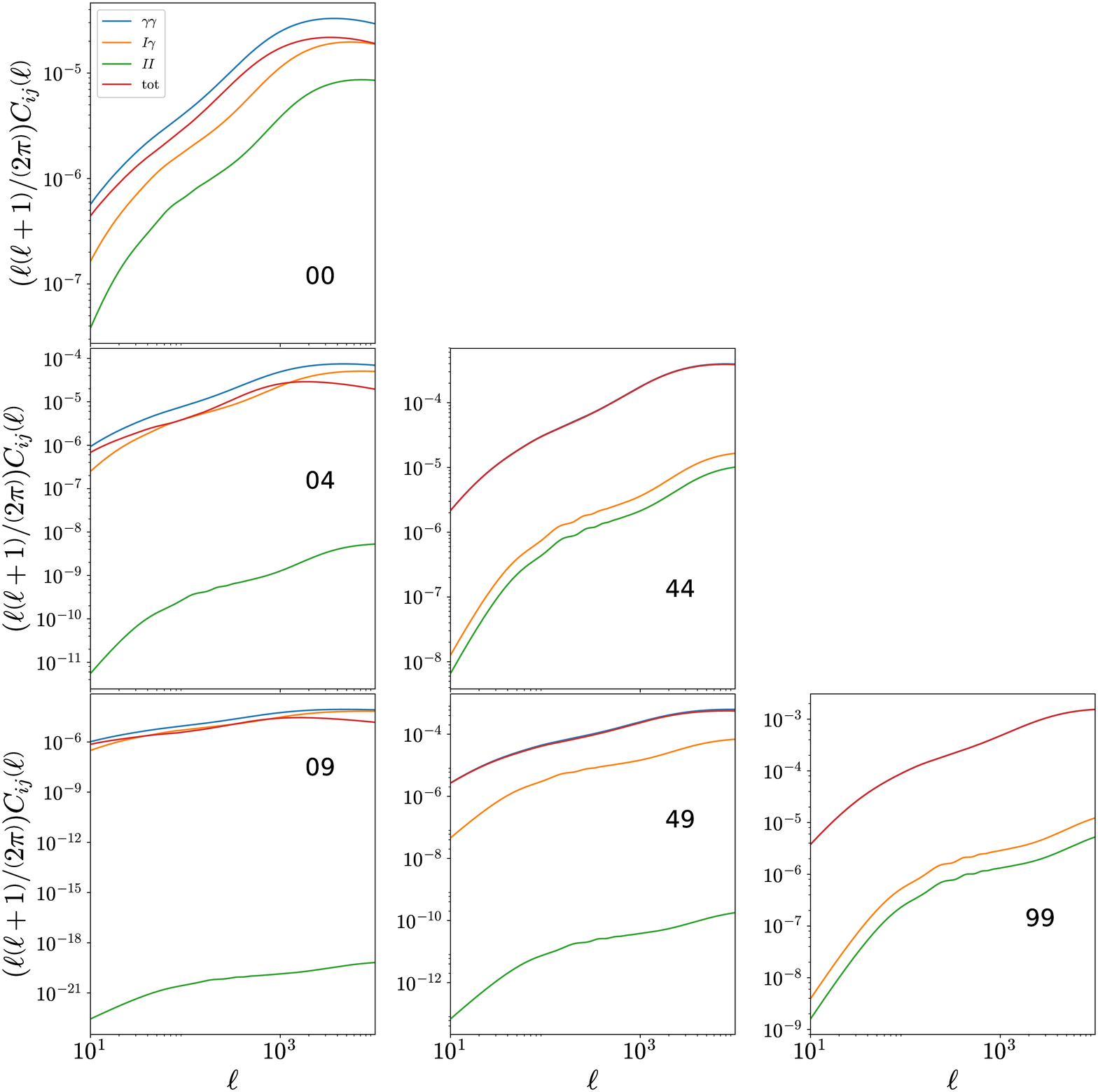}
\caption{Auto- and cross- lensing power spectra in the 00, 44 and 99 redshift bins: $0 < z \leq 0.41$, $0.79 < z \leq 0.89$, $1.52 \leq z \leq 2.0$. Blue: shear-shear power spectrum; orange: absolute value of the IA-shear power spectrum; green: IA-IA power spectrum; red: total power spectrum.}   
\label{fig:lenspower049} 
\end{figure*}

\section{Neutrino Mass Forecasts}
\label{sec:res}
We construct $\left(A_B,\eta_0,\Omega_m,\Omega_b,\Sigma,h,n_s,\sigma_8,w,a_{\rm{IA}}\right)$ Fisher matrices for Euclid-like weak lensing and spectroscopic galaxy surveys. We effectively apply a uniform prior on $\Sigma$ by using the Fisher formalism. This is justified by the fact that different $\Sigma$ values should not significantly change the shape of the multivariate likelihood function. Such a prior can be considered a desirable objective prior that allows the posterior to be determined instead by the likelihood \citep{Heavens&Sellentin18}. The primary goals of this work are to assess the impacts of IA (a phenomenon exclusive to weak lensing) and baryonic effects on neutrino mass forecasts. As the influence of baryons degrades information significantly on scales $k>1 h\, \mathrm{Mpc}^{-1}$ this effect is of far greater concern to weak lensing forecasts than those from galaxy clustering. Hence, our analysis takes the approach of first discussing weak lensing results and the extent of their degradation due to baryons and IA. Information from \textit{Planck} and galaxy clustering is then added to form a more complete picture of the prospects of a Euclid-like survey to measure the neutrino mass with sufficient accuracy to distinguish the hierarchies. 

\vspace{-5mm}

\begin{figure*}
\includegraphics[width=\textwidth]{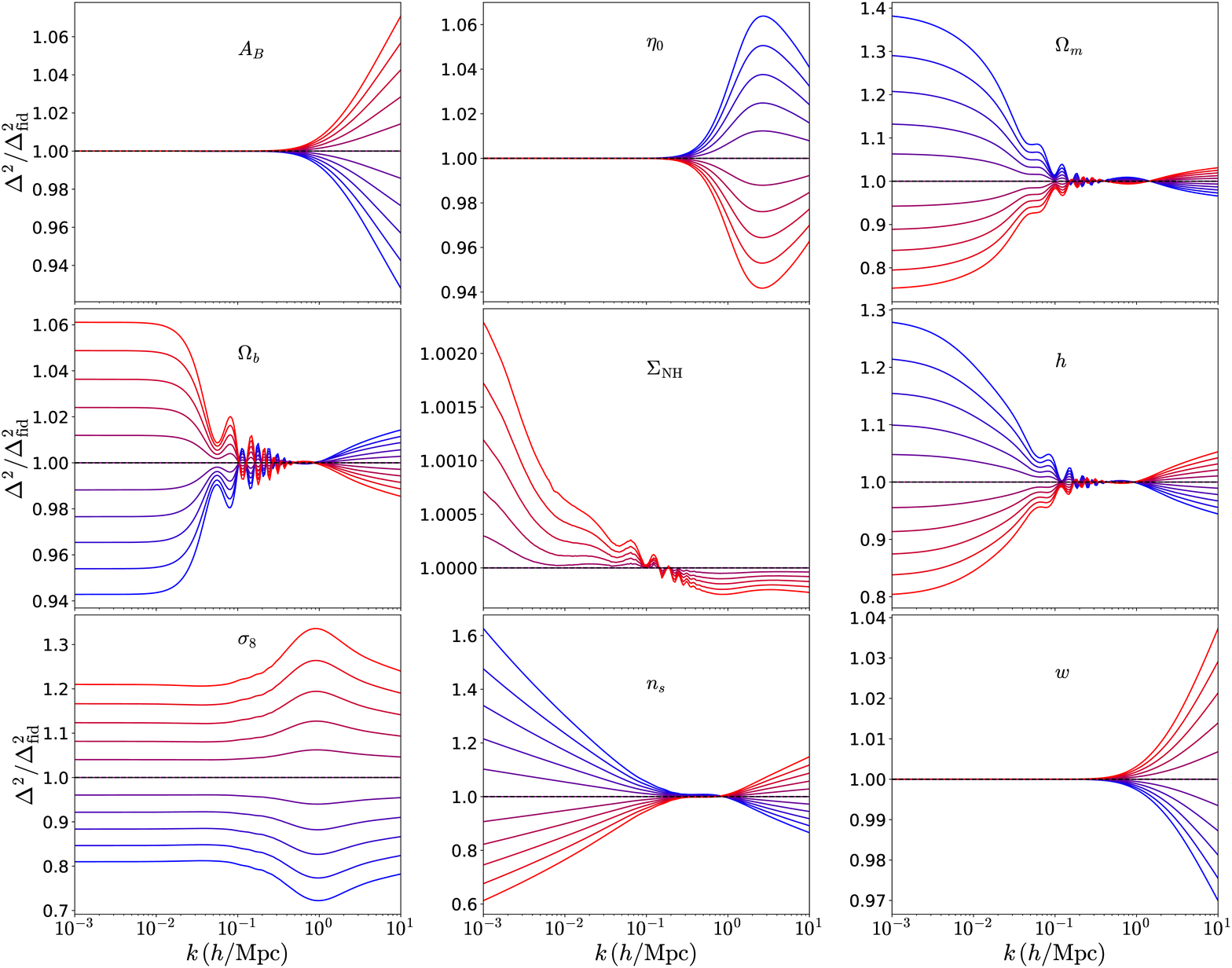}
\caption{The ratio of matter power spectra at $z=0$ using NH for different iterations of parameters in 
$\Theta=\left(A_B,\eta_0,\Omega_m,\Omega_b,\Sigma_{\rm{NH}},h,\sigma_8,n_s,w\right)$, with respect to a fiducial power spectrum computed with parameter values found by \citet{Planck15}. Bluer (redder) curves correspond to lower (higher) values for parameters in the range $0.9\,\Theta_{\rm{fid}} \leq \Theta \leq 1.1\,\Theta_{\rm{fid}}$, except in the case of the neutrino mass parameter, which varies between $\Sigma_{\rm{NH,min}} \leq \Sigma_{\rm{NH}} \leq 1.1\,\Sigma_{\rm{NH,min}}$ with purple curves representing the minimal mass case, $\Sigma_{\rm{NH,min}}=0.06\, \rm{eV}$. \citet{Copeland18} showed similar $\Delta^2\left(k\right)$ response plots but did not include massive neutrinos.}
\label{fig:multipow_matter_z0_nh}
\end{figure*}

\begin{figure*}
\centering
\includegraphics[width=\textwidth]{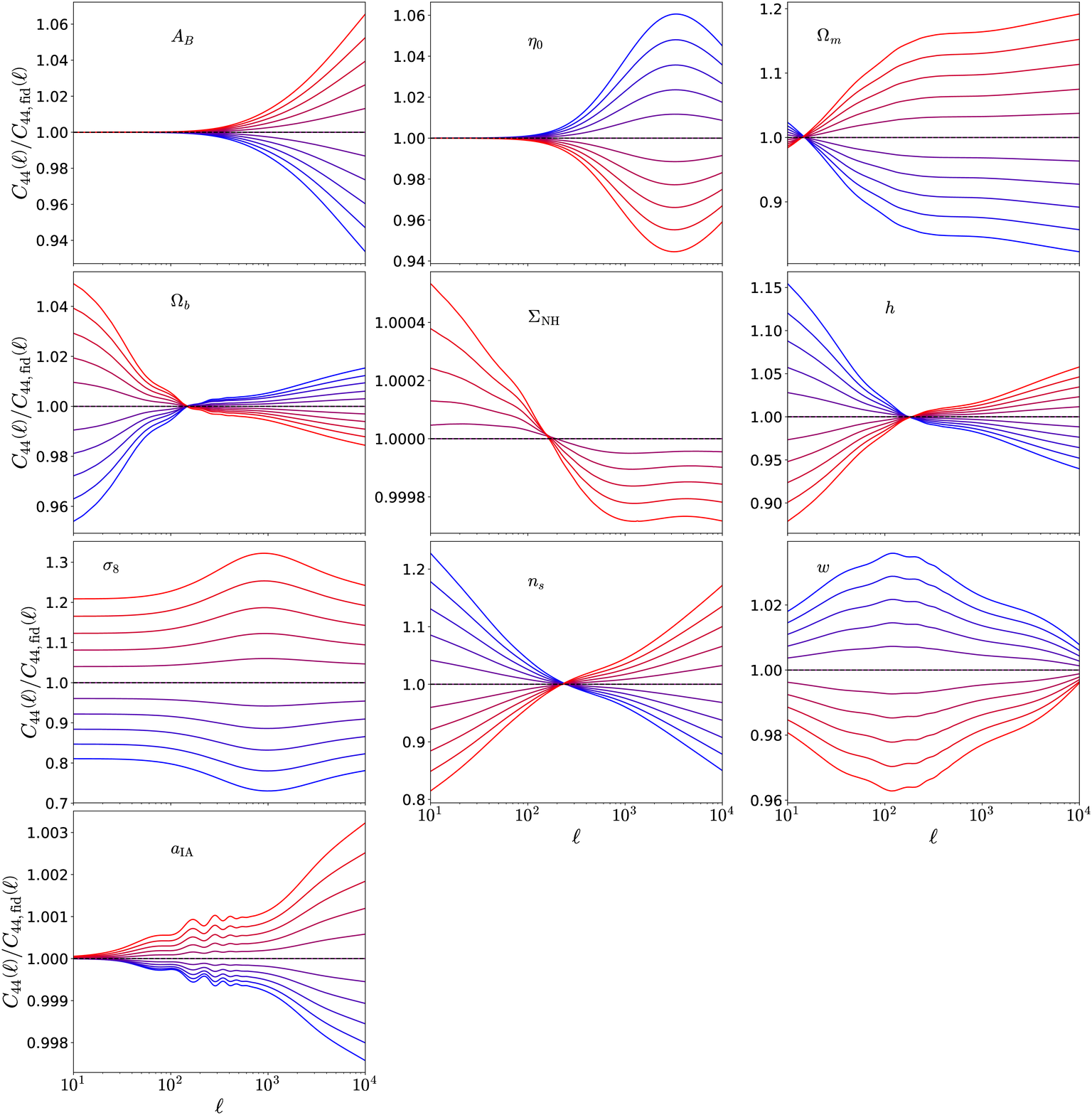}
\caption{Lensing power spectrum responses using the NH in the $ij=44$ redshift bin $0.79 \leq z \leq 0.89$. Blue (red) lines correspond to the lowest (highest) parameter values for $\Theta=\left(A_B,\eta_0,\Omega_m,\Omega_b,\Sigma_{\rm{NH}},h,\sigma_8,n_s,w,a_{\rm{IA}}\right)$ in the range $0.9\Theta_{\rm{fid}} \leq \Theta \leq 1.1\Theta_{\rm{fid}}$, except in the case of the neutrino mass parameter,  which varies between $\Sigma_{\rm{NH,min}} \leq \Sigma_{\rm{NH}} \leq 1.1\,\Sigma_{\rm{NH,min}}$ with purple curves representing the minimal mass case, $\Sigma_{\rm{NH,min}}=0.06\, \rm{eV}$. \citet{Copeland18} showed similar $C_{\ell}$ response plots but did not include massive neutrinos.}
\label{fig:multipow_lensing_nh}
\end{figure*}

\begin{figure*}
\centering
\includegraphics[width=\textwidth]{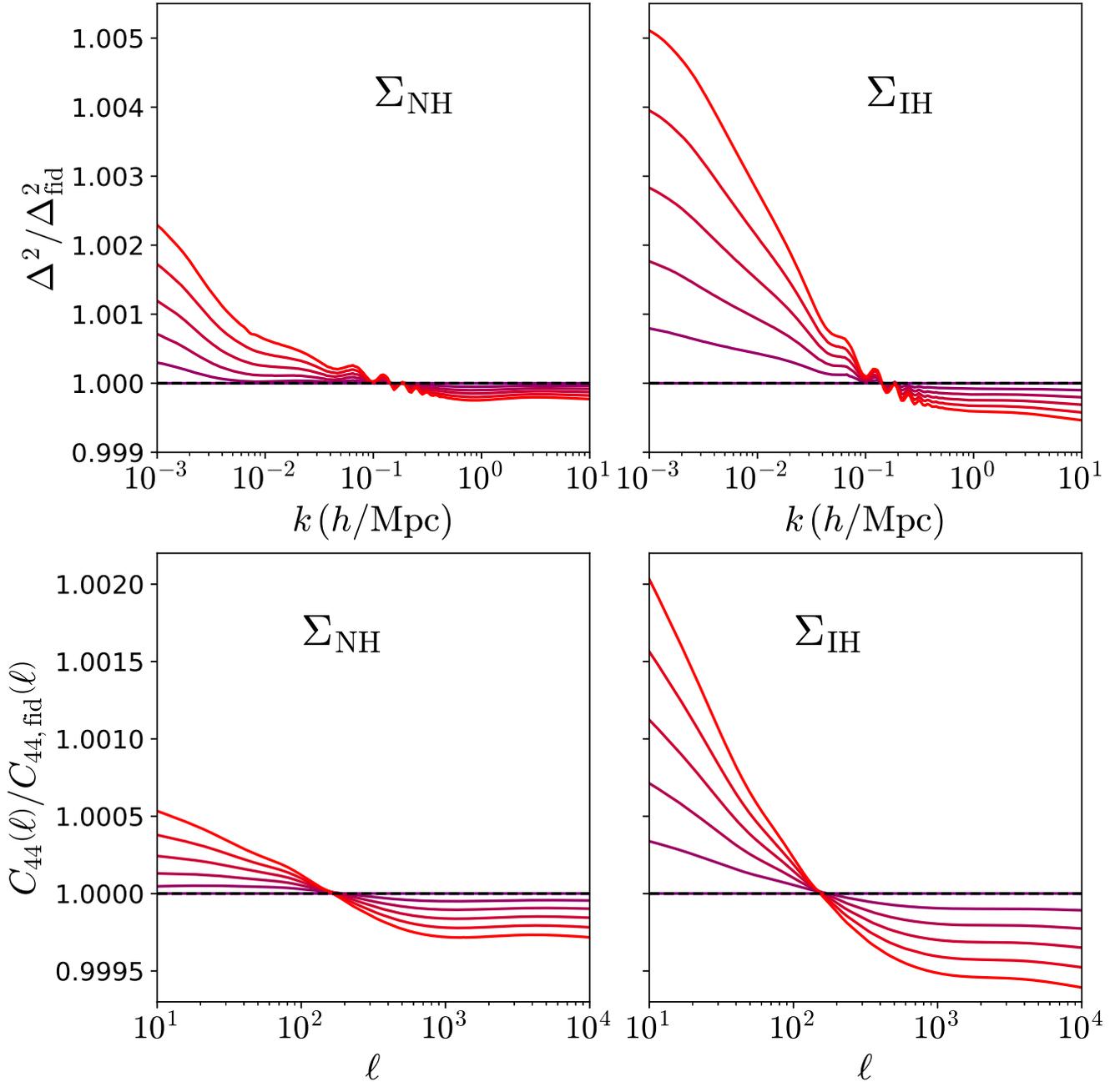}
\caption{Top row: The ratio of matter power spectra at $z=0$ for different iterations of $\Sigma$ using the NH (left) and IH (right), with respect to a fiducial power spectrum computed with parameter values found by \citet{Planck15}. Bottom row: Lensing power spectrum responses for $\Sigma$ using NH (left) and IH (right) in the $ij=44$ redshift auto-bin $0.79 \leq z \leq 0.89$. The mass sum is varied in the range $\Sigma_{\rm{min}} \leq \Sigma \leq 1.1\,\Sigma_{\rm{min}}$ for both hierarchies with purple curves representing the minimal mass cases, $\Sigma_{\rm{NH,min}}=0.06\, \rm{eV}$ and $\Sigma_{\rm{IH,min}}=0.1\, \rm{eV}$.}
\label{fig:multipow_nh_ih_quad}
\end{figure*}

\subsection{Normal and inverted hierarchy results from weak lensing}
\label{subsec:nhihwl}
In Table~\ref{table:cosmicemurange}, we state the weak lensing survey parameters specified by the Euclid survey report \citep{Laurejis11}. $N_{z,\rm{WL}}=10$ redshift bins are chosen in the range $0 \leq z \leq 2$ such that each bin contains an equal number density of galaxies. A large range of scales from $\ell_{\mathrm{min}}=10$ to $\ell_{\mathrm{max}}=5000$ are covered so in practice we compute the summation in equation~\eqref{eq:fishertrace} at logarithmic intervals. We include full confidence ellipse plots for the NH in Appendix~\ref{appendix:completeforecasts}.  The fiducial values of the cosmological parameters in this work are given by the base $\Lambda$CDM Planck TT,TE,EE+lowP likelihood \citep[see Table 4 in][]{Planck15}. For the neutrino masses we use the minimal bounds as fiducial values i.e., $\Sigma_{\rm{NH,fid}}=0.06\, \rm{eV}$ and $\Sigma_{\rm{IH,fid}}=0.1\, \rm{eV}$.
\begin{table}
\centering
\begin{tabular}{|c|c|}
\hline
Parameter & Euclid-like value \\
  \hline 
  $A_{\rm{sky}}$ & $15,000\, \mathrm{deg}^2$ \\
 
  $z_{\rm{min}}$ & $0.$ \\
 
  $z_{\rm{max}}$ & $2.0$ \\
 
$z_{\rm{med}}$ & $0.9$ \\
$N_{z,\rm{WL}}$ & $10$ \\
$n_{\rm{gal}}$ & $30$ gal/arcmin$^2$ \\
$\sigma_z$ & $0.05$ \\
$\sigma_e$ & $0.3$ \\ 
$l_{\rm {min}}$ & $10$ \\  
$l_{\rm {max}}$ & $5000$ \\

  \hline
\end{tabular}
\caption{Survey parameters characterizing a Euclid-like space mission. These include the area of sky covered, $A_{\rm{sky}}$, the survey redshift limits, the median redshift value, $z_{\rm{med}}$, the number of bins, $N_{z,\rm{WL}}$, the number density of surveyed galaxies, $n_{\rm{gal}}$, the photometric redshift error, the intrinsic ellipticity dispersion and the range of accessible wavenumbers.}
\label{table:cosmicemurange}
\end{table}

One can derive significant insight into the results of a Fisher forecast by examining the response of the power spectra to varying parameters across relevant scales. This diagnostic can quickly indicate which scales most Fisher information comes from or is potentially lost due to degenerate power spectrum responses between parameters. In Figures~\ref{fig:multipow_matter_z0_nh} and~\ref{fig:multipow_lensing_nh} we show, for the normal hierarchy, responses of $\Delta^2\left(k\right)$ at $z=0$ and $C_{\ell}$ respectively to varying each parameter with respect to its fiducial value while fixing the other parameters. In \S~\ref{subsec:baryonimpact} we discuss the responses to the baryon parameters. $\Delta^2\left(k\right)$ responses to most parameters exhibit nodes at the scale corresponding to $\sigma_8$, at which we normalize power. For example, the response to $\Sigma$ is an enhancement of power on linear scales in order to satisfy power being damped in the non-linear regime due to free streaming. The impact of neutrinos on these scales is small, determined by the limited sensitivity of the mass function and the halo structure parameters, $\delta_c$ and $\Delta_v$, to $\Sigma$ (discussed in detail in \S~\ref{subsec:modellingneutrinos}). Most of the $C_{\ell}$ responses reflect the corresponding $\Delta^2\left(k\right)$ responses, when accounting for redshift-dependent effects along the line of sight. However, parameters that affect the lensing weight function through their impact on cosmological distances can exhibit significantly different $\Delta^2\left(k\right)$ and $C_{\ell}$ responses. For example, the competing influences of geometry and growth are apparent in the $w$ responses, which \citet{Copeland18} examine in depth.         

We compare the $\Sigma$ responses for $\Delta^2\left(k\right)$ and $C_{\ell}$ for the NH and IH directly in Figure~\ref{fig:multipow_nh_ih_quad}. $\Sigma_{\rm{IH,min}}=0.1\, \rm{eV}$ is less than double $\Sigma_{\rm{NH,min}}=0.06\, \rm{eV}$ but the IH exhibits power responses that are more than twice as large as the NH responses. The leading order cause of this is likely a dependence of power on $\Sigma$ that grows with the mass sum from a quadratic minimum which is reached when $\Sigma$ approaches zero.

An additional effect to account for is the influence of the neutrino free-streaming scale, $k_{\rm{fs}}$, which increases with mass. There are different $k_{\rm{fs}}$ for each mass eigenstate in a hierarchy but, within the approximation of the NH as $\left\{m_1 \approx m_2 \approx 0\, \rm{eV}, m_3 \approx 0.06\, \rm{eV}\right\}$ and the IH as $\left\{m_1 \approx m_2 \approx 0.05\, \rm{eV}, m_3 \approx 0\, \rm{eV}\right\}$, the most relevant free-streaming scales satisfy $k_{\rm{fs,IH}} < k_{\rm{fs,NH}}$. Fixing $\sigma_8$ means that, for increases in $\Sigma$, large-scale power is enhanced relative to the fiducial case to satisfy the suppression of power by neutrinos on small scales. There is a similar contribution from the peak of $P\left(k\right)$ shifting with a decrease in the redshift of matter-radiation equality, which depends on $\Omega_c$ (which decreases to keep $\Omega_m$ constant for increasing $\Sigma$) due to the relativistic nature of neutrinos during this epoch. As the effect of free-streaming neutrinos arises at smaller $k$ for the IH, the relative enhancement of the large-scale IH power is greater than for the NH. Similarly, the responses for the IH are more sensitive to the same fractional change in $\Sigma$ compared to the NH, which we see reflected in Figure~\ref{fig:multipow_nh_ih_quad}.  

When propagating $\Delta^2\left(k\right)$ into $C_{\ell}$, the lowest $\ell$ for a Euclid-like survey correspond to $k$ where the matter power response is weak. As a result, the strongest linear responses for $C_{\ell}$ in Figure~\ref{fig:multipow_nh_ih_quad} are several factors smaller than for $\Delta^2\left(k\right)$. Other parameter responses generally experience significantly more modest changes from one probe to another. It should be noted that the background geometry is not sensitive to increasing $\Sigma$ when neutrinos are non-relativistic; the angular distances entering equation~\eqref{eq:convergencepower} depend on $\Omega_m$, which is kept constant. 

A consequence of the weak lensing probe only being sensitive in the linear regime to scales where the neutrino response is most limited is that the magnitude of the $C_{\ell}$ response is similar between the linear and non-linear regimes. The response for the latter is due mainly to the modest increases in $\delta_c$ and $\Delta_v$. This highlights the importance of pursuing robust physical modelling for neutrino effects on structure at these scales because they are as sensitive to $\Sigma$ as linear scales.   

Weak lensing forecasts of $\Sigma$ are particularly challenging due to this limited sensitivity and the degeneracies apparent between $\Sigma$ and a range of other parameters on non-linear scales. A well-known example is the degeneracy with $\sigma_8$, in which both exhibit a `spoon'-shaped response \citep[see e.g.,][]{Massara14}. An increase in $\Sigma$ most strongly affects high mass haloes, which dominate over small mass haloes on intermediate scales. Figure~\ref{fig:massfunc} shows that in this region higher $\Sigma$ reduces the linear clustering of matter, and therefore the number of high mass haloes that can form, lowering the halo mass function and hence the non-linear matter power. Raising the collapse and virialization density thresholds with $\Sigma$ deepens the resulting dip in the power response and shifts it to larger scales associated with higher masses. At higher $k$ the impact of $\Sigma$ is somewhat decreased and becomes less scale-dependent. 

The bump in the $\sigma_8$ response is qualitatively similar. By altering the clustering amplitude, a halo mass-dependent and hence scale-dependent power response is induced through the change in the peak height, $\nu = \delta_c/\sigma\left(M,z\right)$, a quantity that is also sensitive to $\Sigma$. The response becomes roughly scale-independent on smaller scales. This is similar to the $\Sigma$ response but with the opposite sign, enhancing rather than suppressing power. As most information comes from small scales this degeneracy is significant.   

\begin{figure}
\includegraphics[width=\columnwidth]{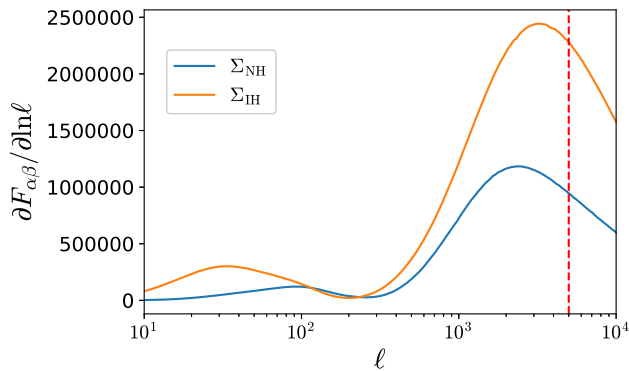}
\caption{Contributions to the $\Sigma\Sigma$ Fisher information element for the NH (blue) and IH (orange) at each $\ell$-mode, weighted by $\ell$. The vertical line (red, dashed) corresponds to the conventional cutoff for Euclid-like surveys at $\ell_{\rm{max}}=5000$. The Fisher derivative with respect to $\ln \ell$ is used in order to convey the integration area for the logarithmically spaced $\ell$-range.}
\label{fig:sensitivity_neu} 
\end{figure}

In Figure~\ref{fig:sensitivity_neu} the contribution per $\ln \ell$ to the Fisher matrix for the mass sum is shown for the NH and IH. This effectively weights the Fisher information by the signal-to-noise in each $\ell$ mode, which increases with $\ell$. The greater number of independent modes in the non-linear regime, and consequently lower noise, results in most of the available information being drawn from these scales. The greater sensitivity of information at high $\ell$ highlights the limitations on achieving strong constraints, given the limited neutrino influence on non-linear power. The IH contributes almost twice as much information as the NH at the peak sensitivity. 

The main difference between the hierarchies in their information contribution is from different logarithmic derivatives of $C_{\ell}$. These are partially predicted by the power responses. On non-linear scales the $C_{\ell}$ response for the IH is larger than double the NH for the same fractional change in $\Sigma$. As $\Sigma_{\rm{IH,min}}=0.1\, \rm{eV}$ is less than double $\Sigma_{\rm{NH,min}}=0.06\, \rm{eV}$, this leads to $\partial\ln C_{\ell}/\partial\Sigma_{\rm{IH}}$ being sufficiently larger than $\partial\ln C_{\ell}/\partial\Sigma_{\rm{NH}}$ that the Fisher sensitivity, which depends quadratically on the logarithmic derivative, is up to twice as large. 

\begin{figure*}
\centering
\includegraphics[width=\textwidth]{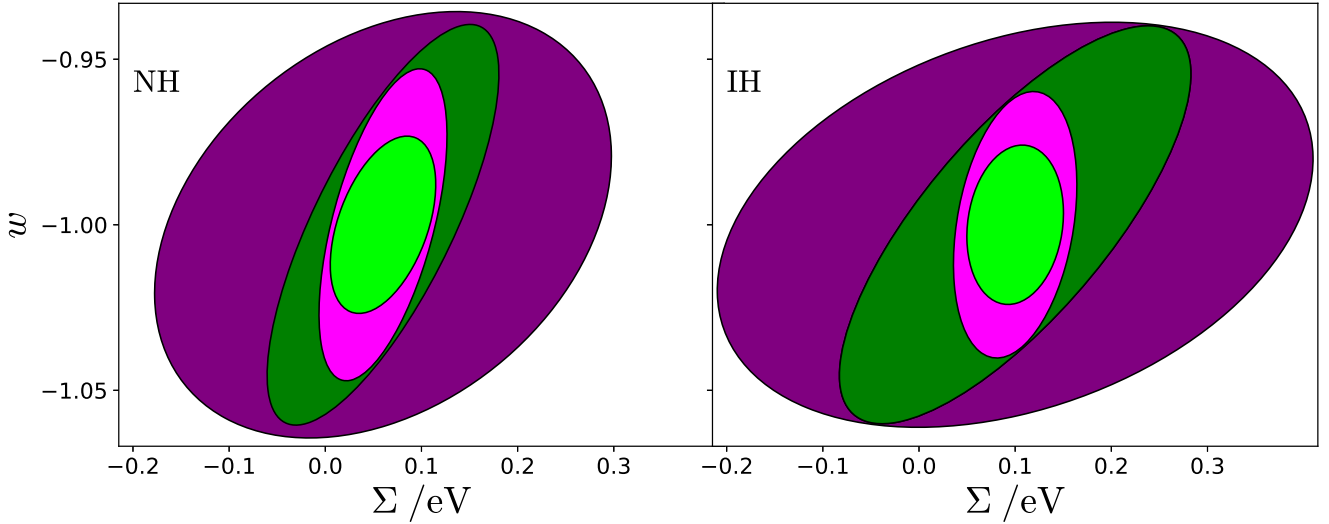}
\caption{1-$\sigma$ $w$-$\Sigma$ confidence ellipses for the normal (left; $\Sigma_{\rm{fid}}=0.06\, \rm{eV}$) and inverted hierarchy (right; $\Sigma_{\rm{fid}}=0.1\, \rm{eV}$) for a Euclid-like weak lensing survey with all parameters in $\Theta=\left(A_B,\eta_0,\Omega_m,\Omega_b, h,\sigma_8,n_s,w,a_{\rm{IA}}\right)$ marginalized over (purple); and with the baryon parameters, $A_B$ and $\eta_0$, fixed to their fiducial values (dark green). \textit{Planck} CMB priors are included for the magenta (light green) ellipses with (without) baryon marginalization.}
\label{fig:barforecasts}
\end{figure*}

\begin{table*}
\centering
\begin{tabular}{c|c|c|c|c|c|c}
  \hline
  & $\sigma_{\rm{WL}}/\rm{eV}$  & $\sigma_{\rm{WL}}/\rm{eV}$ & $R_{B\rm{,WL}}$ & $\sigma_{\rm{WL+CMB}}/\rm{eV}$  & $\sigma_{\rm{WL+CMB}}/\rm{eV}$ & $R_{B\rm{,WL+CMB}}$ \\
& (ex bar.) & (inc. bar.) & & (ex bar.) & (inc. bar.) &   \\
\hline
$\Sigma_{\rm{NH}}$ & 0.079 & 0.156 & 1.97 & 0.036 & 0.044 & 1.21    \\
$\Sigma_{\rm{IH}}$ & 0.120 & 0.204 & 1.70 & 0.033 & 0.042 & 1.28 \\
\hline
\end{tabular}
\caption{1-$\sigma$ error forecasts for a Euclid-like survey of the neutrino mass sum in the normal and inverted hierarchies, without and including marginalization over baryon parameters and the addition of priors on the $\Lambda$CDM cosmological parameters from \textit{Planck} CMB measurements. We also include the response factors, $R_B$, to including baryon marginalization.}
\label{table:priorstable}
\end{table*}

In Figure~\ref{fig:barforecasts} we show NH and IH confidence ellipses for $w$-$\Sigma$ for a Euclid-like weak lensing (WL) survey. We do not expect to constrain $\Sigma$ from WL alone, but we are interested in using it as a baseline probe, through which we can examine both the impact of systematics like baryons and IA and the improvements available from adding CMB and galaxy clustering information.  Before marginalizing over baryons, we find WL errors of $\sigma_{\Sigma_{\rm{NH}}}=0.079\, \mathrm{eV}$ and $\sigma_{\Sigma_{\rm{IH}}}=0.120\, \mathrm{eV}$ (see by Table~\ref{table:priorstable}). We treat the minimal mass for each hierarchy ($\Sigma_{\rm{NH}}=\Sigma_{\rm{NH,min}}=0.06\, \mathrm{eV}$ and $\Sigma_{\rm{IH}}=\Sigma_{\rm{IH,min}}=0.1\, \mathrm{eV}$), as the fiducial cases in this work, so distinguishing the hierarchies requires that error forecasts fall within the threshold, $|\Delta\Sigma_{\rm{min}}| = |\Sigma_{\rm{NH,min}} - \Sigma_{\rm{IH,min}}| = 0.04\, \rm{eV}$. It should be noted that any measurement of $\Sigma \geq 0.1\, \rm{eV}$ permits either hierarchy regardless of the strength of the error constraint. The errors we find are therefore too large for a distinction to be made. They would also be too large to permit any positive detection of neutrinos in the scenarios where the mass sum is given by the minimal mass for each hierarchy.

The latter issue can be alleviated by adding priors on cosmology from sources like galaxy clustering or the early Universe. We focus here on the latter through \textit{Planck} CMB anisotropy measurements, while impact of galaxy clustering is discussed in detail in S~\ref{subsec:gc}. The CMB provides a wealth of information on the matter energy-density and the geometry of the Universe, which is one of the main contributions to LSS signals. We use the publicly available MCMC chains for the base $\nu\Lambda$CDM combined TT, TE and EE power spectra \citep[see][]{Planck15} to construct a covariance matrix which is inverted into a Fisher matrix, $F_{\rm{CMB}}$, that encapsulates the information available from the CMB. This is added to the Euclid-like WL information such that the total Fisher information is
\begin{equation}
F_{\rm{tot}} = F_{\rm{WL}} + F_{\rm{CMB}}. 
\end{equation}
As can be seen in Figure~\ref{fig:barforecasts} there is a substantial improvement to $\Sigma$ forecasts for the NH and IH, with errors reduced below the distinction threshold to $\sigma_{\Sigma_{\rm{NH}}}=0.036\, \mathrm{eV}$ and $\sigma_{\Sigma_{\rm{IH}}}=0.033\, \mathrm{eV}$. This is partly a result of parameter degeneracies being broken, for example the CMB provides strong constraints on $\sigma_8$ which propagate through to our final $\Sigma$ forecasts. It should be noted that this requires CMB experiments to accurately measure the reionization optical depth, $\tau$, as it is degenerate with the primordial power amplitude. We marginalize over $\tau$, along with a range of other parameters, when constructing $F_{\rm{CMB}}$. Cosmological neutrino measurements stand to benefit significantly from future 21 cm experiments that will probe the reionization epoch with high precision \citep{Allison15, Lui16}. However, this work does not extend to examine this in detail.

The CMB prior provides a significant contribution to our forecasts by breaking parameter degeneracies and, as it is the same for both hierarchies, it leads to less difference between the NH and IH for the WL+CMB forecasts. This is reflected in the weaker WL constraint, $\sigma_{\Sigma_{\rm{IH}}}$, which improves by 72\%, benefiting significantly more than $\sigma_{\Sigma_{\rm{NH}}}$, which improves by 54\% (see Table~\ref{table:priorstable}). Our results are comparable to findings from \citet{Audren13} who use an MCMC approach to explore the degenerate hierarchy case. However, even these improvements would only achieve a close to 2$\sigma$ detection of $\Sigma$ in the case of the NH (although the IH case fares better with a 3$\sigma$ detection) and a roughly 1$\sigma$ determination of the hierarchy.   
 
\subsection{Impact of baryons}
\label{subsec:baryonimpact}
When the baryon systematic is accounted for, the prospects of using weak lensing to accurately distinguish hierarchies recede. Figures~\ref{fig:multipow_matter_z0_nh} and~\ref{fig:multipow_lensing_nh} include the $\Delta^2\left(k\right)$ and $C_{\ell}$ responses to the adiabatic contraction parameter, $A_B$, and the baryonic feedback parameter, $\eta_0$. \citet{Copeland18} discuss these in detail while here we highlight key features. The boost to non-linear power from increasing $A_B$ is a reflection of the enhanced halo density profiles in this regime \citep[see Figure 1 of][]{Copeland18}. The response to $\eta_0$ exhibits a peak, which is a more subtle consequence of capturing the effects of baryonic feedback over a range of mass and spatial scales. The bloating of higher-mass haloes is the dominant influence compared to the reduction effect experienced by lower-mass haloes, so decreasing $\eta_0$ generates a net increase of $\Delta^2\left(k\right)$. The peak occurs at deeper non-linear scales for higher redshifts, reflecting the evolution of halo populations. 

Baryons and neutrinos both impact the amplitude of the matter power spectrum on small scales, from which most information is drawn, leading to significant degeneracies between both baryon parameters and $\Sigma$. For example, the mass-dependent response for $\eta_0$ emerges from its effect on the peak height which, as discussed above, determines non-linear halo properties in a similar manner to the neutrino mass sum. In the \citetalias{Mead16} model, the peak height depends on $\Sigma$ directly. The total baryon impact on the $\Sigma$ forecast is shown in Figure~\ref{fig:barforecasts} to be severe. The errors for NH and IH double for the former and increase by 70\% for the latter. The greater sensitivity of $\Sigma\Sigma$ Fisher information for the IH relative to the NH on non-linear scales where the baryons limit information results in this comparatively lesser degradation.  

In \citet{Copeland18} we explored several strategies for mitigating baryon impacts on dark energy forecasts. These included increasing the Euclid-like survey limit from $\ell_{\rm{max}}=5000$ to $\ell_{\rm{max}}=10000$ to access information from modes deeper in the non-linear regime. The relative degradation of constraints on the dark energy equation of state increases, although the absolute constraints do improve. For $\Sigma$ we find that baryon degradation is reduced, to 84\% and 61\% for the NH and IH respectively. Figure~\ref{fig:sensitivity_neu} shows that there is a repository of potential Fisher information on $\Sigma$ at higher $\ell$. Our response plots suggest that the degeneracies between $A_B$, $\eta_0$ and $\Sigma$ may be less severe for very high $\ell$ where the neutrino impact on very small haloes approaches scale-independence \citep{Massara14} in contrast to the baryons. This may explain why the degradation is reduced, despite the increasing influence of baryons on non-linear scales overall, but ultimately this improvement is limited.  

We also explored the gains from adding Fisher information from an external baryon source (e.g., new simulations or improved observations) which in general does mitigate the baryon impact on $\Sigma$, but to remove it as a statistically significant systematic requires more baryon information than could realistically be accessed at present. Therefore, we focus on the most promising source of baryon mitigation from \citet{Copeland18}, the constraints on cosmology from the CMB \citep{Planck15}. These propagate through the Fisher analysis, contributing to the breaking of parameter degeneracies. This limits the baryon degradation, and in the case of neutrinos we find a significant improvement with the impacts of marginalizing over baryons reduced to 21\% and 28\% for the NH and IH. Although the IH still has a smaller $\Sigma$ error it now experiences greater relative baryon degradation than the NH when CMB priors are included. Despite the improvements, the baryon impact is detrimental for the weak lensing probe, raising the WL+CMB errors beyond the 1$\sigma$ hierarchy distinction threshold, $|\Delta\Sigma_{\rm{min}}|$, to $\sigma_{\Sigma,\rm{NH}}=0.044$ and $\sigma_{\Sigma,\rm{IH}}=0.042$. These would constitute confidence in a positive detection of massive neutrinos at the 1$\sigma$ and 2$\sigma$ level for $\Sigma_{\rm{NH}}$ and $\Sigma_{\rm{IH}}$ respectively. It will therefore be important to include additional information from other sources, such as galaxy clustering.

\begin{figure*}
\centering
\includegraphics[width=\textwidth]{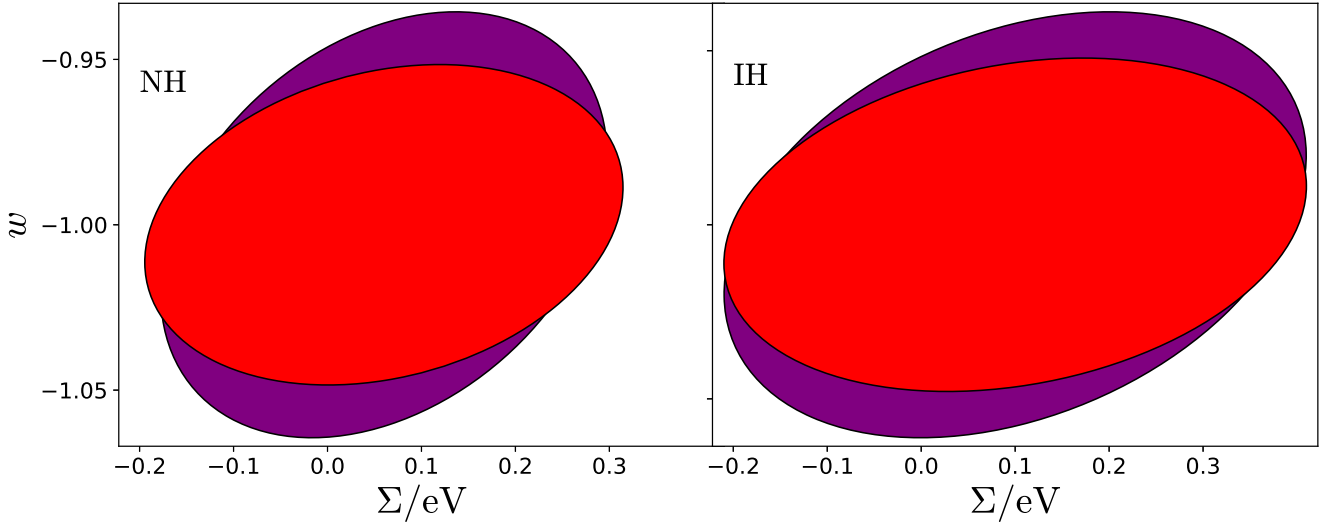}
\caption{1-$\sigma$ $w$-$\Sigma$ confidence ellipses for the normal (left; $\Sigma_{\rm{fid}}=0.06\, \rm{eV}$) and inverted hierarchy (right; $\Sigma_{\rm{fid}}=0.1\, \rm{eV}$) for a Euclid-like weak lensing survey. Purple: intrinsic alignments included in the modelling via the amplitude parameter, $a_{\rm{IA}}$; red: intrinsic alignments not included. All other cosmological parameters in $\Theta=\left(A_B,\eta_0,\Omega_m,\Omega_b, h,\sigma_8,n_s,w\right)$ are marginalized over in both cases.}
\label{fig:aiaforecasts}
\end{figure*}

\vspace*{-5mm}
\subsection{Including intrinsic alignments}
Biased parameter estimations from failing to account for intrinsic alignments have been extensively studied but it is also important to consider the impact on forecasts of including IA in the $C_{\ell}$ signal and then marginalizing over $a_{\rm{IA}}$. Our results are broadly consistent with the predictions of \citet{Krause16} for a Euclid-like survey, with the caveat that we do not extend our analysis to marginalizing over nuisance parameters governing a potential luminosity scaling of the IA amplitude, and so we report slightly less significant impacts. We show, in Figure~\ref{fig:aiaforecasts}, a direct comparison between hierarchies for the $w$-$\Sigma$ forecasts with and without IA. 

Our results indicate that the impact is small for most parameters, including $\Sigma$. This can be understood by considering the $C_{\ell}$ responses with and without IA for different redshift bin combinations. The $C_{\ell}^{\rm{tot}}$ responses for the $ij=44$ $z$-bin ($0.79 \leq z \leq 0.89$) autocorrelation in Figure~\ref{fig:multipow_lensing_nh} are sufficiently similar to the corresponding $C_{\ell}^{\gamma\gamma}$ responses that it is not useful to distinguish between them by explicitly including the latter. This domination of the shear-shear signal is most extreme for auto-correlations in the highest redshift bins, as Figure~\ref{fig:lenspower049} shows. However, for cross-bin correlations the IA-shear signal becomes important as it arises from the alignment of foreground galaxy ellipticities with background lens potentials. This term is negative and large enough to provide significant cancellation to the shear-shear signal in these cases. In Appendix~\ref{appendix:09ia}, we include Figure~\ref{fig:09ia} to illustrate the impact on the $C_{\ell}^{\rm{tot}}$ and $C_{\ell}^{\gamma\gamma}$ responses for the $ij=09$ cross-correlation. As the IA only significantly affect the cross-power between widely separated bins or the auto-power for low redshift bins, and most of the lensing signal comes from the minimally impacted high redshift bin auto-spectra, the $\Sigma$ Fisher information is similar between cases with and without IA. Any degradation of forecasts would therefore come from degeneracies between $\Sigma$ and $a_{\rm{IA}}$. Figures~\ref{fig:multipow_lensing_nh} and~\ref{fig:09ia} show significantly different power responses to these parameters. When combined with the fact that only a small subset of auto- and cross-power spectra are significantly sensitive to the IA, we would therefore expect the impact on constraints to be small.  

For both the NH and IH cases, the IA signal adds limited information on all parameters other than $w$ and $\Omega_m$. The determinant of the Fisher matrix, which effectively quantifies the total available information, is reduced by the inclusion of IA. However, when the inversion to a parameter covariance matrix is performed this manifests as some individual errors (and parameter correlations) decreasing while others increase. Neutrinos have a stronger effect on the power spectrum for the IH compared to the NH, and we find an even smaller, almost negligible, impact from IA on the $\Sigma_{\rm{IH}}$ forecast than for $\Sigma_{\rm{NH}}$. This suggests there is some sensitivity to the fiducial values of the parameters chosen. In turn this implies non-Gaussianity in the posterior, which would limit the validity of the Fisher approximation. However, the impact of IA for neutrino mass forecasts is overall of little concern, with only small improvements and degradations. The main systematic that must be mitigated remains the baryonic phenomena.

\begin{figure*}
\centering
\includegraphics[width=\textwidth]{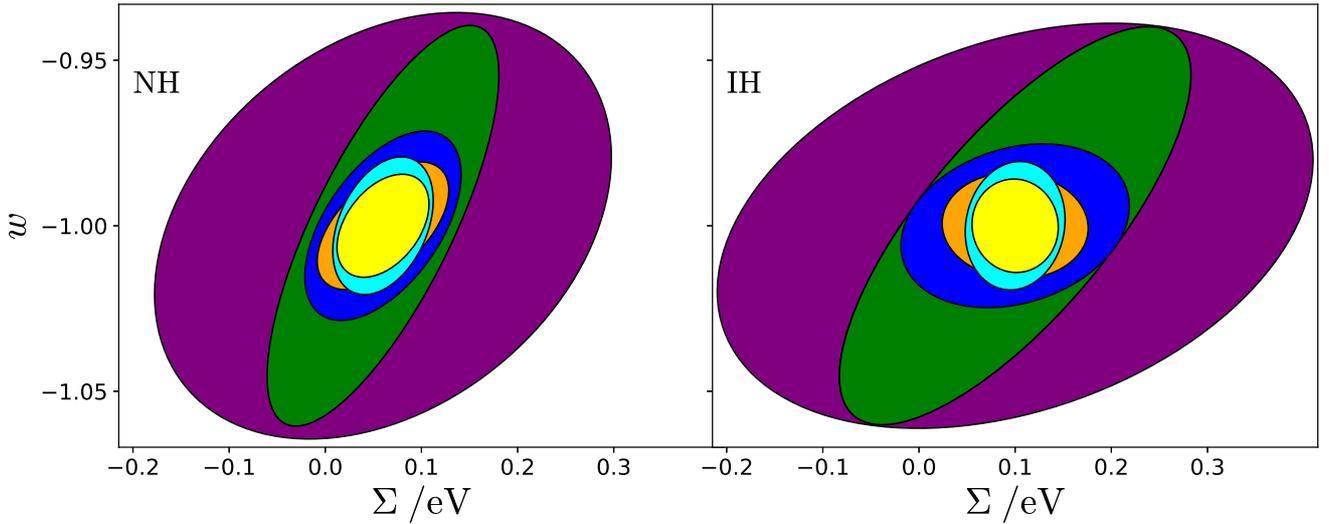}
\caption{1-$\sigma$ $w$-$\Sigma$ confidence ellipses for the normal (left; $\Sigma_{\rm{min}}=0.06\, \rm{eV}$) and inverted hierarchy (right; $\Sigma_{\rm{min}}=0.1\, \rm{eV}$) with all parameters in $\Theta=\left(A_B,\eta_0,\Omega_m,\Omega_b, h,\sigma_8,n_s,w,a_{\rm{IA}}\right)$ marginalized over (purple: weak lensing only; blue: weak lensing and galaxy clustering; cyan: weak lensing, galaxy clustering and CMB priors combined); and with the baryon parameters, $A_B$ and $\eta_0$, fixed to their fiducial values (dark green: WL only; orange: WL+GC; yellow: WL+GC+CMB).}
\label{fig:barforecasts_gc}
\end{figure*}

\begin{table*}
\centering
\begin{tabular}{c|c|c|c|c|c|c|c|c|c|}
  \hline
  & $\sigma_{\rm{WL}}/\rm{eV}$  & $\sigma_{\rm{WL}}/\rm{eV}$ & $R_{B\rm{,WL}}$ & $\sigma_{\rm{WL+GC}}/\rm{eV}$  & $\sigma_{\rm{WL+GC}}/\rm{eV}$ & $R_{B\rm{,WL+GC}}$ & $\sigma_{\rm{WL+GC+CMB}}/\rm{eV}$  & $\sigma_{\rm{WL+GC+CMB}}/\rm{eV}$ & $R_{B\rm{,WL+GC+CMB}}$ \\
& (ex bar.) & (inc. bar.) & & (ex bar.) & (inc. bar.) & & (ex bar.) & (inc. bar.)   \\
\hline

$\Sigma_{\rm{NH}}$ & 0.079 & 0.156 & 1.97 & 0.045 & 0.053 & 1.19 & 0.031 & 0.034 & 1.09   \\
$\Sigma_{\rm{IH}}$ & 0.120 & 0.204 & 1.70 & 0.050 & 0.078 & 1.57 & 0.030 & 0.034 & 1.16 \\
\hline
\end{tabular}
\caption{1-$\sigma$ error forecasts for a Euclid-like survey of the neutrino mass sum in the normal and inverted hierarchies, without and including marginalization over baryon parameters and the addition of priors from galaxy clustering and \textit{Planck} CMB measurements. We also include the response factors, $R_B$, to including baryon marginalization.}
\label{table:priorstable_gc}
\end{table*}     

\subsection{Galaxy clustering}
\label{subsec:gc}
We combine our WL+CMB forecasts with information from galaxy clustering (GC), which is derived from BOSS data at low redshifts ($0.2 < z \leq 0.75$) and from a Euclid-like spectroscopic galaxy survey at higher redshifts ($0.75 < z \leq 2.05$). We show, in Figure~\ref{fig:barforecasts_gc}, confidence ellipses comparing the NH and IH with and without baryon marginalization for three cases: weak lensing only; and weak lensing and galaxy clustering; weak lensing, galaxy clustering and CMB priors combined. Table~\ref{table:priorstable_gc} contains the corresponding 1-$\sigma$ errors and the baryon degradation factors in each case. Neutrinos mainly impact the galaxy power spectrum through the matter power spectrum. The power sensitivity to $\Sigma$ is greater for the IH than the NH, leading to more substantial improvements to forecasts for the former. However, both hierarchies experience a significant reduction in errors, with combined WL+GC+CMB probes achieving $\sigma_{\Sigma_{\rm{NH}}}=0.034\, \rm{eV}$ and $\sigma_{\Sigma_{\rm{IH}}}=0.034\, \rm{eV}$, when accounting for baryons. The degradation due to baryons is also approximately halved in the combined case. Again, even though the IH experiences a far greater improvement overall, the relative degradation is almost double that of the NH at 16\% and 9\% respectively. Interestingly, when examining the impact of adding GC to WL without the inclusion of CMB priors, we see that the NH degradation is reduced by a greater fraction than that of the IH. 

The inclusion of BOSS data on low redshift clustering was found to have a significant impact on galaxy clustering constraints compared to the Euclid-like only case. However, the combined WL+GC errors are only mildly improved by $\sim{5}$-10\% while there is no appreciable change for the WL+GC+CMB errors. Therefore, in our final analysis the dominant information contribution from galaxy clustering is provided by the Euclid-like forecasts rather than BOSS data. Ultimately, our combined forecast constraints represent a nearly 2$\sigma$ (3$\sigma$) detection of $\Sigma_{\rm{NH}}$ ($\Sigma_{\rm{IH}}$) but are not significantly lower than the minimum threshold required to achieve a distinction between hierarchies. 

Our results for a GC+CMB combination are broadly comparable with recent Fisher forecasts by \citet{Boyle18a}, although, as \citet{Boyle18b} notes, making comparisons between results across the literature is difficult. Differences in the implementation of e.g., CMB priors, the use of Fisher versus MCMC methodology, the choice of parameters to be varied, and their fiducial values leads to a range of results. For example, \citet{Audren13} and \citet{Sprenger18} find stronger constraints in studies restricted to the degenerate mass hierarchy. For galaxy clustering, where neutrino information is coming primarily from the shape of the matter power spectrum on linear and mildly non-linear scales, this choice can have a significant impact. We find that the different free-streaming lengths characterizing the NH and IH (discussed in detail in \S~\ref{subsec:nhihwl}) can have a non-negligible effect on the power sensitivity to $\Sigma$. It follows that addressing whether Stage IV surveys will be able to measure $\Sigma$ or distinguish between hierarchies should not simply rely on predicting $\sigma_{\Sigma}$ for an arbitrary model, but should robustly account for the effects of different hierarchy mass splittings. 
This represents a sensitivity to fiducial parameters that will change the multivariate Gaussian likelihood function. When computing the Fisher information, one takes expectation values over the sampling distribution of the data given the true parameter values. Therefore, even for a likelihood that is Gaussian in the parameters, the Fisher matrix can still depend on the fiducial parameter values if the posterior covariance on the parameters entering the likelihood depends on the data. However, even if there is no dependence on the data (which is the case in this work), in practice there should still be a dependence on the true parameter values because we assume that the covariance is computed at these values.

\subsection{Required improvements} 
\begin{figure}
\centering
\includegraphics[width=\columnwidth]{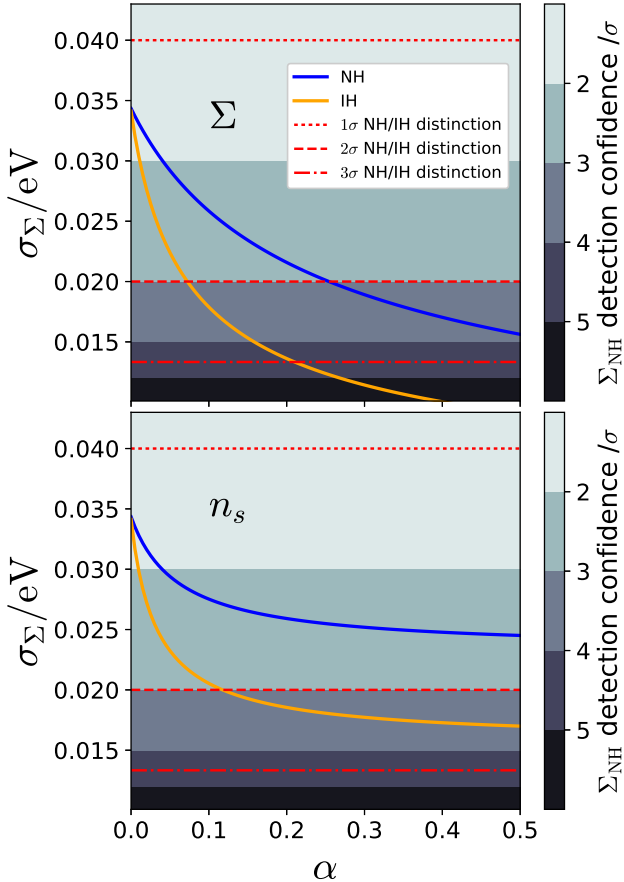}
\caption{1-$\sigma$ WL+GC+CMB errors for $\Sigma$ for the NH (blue) and IH (orange), as a function of arbitrary increases, $\alpha$, to the $\Sigma$ (top panel) and $n_s$ (bottom panel) Fisher information, as defined in equation~\eqref{eq:fishimp}. The shaded regions correspond to the confidence level that $\Sigma$ can be measured at in the case of the NH (we do not include the equivalent regions for the IH because $\Sigma_{\rm{IH}}$ is already constrained to $\sim 3\sigma$ when $\alpha=0$). The red lines mark the thresholds required to distinguish the NH and IH at $1\sigma$ (dotted), $2\sigma$ (dashed) and $3\sigma$ (dash-dotted), assuming that $\Sigma < 0.1\, \rm{eV}$. The difference between the NH and IH fiducial values, $|\Delta\Sigma_{\rm{min}}| = |\Sigma_{\rm{NH,min}} - \Sigma_{\rm{IH,min}}| = 0.04\, \rm{eV}$, represents the 1$\sigma$ limit.}
\label{fig:errorimp}
\end{figure}
To quantify the information required from an arbitrary external source to reduce the errors on $\Sigma_{\rm{NH}}$ and $\Sigma_{\rm{IH}}$ sufficiently for a Euclid-like survey to distinguish the hierarchies beyond the 1$\sigma$ level, we explore the effect of increasing diagonal elements of the (combined WL+GC+CMB) Fisher matrix by a factor, $\left(1+\alpha\right)$, 
\begin{equation}
F_{ii}^{\prime} \longrightarrow F_{ii}\left(1+\alpha\right).
\label{eq:fishimp}
\end{equation}
In Figure~\ref{fig:errorimp} we plot the change in NH and IH errors for increasing $\alpha$ for the parameters $\Sigma$ and $n_s$. These provide the strongest improvements, as the impact of marginalizing over other parameters on $\sigma_{\Sigma}$ is already close to fully mitigated by the WL+GC+CMB combination of probes. 

For $\Sigma$, while the WL+GC+CMB constraints alone constitute close to a 2$\sigma$ detection for the NH (with the IH being constrained to close to 3$\sigma$), this can be improved to between 3$\sigma$ and 4$\sigma$ by adding priors in the range $0.7\,\sigma_{\Sigma,\rm{con}} \lesssim \sigma_{\Sigma,\rm{prior}} \lesssim 2.0\,\sigma_{\Sigma,\rm{con}}$, where $\sigma_{\Sigma,\rm{con}}$ is the conditional error on the mass sum. It may also be possible to achieve a distinction between hierarchies (assuming that $\Sigma < 0.1\, \rm{eV}$) at 2$\sigma$ or even 3$\sigma$ with priors in the range $2.0\,\sigma_{\Sigma,\rm{con}} \lesssim \sigma_{\Sigma,\rm{prior}} \lesssim 2.3\,\sigma_{\Sigma,\rm{con}}$. \citet{Oyama16} find that combining a precise CMB polarization observation such as Simons Array with a 21 cm line observation from phase 2 of the Square kilometer Array and BAO priors could distinguish between hierarchies at the 95\% confidence level provided that $\Sigma$ does not exceed $0.1\, \rm{eV}$. This may indicate scope for similarly significant improvements by combining future CMB polarization data with results from Euclid-like weak lensing surveys. The required information could also be contributed to by e.g., improved data from the Lyman-$\alpha$ forest or possibly further constraints from particle physics. An example of such a particle experiment is measuring the decay rate of neutrinoless double beta decay, as this depends on the sum of mass eigenstates, $m_{\beta\beta}$, when weighted by Majorana phases\footnote{Neutrinoless double beta decay is subject to the assumption that at least one neutrino is a Majorana fermion, i.e. that $\nu \leftrightarrow \bar{\nu}$.} \citep[see e.g.,][]{Capozzi17}. Although neutrinoless double beta decay has not yet been observed, the KamLAND-Zen experiment derives a lower bound of $m_{\beta\beta} \gtrsim 0.061\rm{-}0.165\, \rm{eV}$ \citep{KamLANDZen16}.

There is also considerable scope for improving the $\Sigma$ constraints with additional priors on the spectral index which, if known with absolute certainty, would potentially allow a 2$\sigma$ (5$\sigma$) detection of $\Sigma$ for the NH (IH) and a 2$\sigma$ distinction between the hierarchies. The next generation CMB surveys CMB-S4 \citep{CMBS4_16} predicts constraints on $n_s$ tightening by a factor of 2-3 over \textit{Planck}, which could contribute to the improvements we see required for $\sigma_{\Sigma}$ in Figure~\ref{fig:errorimp}. Note that the greater rate of decrease of $\sigma_{\Sigma_{\rm{IH}}}$ with $\alpha$ compared to $\sigma_{\Sigma_{\rm{IH}}}$ is a consequence of stronger degeneracies, e.g. between $n_s$-$\Sigma$, being broken for the IH compared to the NH.

\section{Model Bias}
\label{sec:modbias}
\begin{figure*}
\centering
\includegraphics[width=\textwidth]{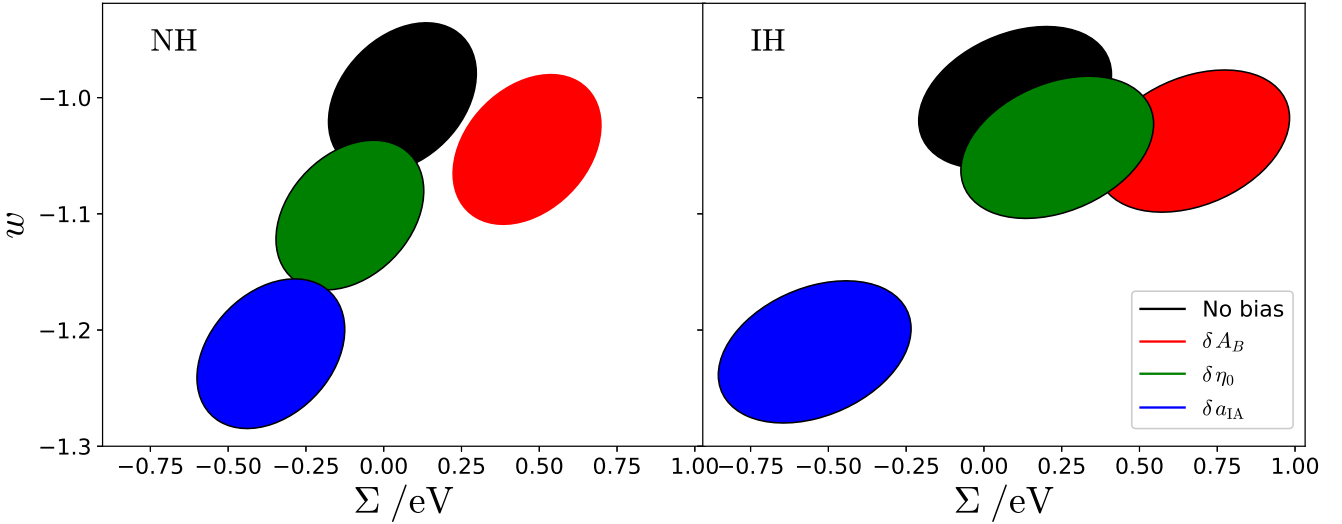}
\caption{Biased 1-$\sigma$ $w$-$\Sigma$ confidence ellipses for the NH and IH with all parameters in $\Theta=\left(A_B,\eta_0,\Omega_m,\Omega_b, h,\sigma_8,n_s,w,a_{\rm{IA}}\right)$ marginalized over. Bias in baryon parameters corresponds to values located in the extremes of the ranges identified in \citetalias{Mead15}, $\delta\, A_B = -1.13$ (red) and $\delta\, \eta_0 = -0.203$ (green). The IA amplitude is biased by 10\% such that $\delta\, a_{\rm{IA}} = 0.1$ in the blue ellipses. The black ellipses represent the unbiased forecasts.}
\label{fig:bias_nh_ih}
\end{figure*}

The parameters controlling the baryon and IA systematics explored in this work are assigned their fiducial values from fits to simulations \citep{Schaye10, Mead15} and shear observations \citep[see e.g.,][]{Hirata04, Bridle&King07} respectively. If the underlying physical models used in simulations is incorrect or if there are inherent limitations in the construction of the simulations, then these fiducial values are biased from the `true' values. It is useful to consider the results of this paper in this context by calculating the degree of calibration bias required to severely bias $\Sigma$ forecasts, for example beyond their 1-$\sigma$ errors. 

To first order, \citet{Taylor07} showed that the bias in a forecast parameter, $\theta$, depends on the bias in a nuisance parameter, $\psi$, such as for baryons or IA, through sub-volumes of the Fisher information according to   
\begin{equation}
\delta\theta_i = -\left[F^{\theta\theta}\right]_{ik}^{-1} F_{kj}^{\theta\psi} \delta\psi_j,
\end{equation}
for which there is an implicit sum over $j$ and $k$. In Figure~\ref{fig:bias_nh_ih} we show the bias in the weak lensing forecasts for $w$-$\Sigma$ when the fiducial values of the baryon and IA parameters have been miscalibrated. We display results for model bias corresponding to the extreme values of the regions $2<A_B<4$ and $0.4<\eta_0<0.8$ that broadly encompass the range of fits to different OWLS simulations made by \citetalias{Mead15}, and for bias in $a_{\rm{IA}}$ of 10\%. 

Figure~\ref{fig:bias_nh_ih} shows that bias in $\Sigma$ from miscalibrating the strength of the IA effect is significant, easily invalidating a measurement for either hierarchy with a 10\% $a_{\rm{IA}}$ bias. This is in contrast to the very small impact on $\Sigma$ forecasts from marginalizing over $a_{\rm{IA}}$. However, as we find percent-level constraints on $a_{\rm{IA}}$ we might expect that a 10\% miscalibration would indeed produce a very large bias. The problem of IA bias for other parameters such as $w$ is well-documented; we do not address it further here beyond reporting the impact on $\Sigma$. The bias due to baryons is less severe, although an extreme miscalibration of the adiabatic contraction parameter would bias $\Sigma$ beyond its 1-$\sigma$ confidence region. However, biases of $\eta_0$, even lying at the extremes of the region of values identified by \citetalias{Mead15}, do not displace the $\Sigma$ estimate beyond the 1-$\sigma$ limit. This difference between $A_B$ and $\eta_0$ may also be reflected in the fact that marginalizing over $A_B$ alone has a significantly greater impact on $\Sigma$ forecasts than marginalizing over $\eta_0$ alone, with almost all of the baryon degradation being due to $A_B$. This highlights the importance of achieving tight calibrations of the adiabatic contraction parameter and the intrinsic alignment amplitude, while a similarly robust understanding of the strength of the baryonic feedback parameter is less urgent. 

A subtlety worth noting is that a bias of $\delta\, \eta_0 = -0.2$ produces biases of opposite sign in $\Sigma$ of different hierarchies. Because the sensitivity to $\eta_0$ is low, it may be that transitioning from one model to the other induces a small change to the bias in $\Sigma$ that is still sufficiently large to change its overall sign. This could complicate attempts to distinguish between hierarchies, but overall the $\eta_0$ bias remains subdominant to other effects. 

As model bias is a product of subgrid limitations, resolving the issue is challenging without resorting to incorporating external data on baryon phenomenology into future simulations. In the case of IA, there has been substantial exploration into self-calibration methods of cleaning IA signals with independent information on IA correlations from deriving scaling relations that predict (and then subtract out) IA-shear signals from shear observables \citep{Zhang10b, Troxel12, Yao17, Yao19}. The IA-IA signal is strongest between close galaxies and so is generally assumed to be more straightforward to treat by purposely using sufficiently high $z$ redshift bins to limit its contribution \citep{Schneider10, Zhang10b}. A future avenue of research into applying these methods to mitigate the bias impact on parameters like the neutrino mass sum is worth pursuing.

\vspace*{-5mm}
\section{Conclusions}
\label{sec:conc}
In this work we have performed detailed Fisher forecast analyses to assess the potential of a Euclid-like Stage IV survey to measure the neutrino mass sum within sufficient accuracy to distinguish between the normal and inverted mass hierarchies. We applied particular focus to the risk of forecast degradation presented by the need to marginalize over baryonic astrophysics, and further explored the impact of intrinsic alignments. To evaluate the widest possible scope of constraints available at present, we combined forecasts for weak lensing with those of spectroscopic galaxy clustering from Euclid and added prior information from \textit{Planck} CMB constraints and BOSS measurements of low redshift clustering.  


The baryon aspect of this paper was an expansion of the baryon degradation analysis of $w_0$-$w_a$ constraints in \citet{Copeland18} to the case of neutrinos. To explore the behaviour of baryons and neutrinos on non-linear scales, we used the halo model presented in \citet{Mead15} and \citet{Mead16}. This captures baryons by incorporating their effect on adiabatic contraction and baryonic feedback into halo structure relations. \citetalias{Mead15} achieve accurate fits for the power spectrum to within a few percent by calibrating the associated parameters, $A_B$ and $\eta_0$, to the OWLS \citep{Schaye10} simulations. The impact of neutrinos is modelled by relating the mass sum to the spherical collapse overdensity and the virial density, which exert a significant influence on halo structure by changing the concentration function and the limits of the density profile respectively. Fits to simulations by \citet{Massara14} reproduce $P\left(k\right)$ to within a few percent up to $k=10\,h\,\mathrm{Mpc}^{-1}$ over different redshifts, improving on the \citet{Bird12} fitting formula.  

We examined both the impact of the $\Sigma$ dependence of the power sensitivity to the mass sum and the effect of different free-streaming scales between hierarchies on the relative sensitivity of the IH  compared to the NH. We illustrated that for both hierarchies, there is low sensitivity on scales relevant to our cosmological probes, $\Delta^2\left(k\right)$ and $C_{\ell}$. The resulting limited Fisher information available for a Euclid-like weak lensing survey leads to forecasts of $\sigma_{\Sigma_{\rm{NH}}}=0.079\, \mathrm{eV}$ and $\sigma_{\Sigma_{\rm{IH}}}=0.120\, \mathrm{eV}$. These are large enough to rule out making any definitive measurement for either hierarchy.    

By incorporating prior CMB information from \citet{Planck15}, these constraints were tightened significantly to $\sigma_{\Sigma_{\rm{NH}}}=0.036\, \mathrm{eV}$ and $\sigma_{\Sigma_{\rm{IH}}}=0.033\, \mathrm{eV}$. We discussed the gains provided by this source of cosmological information in breaking degeneracies between $\Sigma$ and other parameters, such as $\sigma_8$, which has a qualitatively similar effect as $\Sigma$ on non-linear power. We also illustrated the relative advantages in constraining the IH over the NH (72\% and 54\% improvements respectively) when including CMB information.    

This distinction was found to be more nuanced when we expanded our analysis to include marginalization over the baryon parameters. Studying power responses illustrated specific degeneracies, e.g., between $\Sigma$ and $\eta_0$ on non-linear scales, that contributed to the approximate doubling of mass sum error forecasts. While stronger non-linear sensitivity to $\Sigma$ for the IH resulted in a less severe impact than for the NH, when CMB priors were accounted for the degradation for the NH (21\%) was appreciably less than that for the IH (28\%). Despite this, the IH once again exhibited a much greater overall improvement than for the NH.  

Our weak lensing analysis also included an investigation into the impact of including (and marginalizing over) intrinsic alignments in our model on parameter forecasts. Unlike the significant bias induced in parameter estimations by neglecting IA, we found that there is a relatively minimal effect for the separate issue of forecasts because $C_{\ell}$ responses, which determine Fisher information, are very similar in both scenarios (especially for correlations between high redshift bins).  

As the WL+CMB errors including baryons represent only a 1$\sigma$ (2$\sigma$) confidence of measuring $\Sigma_{\rm{NH}}$ ($\Sigma_{\rm{IH}}$) and are greater than $|\Delta\Sigma_{\rm{min}}|=0.04\, \rm{eV}$, we included independent forecasts from a Euclid-like spectroscopic galaxy clustering survey. We showed that this halved the baryon degradation factor for both hierarchies, with the NH still experiencing approximately half the impact of the IH. The galaxy clustering analysis highlighted the importance of specifying the mass hierarchy when making neutrino forecasts as most information was derived from linear matter power responses. The configuration of free-streaming scales corresponding to each neutrino mass played a non-negligible role on these scales. Ultimately, the combination of WL+GC+CMB reduced NH (IH) errors to close to 2$\sigma$ (3$\sigma$) and below the $|\Delta\Sigma_{\rm{min}}|$ threshold. We noted the challenge of navigating the variation in forecast methodologies, and therefore results, made across the literature \citep[e.g.,][]{Carbone11, Audren13, Boyle18a, Sprenger18}. We argue that our approach of simultaneously incorporating the effects of neutrinos and baryons in the weak lensing power spectrum on non-linear scales in a physically well-motivated manner makes our forecasts reliable and robust.

A potentially encouraging finding was that additional information from e.g., neutrinoless double beta decay on the mass sum, or improved constraints on the spectral index from forthcoming Stage IV CMB experiments, could substantially reduce the errors on $\Sigma$. If these priors prove sufficient to increase the Fisher information on their corresponding parameters by 10-50\%, it would achieve a 2$\sigma$ (5$\sigma$) detection of $\Sigma$ for the NH (IH) and distinguish between the hierarchies at the 2$\sigma$ or even 3$\sigma$ level. It is important to emphasize that this conclusion depends on the measured value of the mass sum satisfying $\Sigma < 0.1\, \rm{eV}$. Otherwise, both the NH and IH would be viable regardless of the strength of the error constraint.

Finally, we explored the issue of model bias. We used first order approximations \citep{Taylor07} to determine the bias in $\Sigma$ from miscalibrations of the baryon and intrinsic alignment parameters. We found that large biases in $A_B$ or $a_{\rm{IA}}$ induce a bias in $\Sigma$ larger than $\sigma_{\Sigma}$. By contrast, we found that even significant biases in the baryon feedback parameter, $\eta_0$, did not change the $\Sigma$ estimate by more than $\sigma_{\Sigma}$. Though an understanding of model bias is important to place our analysis in its proper context, it should be noted that it will ultimately require additional baryon information, improved simulations or self-calibrating methods for IA to successfully mitigate the issue.

An important limitation of our analysis to recognize is that the Fisher formalism underpinning it is not ideal for making forecasts in the case of distinct, competing models, as is the case with the normal and inverted mass hierarchies. Fisher information is determined by derivatives of likelihood functions around most-likely values. In cosmology these usually take the form of multi-variable unimodal Gaussians. In this paper we have made assumptions that this is the case when analyzing each hierarchy separately. However, to make a robust forecast of $\Sigma$, given uncertainty over which hierarchy is correct, our analysis should incorporate the resulting bimodal Gaussian (with peaks at $\Sigma_{\rm{NH,min}}=0.06\, \rm{eV}$ and $\Sigma_{\rm{IH,min}}=0.1\, \rm{eV}$). This would require a non-trivial extension of the Fisher formalism, which is beyond the scope of this work but that we intend to pursue in the near future.

In summary, by using the \citetalias{Mead15} and \citetalias{Mead16} prescriptions for baryonic and neutrino effects on haloes, we have been able to perform a full forecast analysis of the neutrino mass sum in the normal and inverted hierarchies. By combining multiple large scale structure probes through Euclid-like weak lensing and spectroscopic galaxy clustering surveys, \textit{Planck} CMB constraints and BOSS low redshift galaxy clustering data, we have shown that degradation due to baryon feedback can be reduced to between 9-16\%. Using these sources of information, and accounting for baryons and intrinsic alignments, Stage IV surveys could be expected to measure $\Sigma$ with 1-$\sigma$ errors of $\sigma_{\Sigma_{\rm{NH}}}=0.034\, \rm{eV}$ and $\sigma_{\Sigma_{\rm{IH}}}=0.034\, \rm{eV}$ for the NH and IH. These approach the confidence level required to meaningfully distinguish the hierarchies, but with additional future priors on $\Sigma$ and $n_s$ there is tentative optimism for achieving more definitive results.

\vspace*{-5mm}
\section*{Acknowledgements}
The authors would like to thank Alexander Mead for several useful discussions about the implementation of massive neutrinos in HMCODE, and John Peacock for valuable input on the approach to galaxy clustering used in this paper. DC acknowledges the support of an STFC studentship. ANT thanks the Royal Society for a Wolfson Research Merit Award, and the STFC for support from a Consolidate Grant. AH is supported by an STFC Consolidated Grant.




\vspace{-5.mm}

\bibliographystyle{mnras}

\bibliography{The_impact_of_baryons_on_the_sensitivity_of_dark_energy_measurements_version2}

\begin{thebibliography}{}
\makeatletter
\relax
\def\mn@urlcharsother{\let\do\@makeother \do\$\do\&\do\#\do\^\do\_\do\%\do\~}
\def\mn@doi{\begingroup\mn@urlcharsother \@ifnextchar [ {\mn@doi@}
  {\mn@doi@[]}}
\def\mn@doi@[#1]#2{\def\@tempa{#1}\ifx\@tempa\@empty \href
  {http://dx.doi.org/#2} {doi:#2}\else \href {http://dx.doi.org/#2} {#1}\fi
  \endgroup}
\def\mn@eprint#1#2{\mn@eprint@#1:#2::\@nil}
\def\mn@eprint@arXiv#1{\href {http://arxiv.org/abs/#1} {{\tt arXiv:#1}}}
\def\mn@eprint@dblp#1{\href {http://dblp.uni-trier.de/rec/bibtex/#1.xml}
  {dblp:#1}}
\def\mn@eprint@#1:#2:#3:#4\@nil{\def\@tempa {#1}\def\@tempb {#2}\def\@tempc
  {#3}\ifx \@tempc \@empty \let \@tempc \@tempb \let \@tempb \@tempa \fi \ifx
  \@tempb \@empty \def\@tempb {arXiv}\fi \@ifundefined
  {mn@eprint@\@tempb}{\@tempb:\@tempc}{\expandafter \expandafter \csname
  mn@eprint@\@tempb\endcsname \expandafter{\@tempc}}}

\bibitem[\protect\citeauthoryear{{Abbott} et~al.,}{{Abbott}
  et~al.}{2016}]{Abbott16}
{Abbott} T.,  et~al., 2016, \mn@doi [\prd] {10.1103/PhysRevD.94.022001}, \href
  {http://adsabs.harvard.edu/abs/2016PhRvD..94b2001A} {94, 022001}

\bibitem[\protect\citeauthoryear{{Albrecht} et~al.,}{{Albrecht}
  et~al.}{2006}]{Albrecht06}
{Albrecht} A.,  et~al., 2006, ArXiv Astrophysics e-prints, \href
  {http://adsabs.harvard.edu/abs/2006astro.ph..9591A} {}

\bibitem[\protect\citeauthoryear{{Bartelmann} \& {Schneider}}{{Bartelmann} \&
  {Schneider}}{2001}]{Bartelmann&Schneider01}
{Bartelmann} M.,  {Schneider} P.,  2001, \mn@doi [\physrep]
  {10.1016/S0370-1573(00)00082-X}, \href
  {http://adsabs.harvard.edu/abs/2001PhR...340..291B} {340, 291}

\bibitem[\protect\citeauthoryear{{Burgess}}{{Burgess}}{2013}]{Burgess13}
{Burgess} C.~P.,  2013, preprint, \href
  {http://adsabs.harvard.edu/abs/2013arXiv1309.4133B} {} (\mn@eprint {arXiv}
  {1309.4133})

\bibitem[\protect\citeauthoryear{{Chevallier} \& {Polarski}}{{Chevallier} \&
  {Polarski}}{2001}]{Chevallier01}
{Chevallier} M.,  {Polarski} D.,  2001, \mn@doi [International Journal of
  Modern Physics D] {10.1142/S0218271801000822}, \href
  {http://adsabs.harvard.edu/abs/2001IJMPD..10..213C} {10, 213}

\bibitem[\protect\citeauthoryear{{Clifton}, {Ferreira}, {Padilla}  \&
  {Skordis}}{{Clifton} et~al.}{2012}]{Clifton11}
{Clifton} T.,  {Ferreira} P.~G.,  {Padilla} A.,   {Skordis} C.,  2012, \mn@doi
  [\physrep] {10.1016/j.physrep.2012.01.001}, \href
  {http://adsabs.harvard.edu/abs/2012PhR...513....1C} {513, 1}

\bibitem[\protect\citeauthoryear{{Cooray} \& {Sheth}}{{Cooray} \&
  {Sheth}}{2002}]{Cooray&Sheth02}
{Cooray} A.,  {Sheth} R.,  2002, \mn@doi [\physrep]
  {10.1016/S0370-1573(02)00276-4}, \href
  {http://adsabs.harvard.edu/abs/2002PhR...372....1C} {372, 1}

\bibitem[\protect\citeauthoryear{{Copeland}, {Sami}  \& {Tsujikawa}}{{Copeland}
  et~al.}{2006}]{Copeland06}
{Copeland} E.~J.,  {Sami} M.,   {Tsujikawa} S.,  2006, \mn@doi [International
  Journal of Modern Physics D] {10.1142/S021827180600942X}, \href
  {http://adsabs.harvard.edu/abs/2006IJMPD..15.1753C} {15, 1753}

\bibitem[\protect\citeauthoryear{{Dubinski} \& {Carlberg}}{{Dubinski} \&
  {Carlberg}}{1991}]{Dubinski&Calberg91}
{Dubinski} J.,  {Carlberg} R.~G.,  1991, \mn@doi [\apj] {10.1086/170451}, \href
  {http://adsabs.harvard.edu/abs/1991ApJ...378..496D} {378, 496}

\bibitem[\protect\citeauthoryear{{Duffy}, {Schaye}, {Kay}, {Dalla Vecchia},
  {Battye}  \& {Booth}}{{Duffy} et~al.}{2010}]{Duffy10}
{Duffy} A.~R.,  {Schaye} J.,  {Kay} S.~T.,  {Dalla Vecchia} C.,  {Battye}
  R.~A.,   {Booth} C.~M.,  2010, \mn@doi [\mnras]
  {10.1111/j.1365-2966.2010.16613.x}, \href
  {http://adsabs.harvard.edu/abs/2010MNRAS.405.2161D} {405, 2161}

\bibitem[\protect\citeauthoryear{{Einasto}}{{Einasto}}{1965}]{Einasto65}
{Einasto} J.,  1965, Trudy Astrofizicheskogo Instituta Alma-Ata, \href
  {http://adsabs.harvard.edu/abs/1965TrAlm...5...87E} {5, 87}

\bibitem[\protect\citeauthoryear{{Fedeli}}{{Fedeli}}{2014}]{Fedeli14}
{Fedeli} C.,  2014, \mn@doi [\jcap] {10.1088/1475-7516/2014/04/028}, \href
  {http://adsabs.harvard.edu/abs/2014JCAP...04..028F} {4, 028}

\bibitem[\protect\citeauthoryear{{Fedeli}, {Semboloni}, {Velliscig}, {Van
  Daalen}, {Schaye}  \& {Hoekstra}}{{Fedeli} et~al.}{2014}]{Fedelietal14}
{Fedeli} C.,  {Semboloni} E.,  {Velliscig} M.,  {Van Daalen} M.,  {Schaye} J.,
   {Hoekstra} H.,  2014, \mn@doi [\jcap] {10.1088/1475-7516/2014/08/028}, \href
  {http://adsabs.harvard.edu/abs/2014JCAP...08..028F} {8, 028}

\bibitem[\protect\citeauthoryear{{Frieman}, {Turner}  \& {Huterer}}{{Frieman}
  et~al.}{2008}]{Frieman08}
{Frieman} J.~A.,  {Turner} M.~S.,   {Huterer} D.,  2008, \mn@doi [\araa]
  {10.1146/annurev.astro.46.060407.145243}, \href
  {http://adsabs.harvard.edu/abs/2008ARA%26A..46..385F} {46, 385}

\bibitem[\protect\citeauthoryear{{Gnedin}, {Kravtsov}, {Klypin}  \&
  {Nagai}}{{Gnedin} et~al.}{2004}]{Gnedin04}
{Gnedin} O.~Y.,  {Kravtsov} A.~V.,  {Klypin} A.~A.,   {Nagai} D.,  2004,
  \mn@doi [\apj] {10.1086/424914}, \href
  {http://adsabs.harvard.edu/abs/2004ApJ...616...16G} {616, 16}

\bibitem[\protect\citeauthoryear{{Gnedin}, {Ceverino}, {Gnedin}, {Klypin},
  {Kravtsov}, {Levine}, {Nagai}  \& {Yepes}}{{Gnedin} et~al.}{2011}]{Gnedin11}
{Gnedin} O.~Y.,  {Ceverino} D.,  {Gnedin} N.~Y.,  {Klypin} A.~A.,  {Kravtsov}
  A.~V.,  {Levine} R.,  {Nagai} D.,   {Yepes} G.,  2011, preprint, \href
  {http://adsabs.harvard.edu/abs/2011arXiv1108.5736G} {} (\mn@eprint {arXiv}
  {1108.5736})

\bibitem[\protect\citeauthoryear{{Governato} et~al.,}{{Governato}
  et~al.}{2012}]{Governato12}
{Governato} F.,  et~al., 2012, \mn@doi [\mnras]
  {10.1111/j.1365-2966.2012.20696.x}, \href
  {http://adsabs.harvard.edu/abs/2012MNRAS.422.1231G} {422, 1231}

\bibitem[\protect\citeauthoryear{{Heitmann}, {Higdon}, {White}, {Habib},
  {Williams}, {Lawrence}  \& {Wagner}}{{Heitmann} et~al.}{2009}]{Heitmann09}
{Heitmann} K.,  {Higdon} D.,  {White} M.,  {Habib} S.,  {Williams} B.~J.,
  {Lawrence} E.,   {Wagner} C.,  2009, \mn@doi [\apj]
  {10.1088/0004-637X/705/1/156}, \href
  {http://adsabs.harvard.edu/abs/2009ApJ...705..156H} {705, 156}

\bibitem[\protect\citeauthoryear{{Heitmann}, {White}, {Wagner}, {Habib}  \&
  {Higdon}}{{Heitmann} et~al.}{2010}]{Heitmann10}
{Heitmann} K.,  {White} M.,  {Wagner} C.,  {Habib} S.,   {Higdon} D.,  2010,
  \mn@doi [\apj] {10.1088/0004-637X/715/1/104}, \href
  {http://adsabs.harvard.edu/abs/2010ApJ...715..104H} {715, 104}

\bibitem[\protect\citeauthoryear{{Heitmann}, {Lawrence}, {Kwan}, {Habib}  \&
  {Higdon}}{{Heitmann} et~al.}{2014}]{Heitmann14}
{Heitmann} K.,  {Lawrence} E.,  {Kwan} J.,  {Habib} S.,   {Higdon} D.,  2014,
  \mn@doi [\apj] {10.1088/0004-637X/780/1/111}, \href
  {http://adsabs.harvard.edu/abs/2014ApJ...780..111H} {780, 111}

\bibitem[\protect\citeauthoryear{{Hildebrandt} et~al.,}{{Hildebrandt}
  et~al.}{2017}]{Hildebrandt17}
{Hildebrandt} H.,  et~al., 2017, \mn@doi [\mnras] {10.1093/mnras/stw2805},
  \href {http://adsabs.harvard.edu/abs/2017MNRAS.465.1454H} {465, 1454}

\bibitem[\protect\citeauthoryear{{Hojjati} et~al.,}{{Hojjati}
  et~al.}{2017}]{Hojjati17}
{Hojjati} A.,  et~al., 2017, \mn@doi [\mnras] {10.1093/mnras/stx1659}, \href
  {http://adsabs.harvard.edu/abs/2017MNRAS.471.1565H} {471, 1565}

\bibitem[\protect\citeauthoryear{{Huterer}}{{Huterer}}{2002}]{Huterer01}
{Huterer} D.,  2002, \mn@doi [\prd] {10.1103/PhysRevD.65.063001}, \href
  {http://adsabs.harvard.edu/abs/2002PhRvD..65f3001H} {65, 063001}

\bibitem[\protect\citeauthoryear{{Jing}, {Zhang}, {Lin}, {Gao}  \&
  {Springel}}{{Jing} et~al.}{2006}]{Jing06}
{Jing} Y.~P.,  {Zhang} P.,  {Lin} W.~P.,  {Gao} L.,   {Springel} V.,  2006,
  \mn@doi [\apjl] {10.1086/503547}, \href
  {http://adsabs.harvard.edu/abs/2006ApJ...640L.119J} {640, L119}

\bibitem[\protect\citeauthoryear{{Kiessling}, {Taylor}  \&
  {Heavens}}{{Kiessling} et~al.}{2011}]{Kiessling11}
{Kiessling} A.,  {Taylor} A.~N.,   {Heavens} A.~F.,  2011, \mn@doi [\mnras]
  {10.1111/j.1365-2966.2011.19108.x}, \href
  {http://adsabs.harvard.edu/abs/2011MNRAS.416.1045K} {416, 1045}

\bibitem[\protect\citeauthoryear{{Kuzio de Naray}, {McGaugh}  \& {de
  Blok}}{{Kuzio de Naray} et~al.}{2008}]{deNaray08}
{Kuzio de Naray} R.,  {McGaugh} S.~S.,   {de Blok} W.~J.~G.,  2008, \mn@doi
  [\apj] {10.1086/527543}, \href
  {http://adsabs.harvard.edu/abs/2008ApJ...676..920K} {676, 920}

\bibitem[\protect\citeauthoryear{{LSST Science Collaboration} et~al.,}{{LSST
  Science Collaboration} et~al.}{2009}]{LSST09}
{LSST Science Collaboration} et~al., 2009, preprint, \href
  {http://adsabs.harvard.edu/abs/2009arXiv0912.0201L} {} (\mn@eprint {arXiv}
  {0912.0201})

\bibitem[\protect\citeauthoryear{{Lagos}, {Lacey}  \& {Baugh}}{{Lagos}
  et~al.}{2013}]{Lagos13}
{Lagos} C.~d.~P.,  {Lacey} C.~G.,   {Baugh} C.~M.,  2013, \mn@doi [\mnras]
  {10.1093/mnras/stt1696}, \href
  {http://adsabs.harvard.edu/abs/2013MNRAS.436.1787L} {436, 1787}

\bibitem[\protect\citeauthoryear{{Laureijs} et~al.,}{{Laureijs}
  et~al.}{2011}]{Laurejis11}
{Laureijs} R.,  et~al., 2011, preprint, \href
  {http://adsabs.harvard.edu/abs/2011arXiv1110.3193L} {} (\mn@eprint {arXiv}
  {1110.3193})

\bibitem[\protect\citeauthoryear{{Lawrence}, {Heitmann}, {White}, {Higdon},
  {Wagner}, {Habib}  \& {Williams}}{{Lawrence} et~al.}{2010}]{Lawrence10}
{Lawrence} E.,  {Heitmann} K.,  {White} M.,  {Higdon} D.,  {Wagner} C.,
  {Habib} S.,   {Williams} B.,  2010, \mn@doi [\apj]
  {10.1088/0004-637X/713/2/1322}, \href
  {http://adsabs.harvard.edu/abs/2010ApJ...713.1322L} {713, 1322}

\bibitem[\protect\citeauthoryear{{Linder}}{{Linder}}{2003}]{Linder03}
{Linder} E.~V.,  2003, \mn@doi [Physical Review Letters]
  {10.1103/PhysRevLett.90.091301}, \href
  {http://adsabs.harvard.edu/abs/2003PhRvL..90i1301L} {90, 091301}

\bibitem[\protect\citeauthoryear{{Ma}, {Hu}  \& {Huterer}}{{Ma}
  et~al.}{2006}]{Ma06}
{Ma} Z.,  {Hu} W.,   {Huterer} D.,  2006, \mn@doi [\apj] {10.1086/497068},
  \href {http://adsabs.harvard.edu/abs/2006ApJ...636...21M} {636, 21}

\bibitem[\protect\citeauthoryear{{Ma}, {Van Waerbeke}, {Hinshaw}, {Hojjati},
  {Scott}  \& {Zuntz}}{{Ma} et~al.}{2015}]{Ma15}
{Ma} Y.-Z.,  {Van Waerbeke} L.,  {Hinshaw} G.,  {Hojjati} A.,  {Scott} D.,
  {Zuntz} J.,  2015, \mn@doi [\jcap] {10.1088/1475-7516/2015/09/046}, \href
  {http://adsabs.harvard.edu/abs/2015JCAP...09..046M} {9, 046}

\bibitem[\protect\citeauthoryear{{MacCrann} et~al.,}{{MacCrann}
  et~al.}{2017}]{MacCrann17}
{MacCrann} N.,  et~al., 2017, \mn@doi [\mnras] {10.1093/mnras/stw2849}, \href
  {http://adsabs.harvard.edu/abs/2017MNRAS.465.2567M} {465, 2567}

\bibitem[\protect\citeauthoryear{{Marsh} \& {Pop}}{{Marsh} \&
  {Pop}}{2015}]{Marsh15}
{Marsh} D.~J.~E.,  {Pop} A.-R.,  2015, \mn@doi [\mnras]
  {10.1093/mnras/stv1050}, \href
  {http://adsabs.harvard.edu/abs/2015MNRAS.451.2479M} {451, 2479}

\bibitem[\protect\citeauthoryear{{Martizzi}, {Teyssier}, {Moore}  \&
  {Wentz}}{{Martizzi} et~al.}{2012}]{Martizzi12}
{Martizzi} D.,  {Teyssier} R.,  {Moore} B.,   {Wentz} T.,  2012, \mn@doi
  [\mnras] {10.1111/j.1365-2966.2012.20879.x}, \href
  {http://adsabs.harvard.edu/abs/2012MNRAS.422.3081M} {422, 3081}

\bibitem[\protect\citeauthoryear{{Martizzi}, {Teyssier}  \& {Moore}}{{Martizzi}
  et~al.}{2013}]{Martizzi13}
{Martizzi} D.,  {Teyssier} R.,   {Moore} B.,  2013, \mn@doi [\mnras]
  {10.1093/mnras/stt297}, \href
  {http://adsabs.harvard.edu/abs/2013MNRAS.432.1947M} {432, 1947}

\bibitem[\protect\citeauthoryear{{Martizzi}, {Mohammed}, {Teyssier}  \&
  {Moore}}{{Martizzi} et~al.}{2014}]{Martizzi14}
{Martizzi} D.,  {Mohammed} I.,  {Teyssier} R.,   {Moore} B.,  2014, \mn@doi
  [\mnras] {10.1093/mnras/stu440}, \href
  {http://adsabs.harvard.edu/abs/2014MNRAS.440.2290M} {440, 2290}

\bibitem[\protect\citeauthoryear{{Mead}, {Peacock}, {Heymans}, {Joudaki}  \&
  {Heavens}}{{Mead} et~al.}{2015}]{Mead15}
{Mead} A.~J.,  {Peacock} J.~A.,  {Heymans} C.,  {Joudaki} S.,   {Heavens}
  A.~F.,  2015, \mn@doi [\mnras] {10.1093/mnras/stv2036}, \href
  {http://adsabs.harvard.edu/abs/2015MNRAS.454.1958M} {454, 1958}

\bibitem[\protect\citeauthoryear{{Mead}, {Heymans}, {Lombriser}, {Peacock},
  {Steele}  \& {Winther}}{{Mead} et~al.}{2016}]{Mead16}
{Mead} A.~J.,  {Heymans} C.,  {Lombriser} L.,  {Peacock} J.~A.,  {Steele}
  O.~I.,   {Winther} H.~A.,  2016, \mn@doi [\mnras] {10.1093/mnras/stw681},
  \href {http://adsabs.harvard.edu/abs/2016MNRAS.459.1468M} {459, 1468}

\bibitem[\protect\citeauthoryear{{Mohammed}, {Martizzi}, {Teyssier}  \&
  {Amara}}{{Mohammed} et~al.}{2014}]{Mohammed14}
{Mohammed} I.,  {Martizzi} D.,  {Teyssier} R.,   {Amara} A.,  2014, preprint,
  \href {http://adsabs.harvard.edu/abs/2014arXiv1410.6826M} {} (\mn@eprint
  {arXiv} {1410.6826})

\bibitem[\protect\citeauthoryear{{Navarro}, {Frenk}  \& {White}}{{Navarro}
  et~al.}{1997}]{NFW97}
{Navarro} J.~F.,  {Frenk} C.~S.,   {White} S.~D.~M.,  1997, \mn@doi [\apj]
  {10.1086/304888}, \href {http://adsabs.harvard.edu/abs/1997ApJ...490..493N}
  {490, 493}

\bibitem[\protect\citeauthoryear{{Navarro} et~al.,}{{Navarro}
  et~al.}{2004}]{Navarro04}
{Navarro} J.~F.,  et~al., 2004, \mn@doi [\mnras]
  {10.1111/j.1365-2966.2004.07586.x}, \href
  {http://adsabs.harvard.edu/abs/2004MNRAS.349.1039N} {349, 1039}

\bibitem[\protect\citeauthoryear{{Oh}, {Brook}, {Governato}, {Brinks}, {Mayer},
  {de Blok}, {Brooks}  \& {Walter}}{{Oh} et~al.}{2011}]{Oh11b}
{Oh} S.-H.,  {Brook} C.,  {Governato} F.,  {Brinks} E.,  {Mayer} L.,  {de Blok}
  W.~J.~G.,  {Brooks} A.,   {Walter} F.,  2011, \mn@doi [\aj]
  {10.1088/0004-6256/142/1/24}, \href
  {http://adsabs.harvard.edu/abs/2011AJ....142...24O} {142, 24}

\bibitem[\protect\citeauthoryear{{Pe{\~n}arrubia}, {Pontzen}, {Walker}  \&
  {Koposov}}{{Pe{\~n}arrubia} et~al.}{2012}]{Pen12}
{Pe{\~n}arrubia} J.,  {Pontzen} A.,  {Walker} M.~G.,   {Koposov} S.~E.,  2012,
  \mn@doi [\apjl] {10.1088/2041-8205/759/2/L42}, \href
  {http://adsabs.harvard.edu/abs/2012ApJ...759L..42P} {759, L42}

\bibitem[\protect\citeauthoryear{{Peacock} \& {Smith}}{{Peacock} \&
  {Smith}}{2000}]{Peacock00}
{Peacock} J.~A.,  {Smith} R.~E.,  2000, \mn@doi [\mnras]
  {10.1046/j.1365-8711.2000.03779.x}, \href
  {http://adsabs.harvard.edu/abs/2000MNRAS.318.1144P} {318, 1144}

\bibitem[\protect\citeauthoryear{{Peacock}, {Schneider}, {Efstathiou}, {Ellis},
  {Leibundgut}, {Lilly}  \& {Mellier}}{{Peacock} et~al.}{2006}]{Peacock06}
{Peacock} J.~A.,  {Schneider} P.,  {Efstathiou} G.,  {Ellis} J.~R.,
  {Leibundgut} B.,  {Lilly} S.~J.,   {Mellier} Y.,  2006, Technical report,
  {ESA-ESO Working Group on ``Fundamental Cosmology''}.
 (\mn@eprint {} {astro-ph/0610906})

\bibitem[\protect\citeauthoryear{{Perlmutter} et~al.,}{{Perlmutter}
  et~al.}{1999}]{Perlmutter99}
{Perlmutter} S.,  et~al., 1999, \mn@doi [\apj] {10.1086/307221}, \href
  {http://adsabs.harvard.edu/abs/1999ApJ...517..565P} {517, 565}

\bibitem[\protect\citeauthoryear{{Planck Collaboration} et~al.,}{{Planck
  Collaboration} et~al.}{2016a}]{Planck15}
{Planck Collaboration} et~al., 2016a, \mn@doi [\aap]
  {10.1051/0004-6361/201525830}, \href
  {http://adsabs.harvard.edu/abs/2016A%26A...594A..13P} {594, A13}

\bibitem[\protect\citeauthoryear{{Planck Collaboration} et~al.,}{{Planck
  Collaboration} et~al.}{2016b}]{Planck15XVI}
{Planck Collaboration} et~al., 2016b, \mn@doi [\aap]
  {10.1051/0004-6361/201525814}, \href
  {http://adsabs.harvard.edu/abs/2016A%26A...594A..14P} {594, A14}

\bibitem[\protect\citeauthoryear{{Pontzen} \& {Governato}}{{Pontzen} \&
  {Governato}}{2012}]{Pontzen&Governato12}
{Pontzen} A.,  {Governato} F.,  2012, \mn@doi [\mnras]
  {10.1111/j.1365-2966.2012.20571.x}, \href
  {http://adsabs.harvard.edu/abs/2012MNRAS.421.3464P} {421, 3464}

\bibitem[\protect\citeauthoryear{{Read} \& {Gilmore}}{{Read} \&
  {Gilmore}}{2005}]{Read&Gilmore05}
{Read} J.~I.,  {Gilmore} G.,  2005, \mn@doi [\mnras]
  {10.1111/j.1365-2966.2004.08424.x}, \href
  {http://adsabs.harvard.edu/abs/2005MNRAS.356..107R} {356, 107}

\bibitem[\protect\citeauthoryear{Riess, Bob, Burns  \& Dody}{Riess
  et~al.}{1998}]{jr:riess98}
Riess A.~G.,  Bob B.,  Burns C.,   Dody D.,  1998, Astron. J., 116, 1009

\bibitem[\protect\citeauthoryear{{Rudd}, {Zentner}  \& {Kravtsov}}{{Rudd}
  et~al.}{2008}]{Rudd08}
{Rudd} D.~H.,  {Zentner} A.~R.,   {Kravtsov} A.~V.,  2008, \mn@doi [\apj]
  {10.1086/523836}, \href {http://adsabs.harvard.edu/abs/2008ApJ...672...19R}
  {672, 19}

\bibitem[\protect\citeauthoryear{{Schaye} et~al.,}{{Schaye}
  et~al.}{2010}]{Schaye10}
{Schaye} J.,  et~al., 2010, \mn@doi [\mnras]
  {10.1111/j.1365-2966.2009.16029.x}, \href
  {http://adsabs.harvard.edu/abs/2010MNRAS.402.1536S} {402, 1536}

\bibitem[\protect\citeauthoryear{{Schneider} \& {Teyssier}}{{Schneider} \&
  {Teyssier}}{2015}]{Schneider&Teyssier15}
{Schneider} A.,  {Teyssier} R.,  2015, \mn@doi [\jcap]
  {10.1088/1475-7516/2015/12/049}, \href
  {http://adsabs.harvard.edu/abs/2015JCAP...12..049S} {12, 049}

\bibitem[\protect\citeauthoryear{{Seljak}}{{Seljak}}{2000}]{Seljak00}
{Seljak} U.,  2000, \mn@doi [\mnras] {10.1046/j.1365-8711.2000.03715.x}, \href
  {http://adsabs.harvard.edu/abs/2000MNRAS.318..203S} {318, 203}

\bibitem[\protect\citeauthoryear{{Semboloni}, {Hoekstra}, {Schaye}, {van
  Daalen}  \& {McCarthy}}{{Semboloni} et~al.}{2011}]{Semboloni11}
{Semboloni} E.,  {Hoekstra} H.,  {Schaye} J.,  {van Daalen} M.~P.,   {McCarthy}
  I.~G.,  2011, \mn@doi [\mnras] {10.1111/j.1365-2966.2011.19385.x}, \href
  {http://adsabs.harvard.edu/abs/2011MNRAS.417.2020S} {417, 2020}

\bibitem[\protect\citeauthoryear{{Semboloni}, {Hoekstra}  \&
  {Schaye}}{{Semboloni} et~al.}{2013}]{Semboloni13}
{Semboloni} E.,  {Hoekstra} H.,   {Schaye} J.,  2013, \mn@doi [\mnras]
  {10.1093/mnras/stt1013}, \href
  {http://adsabs.harvard.edu/abs/2013MNRAS.434..148S} {434, 148}

\bibitem[\protect\citeauthoryear{{Takada} \& {Jain}}{{Takada} \&
  {Jain}}{2004}]{Takada&Jain04}
{Takada} M.,  {Jain} B.,  2004, \mn@doi [\mnras]
  {10.1111/j.1365-2966.2004.07410.x}, \href
  {http://adsabs.harvard.edu/abs/2004MNRAS.348..897T} {348, 897}

\bibitem[\protect\citeauthoryear{{Taylor}, {Kitching}, {Bacon}  \&
  {Heavens}}{{Taylor} et~al.}{2007}]{Taylor07}
{Taylor} A.~N.,  {Kitching} T.~D.,  {Bacon} D.~J.,   {Heavens} A.~F.,  2007,
  \mn@doi [\mnras] {10.1111/j.1365-2966.2006.11257.x}, \href
  {http://adsabs.harvard.edu/abs/2007MNRAS.374.1377T} {374, 1377}

\bibitem[\protect\citeauthoryear{{Tegmark}, {Taylor}  \& {Heavens}}{{Tegmark}
  et~al.}{1997}]{Tegmark97}
{Tegmark} M.,  {Taylor} A.~N.,   {Heavens} A.~F.,  1997, \mn@doi [\apj]
  {10.1086/303939}, \href {http://adsabs.harvard.edu/abs/1997ApJ...480...22T}
  {480, 22}

\bibitem[\protect\citeauthoryear{{Teyssier}, {Pontzen}, {Dubois}  \&
  {Read}}{{Teyssier} et~al.}{2013}]{Teyssier13}
{Teyssier} R.,  {Pontzen} A.,  {Dubois} Y.,   {Read} J.~I.,  2013, \mn@doi
  [\mnras] {10.1093/mnras/sts563}, \href
  {http://adsabs.harvard.edu/abs/2013MNRAS.429.3068T} {429, 3068}

\bibitem[\protect\citeauthoryear{{The Dark Energy Survey Collaboration}}{{The
  Dark Energy Survey Collaboration}}{2005}]{DES05}
{The Dark Energy Survey Collaboration} 2005, ArXiv Astrophysics e-prints, \href
  {http://adsabs.harvard.edu/abs/2005astro.ph.10346T} {}

\bibitem[\protect\citeauthoryear{{Walker} \& {Pe{\~n}arrubia}}{{Walker} \&
  {Pe{\~n}arrubia}}{2011}]{Walker&Pen11}
{Walker} M.~G.,  {Pe{\~n}arrubia} J.,  2011, \mn@doi [\apj]
  {10.1088/0004-637X/742/1/20}, \href
  {http://adsabs.harvard.edu/abs/2011ApJ...742...20W} {742, 20}

\bibitem[\protect\citeauthoryear{Weinberg}{Weinberg}{1989}]{Weinberg89}
Weinberg S.,  1989, \mn@doi [Rev. Mod. Phys.] {10.1103/RevModPhys.61.1}, 61, 1

\bibitem[\protect\citeauthoryear{{Wu}}{{Wu}}{2018}]{Wu18}
{Wu} P.-F.,  2018, \mn@doi [\mnras] {10.1093/mnras/stx2745}, \href
  {http://adsabs.harvard.edu/abs/2018MNRAS.473.5468W} {473, 5468}

\bibitem[\protect\citeauthoryear{{Zentner}, {Semboloni}, {Dodelson}, {Eifler},
  {Krause}  \& {Hearin}}{{Zentner} et~al.}{2013}]{Zentner12}
{Zentner} A.~R.,  {Semboloni} E.,  {Dodelson} S.,  {Eifler} T.,  {Krause} E.,
  {Hearin} A.~P.,  2013, \mn@doi [\prd] {10.1103/PhysRevD.87.043509}, \href
  {http://adsabs.harvard.edu/abs/2013PhRvD..87d3509Z} {87, 043509}

\bibitem[\protect\citeauthoryear{{Zhang}, {Sming Tsai}, {Cheung}  \&
  {Chu}}{{Zhang} et~al.}{2016}]{Zhang16}
{Zhang} J.,  {Sming Tsai} Y.-L.,  {Cheung} K.,   {Chu} M.-C.,  2016, preprint,
  \href {http://adsabs.harvard.edu/abs/2016arXiv161100892Z} {} (\mn@eprint
  {arXiv} {1611.00892})

\bibitem[\protect\citeauthoryear{{Zhao}}{{Zhao}}{1996}]{Zhao96}
{Zhao} H.,  1996, \mn@doi [\mnras] {10.1093/mnras/278.2.488}, \href
  {http://adsabs.harvard.edu/abs/1996MNRAS.278..488Z} {278, 488}

\bibitem[\protect\citeauthoryear{{Zorrilla Matilla}, {Haiman}, {Petri}  \&
  {Namikawa}}{{Zorrilla Matilla} et~al.}{2017}]{Matilla17}
{Zorrilla Matilla} J.~M.,  {Haiman} Z.,  {Petri} A.,   {Namikawa} T.,  2017,
  preprint, \href {http://adsabs.harvard.edu/abs/2017arXiv170605133Z} {}
  (\mn@eprint {arXiv} {1706.05133})

\bibitem[\protect\citeauthoryear{{de Blok}, {Bosma}  \& {McGaugh}}{{de Blok}
  et~al.}{2003}]{deBlok02}
{de Blok} W.~J.~G.,  {Bosma} A.,   {McGaugh} S.,  2003, \mn@doi [\mnras]
  {10.1046/j.1365-8711.2003.06330.x}, \href
  {http://adsabs.harvard.edu/abs/2003MNRAS.340..657D} {340, 657}

\bibitem[\protect\citeauthoryear{{van Daalen}, {Schaye}, {Booth}  \& {Dalla
  Vecchia}}{{van Daalen} et~al.}{2011}]{vandaalen11}
{van Daalen} M.~P.,  {Schaye} J.,  {Booth} C.~M.,   {Dalla Vecchia} C.,  2011,
  \mn@doi [\mnras] {10.1111/j.1365-2966.2011.18981.x}, \href
  {http://adsabs.harvard.edu/abs/2011MNRAS.415.3649V} {415, 3649}

\bibitem[\protect\citeauthoryear{{van Daalen}, {Schaye}, {McCarthy}, {Booth}
  \& {Dalla Vecchia}}{{van Daalen} et~al.}{2014}]{vandaalen14}
{van Daalen} M.~P.,  {Schaye} J.,  {McCarthy} I.~G.,  {Booth} C.~M.,   {Dalla
  Vecchia} C.,  2014, \mn@doi [\mnras] {10.1093/mnras/stu482}, \href
  {http://adsabs.harvard.edu/abs/2014MNRAS.440.2997V} {440, 2997}

\makeatother
\end{thebibliography}


\begin{thebibliography}{}
\makeatletter
\relax
\def\mn@urlcharsother{\let\do\@makeother \do\$\do\&\do\#\do\^\do\_\do\%\do\~}
\def\mn@doi{\begingroup\mn@urlcharsother \@ifnextchar [ {\mn@doi@}
  {\mn@doi@[]}}
\def\mn@doi@[#1]#2{\def\@tempa{#1}\ifx\@tempa\@empty \href
  {http://dx.doi.org/#2} {doi:#2}\else \href {http://dx.doi.org/#2} {#1}\fi
  \endgroup}
\def\mn@eprint#1#2{\mn@eprint@#1:#2::\@nil}
\def\mn@eprint@arXiv#1{\href {http://arxiv.org/abs/#1} {{\tt arXiv:#1}}}
\def\mn@eprint@dblp#1{\href {http://dblp.uni-trier.de/rec/bibtex/#1.xml}
  {dblp:#1}}
\def\mn@eprint@#1:#2:#3:#4\@nil{\def\@tempa {#1}\def\@tempb {#2}\def\@tempc
  {#3}\ifx \@tempc \@empty \let \@tempc \@tempb \let \@tempb \@tempa \fi \ifx
  \@tempb \@empty \def\@tempb {arXiv}\fi \@ifundefined
  {mn@eprint@\@tempb}{\@tempb:\@tempc}{\expandafter \expandafter \csname
  mn@eprint@\@tempb\endcsname \expandafter{\@tempc}}}

\bibitem[\protect\citeauthoryear{{Abazajian} et~al.,}{{Abazajian}
  et~al.}{2016}]{CMBS4_16}
{Abazajian} K.~N.,  et~al., 2016, arXiv e-prints, \href
  {https://ui.adsabs.harvard.edu/\#abs/2016arXiv161002743A} {p.
  arXiv:1610.02743}

\bibitem[\protect\citeauthoryear{{Alam} et~al.,}{{Alam} et~al.}{2017}]{BOSS17}
{Alam} S.,  et~al., 2017, \mn@doi [\mnras] {10.1093/mnras/stx721}, \href
  {https://ui.adsabs.harvard.edu/\#abs/2017MNRAS.470.2617A} {470, 2617}

\bibitem[\protect\citeauthoryear{{Albrecht} et~al.,}{{Albrecht}
  et~al.}{2006}]{Albrecht06}
{Albrecht} A.,  et~al., 2006, ArXiv Astrophysics e-prints, \href
  {http://adsabs.harvard.edu/abs/2006astro.ph..9591A} {}

\bibitem[\protect\citeauthoryear{{Alcock} \& {Paczynski}}{{Alcock} \&
  {Paczynski}}{1979}]{Alcock&Paczynski79}
{Alcock} C.,  {Paczynski} B.,  1979, \mn@doi [\nat] {10.1038/281358a0}, \href
  {http://adsabs.harvard.edu/abs/1979Natur.281..358A} {281, 358}

\bibitem[\protect\citeauthoryear{{Allison}, {Caucal}, {Calabrese}, {Dunkley}
  \& {Louis}}{{Allison} et~al.}{2015}]{Allison15}
{Allison} R.,  {Caucal} P.,  {Calabrese} E.,  {Dunkley} J.,   {Louis} T.,
  2015, \mn@doi [\prd] {10.1103/PhysRevD.92.123535}, \href
  {https://ui.adsabs.harvard.edu/\#abs/2015PhRvD..92l3535A} {92, 123535}

\bibitem[\protect\citeauthoryear{{Amendola} et~al.,}{{Amendola}
  et~al.}{2018}]{Amendola18}
{Amendola} L.,  et~al., 2018, \mn@doi [Living Reviews in Relativity]
  {10.1007/s41114-017-0010-3}, \href
  {https://ui.adsabs.harvard.edu/\#abs/2018LRR....21....2A} {21, 2}

\bibitem[\protect\citeauthoryear{{Audren}, {Lesgourgues}, {Bird}, {Haehnelt}
  \& {Viel}}{{Audren} et~al.}{2013}]{Audren13}
{Audren} B.,  {Lesgourgues} J.,  {Bird} S.,  {Haehnelt} M.~G.,   {Viel} M.,
  2013, \mn@doi [Journal of Cosmology and Astro-Particle Physics]
  {10.1088/1475-7516/2013/01/026}, \href
  {https://ui.adsabs.harvard.edu/\#abs/2013JCAP...01..026A} {2013, 026}

\bibitem[\protect\citeauthoryear{{Ballinger}, {Peacock}  \&
  {Heavens}}{{Ballinger} et~al.}{1996}]{Ballinger96}
{Ballinger} W.~E.,  {Peacock} J.~A.,   {Heavens} A.~F.,  1996, \mn@doi [\mnras]
  {10.1093/mnras/282.3.877}, \href
  {https://ui.adsabs.harvard.edu/#abs/1996MNRAS.282..877B} {282, 877}

\bibitem[\protect\citeauthoryear{{Bartelmann} \& {Schneider}}{{Bartelmann} \&
  {Schneider}}{2001}]{Bartelmann&Schneider01}
{Bartelmann} M.,  {Schneider} P.,  2001, \mn@doi [\physrep]
  {10.1016/S0370-1573(00)00082-X}, \href
  {http://adsabs.harvard.edu/abs/2001PhR...340..291B} {340, 291}

\bibitem[\protect\citeauthoryear{{Bird}, {Viel}  \& {Haehnelt}}{{Bird}
  et~al.}{2012}]{Bird12}
{Bird} S.,  {Viel} M.,   {Haehnelt} M.~G.,  2012, \mn@doi [\mnras]
  {10.1111/j.1365-2966.2011.20222.x}, \href
  {https://ui.adsabs.harvard.edu/#abs/2012MNRAS.420.2551B} {420, 2551}

\bibitem[\protect\citeauthoryear{{Boyle}}{{Boyle}}{2018}]{Boyle18b}
{Boyle} A.,  2018, arXiv e-prints, \href
  {https://ui.adsabs.harvard.edu/\#abs/2018arXiv181107636B} {p.
  arXiv:1811.07636}

\bibitem[\protect\citeauthoryear{{Boyle} \& {Komatsu}}{{Boyle} \&
  {Komatsu}}{2018}]{Boyle18a}
{Boyle} A.,  {Komatsu} E.,  2018, \mn@doi [Journal of Cosmology and
  Astro-Particle Physics] {10.1088/1475-7516/2018/03/035}, \href
  {https://ui.adsabs.harvard.edu/\#abs/2018JCAP...03..035B} {2018, 035}

\bibitem[\protect\citeauthoryear{{Bridle} \& {King}}{{Bridle} \&
  {King}}{2007}]{Bridle&King07}
{Bridle} S.,  {King} L.,  2007, \mn@doi [New Journal of Physics]
  {10.1088/1367-2630/9/12/444}, \href
  {https://ui.adsabs.harvard.edu/\#abs/2007NJPh....9..444B} {9, 444}

\bibitem[\protect\citeauthoryear{{Brown}, {Taylor}, {Hambly}  \& {Dye}}{{Brown}
  et~al.}{2002}]{Brown02}
{Brown} M.~L.,  {Taylor} A.~N.,  {Hambly} N.~C.,   {Dye} S.,  2002, \mn@doi
  [\mnras] {10.1046/j.1365-8711.2002.05354.x}, \href
  {https://ui.adsabs.harvard.edu/\#abs/2002MNRAS.333..501B} {333, 501}

\bibitem[\protect\citeauthoryear{{Bullock}, {Kolatt}, {Sigad}, {Somerville},
  {Kravtsov}, {Klypin}, {Primack}  \& {Dekel}}{{Bullock}
  et~al.}{2001}]{Bullock01}
{Bullock} J.~S.,  {Kolatt} T.~S.,  {Sigad} Y.,  {Somerville} R.~S.,  {Kravtsov}
  A.~V.,  {Klypin} A.~A.,  {Primack} J.~R.,   {Dekel} A.,  2001, \mn@doi
  [\mnras] {10.1046/j.1365-8711.2001.04068.x}, \href
  {https://ui.adsabs.harvard.edu/\#abs/2001MNRAS.321..559B} {321, 559}

\bibitem[\protect\citeauthoryear{{Capozzi}, {Di Valentino}, {Lisi}, {Marrone},
  {Melchiorri}  \& {Palazzo}}{{Capozzi} et~al.}{2017}]{Capozzi17}
{Capozzi} F.,  {Di Valentino} E.,  {Lisi} E.,  {Marrone} A.,  {Melchiorri} A.,
   {Palazzo} A.,  2017, \mn@doi [\prd] {10.1103/PhysRevD.95.096014}, \href
  {https://ui.adsabs.harvard.edu/\#abs/2017PhRvD..95i6014C} {95, 096014}

\bibitem[\protect\citeauthoryear{{Carbone}, {Verde}, {Wang}  \&
  {Cimatti}}{{Carbone} et~al.}{2011}]{Carbone11}
{Carbone} C.,  {Verde} L.,  {Wang} Y.,   {Cimatti} A.,  2011, \mn@doi [Journal
  of Cosmology and Astro-Particle Physics] {10.1088/1475-7516/2011/03/030},
  \href {https://ui.adsabs.harvard.edu/\#abs/2011JCAP...03..030C} {2011, 030}

\bibitem[\protect\citeauthoryear{{Cataneo}, {Lombriser}, {Heymans}, {Mead},
  {Barreira}, {Bose}  \& {Li}}{{Cataneo} et~al.}{2018}]{Cataneo18}
{Cataneo} M.,  {Lombriser} L.,  {Heymans} C.,  {Mead} A.,  {Barreira} A.,
  {Bose} S.,   {Li} B.,  2018, arXiv e-prints, \href
  {https://ui.adsabs.harvard.edu/\#abs/2018arXiv181205594C} {p.
  arXiv:1812.05594}

\bibitem[\protect\citeauthoryear{{Copeland}, {Taylor}  \& {Hall}}{{Copeland}
  et~al.}{2018}]{Copeland18}
{Copeland} D.,  {Taylor} A.,   {Hall} A.,  2018, \mn@doi [\mnras]
  {10.1093/mnras/sty2001}, \href
  {https://ui.adsabs.harvard.edu/#abs/2018MNRAS.480.2247C} {480, 2247}

\bibitem[\protect\citeauthoryear{{Davis} \& {Peebles}}{{Davis} \&
  {Peebles}}{1983}]{Davis83}
{Davis} M.,  {Peebles} P.~J.~E.,  1983, \mn@doi [\apj] {10.1086/160884}, \href
  {http://adsabs.harvard.edu/abs/1983ApJ...267..465D} {267, 465}

\bibitem[\protect\citeauthoryear{{Desjacques}, {Jeong}  \&
  {Schmidt}}{{Desjacques} et~al.}{2018}]{Desjacques18}
{Desjacques} V.,  {Jeong} D.,   {Schmidt} F.,  2018, \mn@doi [\physrep]
  {10.1016/j.physrep.2017.12.002}, \href
  {https://ui.adsabs.harvard.edu/#abs/2018PhR...733....1D} {733, 1}

\bibitem[\protect\citeauthoryear{{Duffy}, {Schaye}, {Kay}, {Dalla Vecchia},
  {Battye}  \& {Booth}}{{Duffy} et~al.}{2010}]{Duffy10}
{Duffy} A.~R.,  {Schaye} J.,  {Kay} S.~T.,  {Dalla Vecchia} C.,  {Battye}
  R.~A.,   {Booth} C.~M.,  2010, \mn@doi [\mnras]
  {10.1111/j.1365-2966.2010.16613.x}, \href
  {http://adsabs.harvard.edu/abs/2010MNRAS.405.2161D} {405, 2161}

\bibitem[\protect\citeauthoryear{{Eisenstein}, {Seo}  \& {White}}{{Eisenstein}
  et~al.}{2007}]{Eisenstein07}
{Eisenstein} D.~J.,  {Seo} H.-J.,   {White} M.,  2007, \mn@doi [\apj]
  {10.1086/518755}, \href
  {https://ui.adsabs.harvard.edu/\#abs/2007ApJ...664..660E} {664, 660}

\bibitem[\protect\citeauthoryear{{Fedeli}}{{Fedeli}}{2014}]{Fedeli14}
{Fedeli} C.,  2014, \mn@doi [\jcap] {10.1088/1475-7516/2014/04/028}, \href
  {http://adsabs.harvard.edu/abs/2014JCAP...04..028F} {4, 028}

\bibitem[\protect\citeauthoryear{{Fogli}, {Lisi}, {Marrone}  \&
  {Palazzo}}{{Fogli} et~al.}{2006}]{Fogli06}
{Fogli} G.~L.,  {Lisi} E.,  {Marrone} A.,   {Palazzo} A.,  2006, \mn@doi
  [Progress in Particle and Nuclear Physics] {10.1016/j.ppnp.2005.08.002},
  \href {http://adsabs.harvard.edu/abs/2006PrPNP..57..742F} {57, 742}

\bibitem[\protect\citeauthoryear{{Gnedin}, {Kravtsov}, {Klypin}  \&
  {Nagai}}{{Gnedin} et~al.}{2004}]{Gnedin04}
{Gnedin} O.~Y.,  {Kravtsov} A.~V.,  {Klypin} A.~A.,   {Nagai} D.,  2004,
  \mn@doi [\apj] {10.1086/424914}, \href
  {http://adsabs.harvard.edu/abs/2004ApJ...616...16G} {616, 16}

\bibitem[\protect\citeauthoryear{{Gnedin}, {Ceverino}, {Gnedin}, {Klypin},
  {Kravtsov}, {Levine}, {Nagai}  \& {Yepes}}{{Gnedin} et~al.}{2011}]{Gnedin11}
{Gnedin} O.~Y.,  {Ceverino} D.,  {Gnedin} N.~Y.,  {Klypin} A.~A.,  {Kravtsov}
  A.~V.,  {Levine} R.,  {Nagai} D.,   {Yepes} G.,  2011, preprint, \href
  {http://adsabs.harvard.edu/abs/2011arXiv1108.5736G} {} (\mn@eprint {arXiv}
  {1108.5736})

\bibitem[\protect\citeauthoryear{{Hall} \& {Challinor}}{{Hall} \&
  {Challinor}}{2012}]{Hall&Challinor12}
{Hall} A.~C.,  {Challinor} A.,  2012, \mn@doi [\mnras]
  {10.1111/j.1365-2966.2012.21493.x}, \href
  {https://ui.adsabs.harvard.edu/\#abs/2012MNRAS.425.1170H} {425, 1170}

\bibitem[\protect\citeauthoryear{{Hamilton}}{{Hamilton}}{1998}]{Hamilton98}
{Hamilton} A.~J.~S.,  1998, in {Hamilton} D.,  ed.,  Vol. 231, The Evolving
  Universe. p.~185 (\mn@eprint {arXiv} {astro-ph/9708102}),
  \mn@doi{10.1007/978-94-011-4960-0_17}

\bibitem[\protect\citeauthoryear{{Hauser} \& {Peebles}}{{Hauser} \&
  {Peebles}}{1973}]{Hauser73}
{Hauser} M.~G.,  {Peebles} P.~J.~E.,  1973, \mn@doi [\apj] {10.1086/152453},
  \href {http://adsabs.harvard.edu/abs/1973ApJ...185..757H} {185, 757}

\bibitem[\protect\citeauthoryear{{Heavens} \& {Sellentin}}{{Heavens} \&
  {Sellentin}}{2018}]{Heavens&Sellentin18}
{Heavens} A.~F.,  {Sellentin} E.,  2018, \mn@doi [\jcap]
  {10.1088/1475-7516/2018/04/047}, \href
  {https://ui.adsabs.harvard.edu/abs/2018JCAP...04..047H} {2018, 047}

\bibitem[\protect\citeauthoryear{{Heitmann}, {Higdon}, {White}, {Habib},
  {Williams}, {Lawrence}  \& {Wagner}}{{Heitmann} et~al.}{2009}]{Heitmann09}
{Heitmann} K.,  {Higdon} D.,  {White} M.,  {Habib} S.,  {Williams} B.~J.,
  {Lawrence} E.,   {Wagner} C.,  2009, \mn@doi [\apj]
  {10.1088/0004-637X/705/1/156}, \href
  {http://adsabs.harvard.edu/abs/2009ApJ...705..156H} {705, 156}

\bibitem[\protect\citeauthoryear{{Heitmann}, {White}, {Wagner}, {Habib}  \&
  {Higdon}}{{Heitmann} et~al.}{2010}]{Heitmann10}
{Heitmann} K.,  {White} M.,  {Wagner} C.,  {Habib} S.,   {Higdon} D.,  2010,
  \mn@doi [\apj] {10.1088/0004-637X/715/1/104}, \href
  {http://adsabs.harvard.edu/abs/2010ApJ...715..104H} {715, 104}

\bibitem[\protect\citeauthoryear{{Heitmann}, {Lawrence}, {Kwan}, {Habib}  \&
  {Higdon}}{{Heitmann} et~al.}{2014}]{Heitmann14}
{Heitmann} K.,  {Lawrence} E.,  {Kwan} J.,  {Habib} S.,   {Higdon} D.,  2014,
  \mn@doi [\apj] {10.1088/0004-637X/780/1/111}, \href
  {http://adsabs.harvard.edu/abs/2014ApJ...780..111H} {780, 111}

\bibitem[\protect\citeauthoryear{{Hildebrandt} et~al.,}{{Hildebrandt}
  et~al.}{2017}]{Hildebrandt17}
{Hildebrandt} H.,  et~al., 2017, \mn@doi [\mnras] {10.1093/mnras/stw2805},
  \href {http://adsabs.harvard.edu/abs/2017MNRAS.465.1454H} {465, 1454}

\bibitem[\protect\citeauthoryear{{Hirata} \& {Seljak}}{{Hirata} \&
  {Seljak}}{2004}]{Hirata04}
{Hirata} C.~M.,  {Seljak} U.,  2004, \mn@doi [\prd]
  {10.1103/PhysRevD.70.063526}, \href
  {https://ui.adsabs.harvard.edu/\#abs/2004PhRvD..70f3526H} {70, 063526}

\bibitem[\protect\citeauthoryear{{Hu}, {Eisenstein}  \& {Tegmark}}{{Hu}
  et~al.}{1998}]{Hu98}
{Hu} W.,  {Eisenstein} D.~J.,   {Tegmark} M.,  1998, \mn@doi [\prl]
  {10.1103/PhysRevLett.80.5255}, \href
  {https://ui.adsabs.harvard.edu/#abs/1998PhRvL..80.5255H} {80, 5255}

\bibitem[\protect\citeauthoryear{{Ichiki} \& {Takada}}{{Ichiki} \&
  {Takada}}{2012}]{Ichiki12}
{Ichiki} K.,  {Takada} M.,  2012, \mn@doi [\prd] {10.1103/PhysRevD.85.063521},
  \href {https://ui.adsabs.harvard.edu/\#abs/2012PhRvD..85f3521I} {85, 063521}

\bibitem[\protect\citeauthoryear{{Jing}, {Zhang}, {Lin}, {Gao}  \&
  {Springel}}{{Jing} et~al.}{2006}]{Jing06}
{Jing} Y.~P.,  {Zhang} P.,  {Lin} W.~P.,  {Gao} L.,   {Springel} V.,  2006,
  \mn@doi [\apjl] {10.1086/503547}, \href
  {http://adsabs.harvard.edu/abs/2006ApJ...640L.119J} {640, L119}

\bibitem[\protect\citeauthoryear{{Joachimi}, {Mandelbaum}, {Abdalla}  \&
  {Bridle}}{{Joachimi} et~al.}{2011}]{Joachimi11}
{Joachimi} B.,  {Mandelbaum} R.,  {Abdalla} F.~B.,   {Bridle} S.~L.,  2011,
  \mn@doi [\aap] {10.1051/0004-6361/201015621}, \href
  {https://ui.adsabs.harvard.edu/\#abs/2011A&A...527A..26J} {527, A26}

\bibitem[\protect\citeauthoryear{{Kaiser}}{{Kaiser}}{1987}]{Kaiser87}
{Kaiser} N.,  1987, \mn@doi [\mnras] {10.1093/mnras/227.1.1}, \href
  {http://adsabs.harvard.edu/abs/1987MNRAS.227....1K} {227, 1}

\bibitem[\protect\citeauthoryear{{KamLAND-Zen Collaboration}}{{KamLAND-Zen
  Collaboration}}{2016}]{KamLANDZen16}
{KamLAND-Zen Collaboration} 2016, arXiv e-prints, \href
  {https://ui.adsabs.harvard.edu/\#abs/2016arXiv160502889K} {p.
  arXiv:1605.02889}

\bibitem[\protect\citeauthoryear{{Kiessling}, {Taylor}  \&
  {Heavens}}{{Kiessling} et~al.}{2011}]{Kiessling11}
{Kiessling} A.,  {Taylor} A.~N.,   {Heavens} A.~F.,  2011, \mn@doi [\mnras]
  {10.1111/j.1365-2966.2011.19108.x}, \href
  {http://adsabs.harvard.edu/abs/2011MNRAS.416.1045K} {416, 1045}

\bibitem[\protect\citeauthoryear{{Krause}, {Eifler}  \& {Blazek}}{{Krause}
  et~al.}{2016}]{Krause16}
{Krause} E.,  {Eifler} T.,   {Blazek} J.,  2016, \mn@doi [\mnras]
  {10.1093/mnras/stv2615}, \href
  {https://ui.adsabs.harvard.edu/\#abs/2016MNRAS.456..207K} {456, 207}

\bibitem[\protect\citeauthoryear{{LSST Science Collaboration} et~al.,}{{LSST
  Science Collaboration} et~al.}{2009}]{LSST09}
{LSST Science Collaboration} et~al., 2009, preprint, \href
  {http://adsabs.harvard.edu/abs/2009arXiv0912.0201L} {} (\mn@eprint {arXiv}
  {0912.0201})

\bibitem[\protect\citeauthoryear{{Lagos}, {Lacey}  \& {Baugh}}{{Lagos}
  et~al.}{2013}]{Lagos13}
{Lagos} C.~d.~P.,  {Lacey} C.~G.,   {Baugh} C.~M.,  2013, \mn@doi [\mnras]
  {10.1093/mnras/stt1696}, \href
  {http://adsabs.harvard.edu/abs/2013MNRAS.436.1787L} {436, 1787}

\bibitem[\protect\citeauthoryear{{Laureijs} et~al.,}{{Laureijs}
  et~al.}{2011}]{Laurejis11}
{Laureijs} R.,  et~al., 2011, preprint, \href
  {http://adsabs.harvard.edu/abs/2011arXiv1110.3193L} {} (\mn@eprint {arXiv}
  {1110.3193})

\bibitem[\protect\citeauthoryear{{Lawrence}, {Heitmann}, {White}, {Higdon},
  {Wagner}, {Habib}  \& {Williams}}{{Lawrence} et~al.}{2010}]{Lawrence10}
{Lawrence} E.,  {Heitmann} K.,  {White} M.,  {Higdon} D.,  {Wagner} C.,
  {Habib} S.,   {Williams} B.,  2010, \mn@doi [\apj]
  {10.1088/0004-637X/713/2/1322}, \href
  {http://adsabs.harvard.edu/abs/2010ApJ...713.1322L} {713, 1322}

\bibitem[\protect\citeauthoryear{{Lesgourgues} \& {Pastor}}{{Lesgourgues} \&
  {Pastor}}{2006}]{Lesgourges06}
{Lesgourgues} J.,  {Pastor} S.,  2006, \mn@doi [\physrep]
  {10.1016/j.physrep.2006.04.001}, \href
  {http://adsabs.harvard.edu/abs/2006PhR...429..307L} {429, 307}

\bibitem[\protect\citeauthoryear{{Lesgourgues}, {Matarrese}, {Pietroni}  \&
  {Riotto}}{{Lesgourgues} et~al.}{2009}]{Lesgourgues09}
{Lesgourgues} J.,  {Matarrese} S.,  {Pietroni} M.,   {Riotto} A.,  2009,
  \mn@doi [Journal of Cosmology and Astro-Particle Physics]
  {10.1088/1475-7516/2009/06/017}, \href
  {https://ui.adsabs.harvard.edu/#abs/2009JCAP...06..017L} {2009, 017}

\bibitem[\protect\citeauthoryear{{Levi} \& {Vlah}}{{Levi} \&
  {Vlah}}{2016}]{Levi16}
{Levi} M.,  {Vlah} Z.,  2016, preprint, \href
  {https://ui.adsabs.harvard.edu/#abs/2016arXiv160509417L} {p.
  arXiv:1605.09417} (\mn@eprint {arXiv} {1605.09417})

\bibitem[\protect\citeauthoryear{{Lewis}, {Challinor}  \& {Lasenby}}{{Lewis}
  et~al.}{2000}]{Lewis00}
{Lewis} A.,  {Challinor} A.,   {Lasenby} A.,  2000, \mn@doi [\apj]
  {10.1086/309179}, \href
  {https://ui.adsabs.harvard.edu/\#abs/2000ApJ...538..473L} {538, 473}

\bibitem[\protect\citeauthoryear{{Liu}, {Pritchard}, {Allison}, {Parsons},
  {Seljak}  \& {Sherwin}}{{Liu} et~al.}{2016}]{Lui16}
{Liu} A.,  {Pritchard} J.~R.,  {Allison} R.,  {Parsons} A.~R.,  {Seljak} U.,
  {Sherwin} B.~D.,  2016, \mn@doi [\prd] {10.1103/PhysRevD.93.043013}, \href
  {https://ui.adsabs.harvard.edu/\#abs/2016PhRvD..93d3013L} {93, 043013}

\bibitem[\protect\citeauthoryear{{Liu}, {Bird}, {Zorrilla Matilla}, {Hill},
  {Haiman}, {Madhavacheril}, {Petri}  \& {Spergel}}{{Liu} et~al.}{2018}]{Liu18}
{Liu} J.,  {Bird} S.,  {Zorrilla Matilla} J.~M.,  {Hill} J.~C.,  {Haiman} Z.,
  {Madhavacheril} M.~S.,  {Petri} A.,   {Spergel} D.~N.,  2018, \mn@doi
  [Journal of Cosmology and Astro-Particle Physics]
  {10.1088/1475-7516/2018/03/049}, \href
  {https://ui.adsabs.harvard.edu/#abs/2018JCAP...03..049L} {2018, 049}

\bibitem[\protect\citeauthoryear{{LoVerde}}{{LoVerde}}{2014}]{LoVerde14}
{LoVerde} M.,  2014, \mn@doi [\prd] {10.1103/PhysRevD.90.083518}, \href
  {https://ui.adsabs.harvard.edu/\#abs/2014PhRvD..90h3518L} {90, 083518}

\bibitem[\protect\citeauthoryear{{Ma}, {Hu}  \& {Huterer}}{{Ma}
  et~al.}{2006}]{Ma06}
{Ma} Z.,  {Hu} W.,   {Huterer} D.,  2006, \mn@doi [\apj] {10.1086/497068},
  \href {http://adsabs.harvard.edu/abs/2006ApJ...636...21M} {636, 21}

\bibitem[\protect\citeauthoryear{{Maltoni}, {Schwetz}, {T{\'o}rtola}  \&
  {Valle}}{{Maltoni} et~al.}{2004}]{Maltoni04}
{Maltoni} M.,  {Schwetz} T.,  {T{\'o}rtola} M.,   {Valle} J.~W.~F.,  2004,
  \mn@doi [New Journal of Physics] {10.1088/1367-2630/6/1/122}, \href
  {http://adsabs.harvard.edu/abs/2004NJPh....6..122M} {6, 122}

\bibitem[\protect\citeauthoryear{{Martizzi}, {Teyssier}  \& {Moore}}{{Martizzi}
  et~al.}{2013}]{Martizzi13}
{Martizzi} D.,  {Teyssier} R.,   {Moore} B.,  2013, \mn@doi [\mnras]
  {10.1093/mnras/stt297}, \href
  {http://adsabs.harvard.edu/abs/2013MNRAS.432.1947M} {432, 1947}

\bibitem[\protect\citeauthoryear{{Martizzi}, {Mohammed}, {Teyssier}  \&
  {Moore}}{{Martizzi} et~al.}{2014}]{Martizzi14}
{Martizzi} D.,  {Mohammed} I.,  {Teyssier} R.,   {Moore} B.,  2014, \mn@doi
  [\mnras] {10.1093/mnras/stu440}, \href
  {http://adsabs.harvard.edu/abs/2014MNRAS.440.2290M} {440, 2290}

\bibitem[\protect\citeauthoryear{{Massara}, {Villaescusa-Navarro}  \&
  {Viel}}{{Massara} et~al.}{2014}]{Massara14}
{Massara} E.,  {Villaescusa-Navarro} F.,   {Viel} M.,  2014, \mn@doi [\jcap]
  {10.1088/1475-7516/2014/12/053}, \href
  {http://adsabs.harvard.edu/abs/2014JCAP...12..053M} {12, 053}

\bibitem[\protect\citeauthoryear{{Mead}}{{Mead}}{2017}]{Mead17}
{Mead} A.~J.,  2017, \mn@doi [\mnras] {10.1093/mnras/stw2312}, \href
  {https://ui.adsabs.harvard.edu/\#abs/2017MNRAS.464.1282M} {464, 1282}

\bibitem[\protect\citeauthoryear{{Mead}, {Peacock}, {Heymans}, {Joudaki}  \&
  {Heavens}}{{Mead} et~al.}{2015}]{Mead15}
{Mead} A.~J.,  {Peacock} J.~A.,  {Heymans} C.,  {Joudaki} S.,   {Heavens}
  A.~F.,  2015, \mn@doi [\mnras] {10.1093/mnras/stv2036}, \href
  {http://adsabs.harvard.edu/abs/2015MNRAS.454.1958M} {454, 1958}

\bibitem[\protect\citeauthoryear{{Mead}, {Heymans}, {Lombriser}, {Peacock},
  {Steele}  \& {Winther}}{{Mead} et~al.}{2016}]{Mead16}
{Mead} A.~J.,  {Heymans} C.,  {Lombriser} L.,  {Peacock} J.~A.,  {Steele}
  O.~I.,   {Winther} H.~A.,  2016, \mn@doi [\mnras] {10.1093/mnras/stw681},
  \href {http://adsabs.harvard.edu/abs/2016MNRAS.459.1468M} {459, 1468}

\bibitem[\protect\citeauthoryear{{Mohammed}, {Martizzi}, {Teyssier}  \&
  {Amara}}{{Mohammed} et~al.}{2014}]{Mohammed14}
{Mohammed} I.,  {Martizzi} D.,  {Teyssier} R.,   {Amara} A.,  2014, preprint,
  \href {http://adsabs.harvard.edu/abs/2014arXiv1410.6826M} {} (\mn@eprint
  {arXiv} {1410.6826})

\bibitem[\protect\citeauthoryear{{Natarajan}, {Zentner}, {Battaglia}  \&
  {Trac}}{{Natarajan} et~al.}{2014}]{Natarajan14}
{Natarajan} A.,  {Zentner} A.~R.,  {Battaglia} N.,   {Trac} H.,  2014, \mn@doi
  [\prd] {10.1103/PhysRevD.90.063516}, \href
  {https://ui.adsabs.harvard.edu/\#abs/2014PhRvD..90f3516N} {90, 063516}

\bibitem[\protect\citeauthoryear{{Navarro}, {Frenk}  \& {White}}{{Navarro}
  et~al.}{1997}]{NFW97}
{Navarro} J.~F.,  {Frenk} C.~S.,   {White} S.~D.~M.,  1997, \mn@doi [\apj]
  {10.1086/304888}, \href {http://adsabs.harvard.edu/abs/1997ApJ...490..493N}
  {490, 493}

\bibitem[\protect\citeauthoryear{{Oyama}, {Kohri}  \& {Hazumi}}{{Oyama}
  et~al.}{2016}]{Oyama16}
{Oyama} Y.,  {Kohri} K.,   {Hazumi} M.,  2016, \mn@doi [\jcap]
  {10.1088/1475-7516/2016/02/008}, \href
  {https://ui.adsabs.harvard.edu/abs/2016JCAP...02..008O} {2016, 008}

\bibitem[\protect\citeauthoryear{{Palanque-Delabrouille}
  et~al.,}{{Palanque-Delabrouille} et~al.}{2015}]{Palanque-Delabrouille15}
{Palanque-Delabrouille} N.,  et~al., 2015, \mn@doi [\jcap]
  {10.1088/1475-7516/2015/11/011}, \href
  {http://adsabs.harvard.edu/abs/2015JCAP...11..011P} {11, 011}

\bibitem[\protect\citeauthoryear{{Parimbelli}, {Viel}  \&
  {Sefusatti}}{{Parimbelli} et~al.}{2018}]{Parimbelli18}
{Parimbelli} G.,  {Viel} M.,   {Sefusatti} E.,  2018, preprint, \href
  {https://ui.adsabs.harvard.edu/#abs/2018arXiv180906634P} {p.
  arXiv:1809.06634} (\mn@eprint {arXiv} {1809.06634})

\bibitem[\protect\citeauthoryear{{Peacock} \& {Smith}}{{Peacock} \&
  {Smith}}{2000}]{Peacock00}
{Peacock} J.~A.,  {Smith} R.~E.,  2000, \mn@doi [\mnras]
  {10.1046/j.1365-8711.2000.03779.x}, \href
  {http://adsabs.harvard.edu/abs/2000MNRAS.318.1144P} {318, 1144}

\bibitem[\protect\citeauthoryear{{Peacock}, {Schneider}, {Efstathiou}, {Ellis},
  {Leibundgut}, {Lilly}  \& {Mellier}}{{Peacock} et~al.}{2006}]{Peacock06}
{Peacock} J.~A.,  {Schneider} P.,  {Efstathiou} G.,  {Ellis} J.~R.,
  {Leibundgut} B.,  {Lilly} S.~J.,   {Mellier} Y.,  2006, Technical report,
  {ESA-ESO Working Group on ``Fundamental Cosmology''}.
 (\mn@eprint {} {astro-ph/0610906})

\bibitem[\protect\citeauthoryear{{Pietroni}}{{Pietroni}}{2008}]{Pietroni08}
{Pietroni} M.,  2008, \mn@doi [Journal of Cosmology and Astro-Particle Physics]
  {10.1088/1475-7516/2008/10/036}, \href
  {https://ui.adsabs.harvard.edu/#abs/2008JCAP...10..036P} {2008, 036}

\bibitem[\protect\citeauthoryear{{Planck Collaboration} et~al.,}{{Planck
  Collaboration} et~al.}{2016}]{Planck15}
{Planck Collaboration} et~al., 2016, \mn@doi [\aap]
  {10.1051/0004-6361/201525830}, \href
  {http://adsabs.harvard.edu/abs/2016A%26A...594A..13P} {594, A13}

\bibitem[\protect\citeauthoryear{{Planck Collaboration} et~al.,}{{Planck
  Collaboration} et~al.}{2018}]{Planck18}
{Planck Collaboration} et~al., 2018, preprint, \href
  {http://adsabs.harvard.edu/abs/2018arXiv180706209P} {} (\mn@eprint {arXiv}
  {1807.06209})

\bibitem[\protect\citeauthoryear{{Pontzen} \& {Governato}}{{Pontzen} \&
  {Governato}}{2012}]{Pontzen&Governato12}
{Pontzen} A.,  {Governato} F.,  2012, \mn@doi [\mnras]
  {10.1111/j.1365-2966.2012.20571.x}, \href
  {http://adsabs.harvard.edu/abs/2012MNRAS.421.3464P} {421, 3464}

\bibitem[\protect\citeauthoryear{{Press} \& {Schechter}}{{Press} \&
  {Schechter}}{1974}]{Press&Schechter74}
{Press} W.~H.,  {Schechter} P.,  1974, \mn@doi [\apj] {10.1086/152650}, \href
  {http://adsabs.harvard.edu/abs/1974ApJ...187..425P} {187, 425}

\bibitem[\protect\citeauthoryear{{Rudd}, {Zentner}  \& {Kravtsov}}{{Rudd}
  et~al.}{2008}]{Rudd08}
{Rudd} D.~H.,  {Zentner} A.~R.,   {Kravtsov} A.~V.,  2008, \mn@doi [\apj]
  {10.1086/523836}, \href {http://adsabs.harvard.edu/abs/2008ApJ...672...19R}
  {672, 19}

\bibitem[\protect\citeauthoryear{{Ruggeri}, {Percival}, {Gil-Mar{\'\i}n},
  {Zhu}, {Zhao}  \& {Wang}}{{Ruggeri} et~al.}{2017}]{Ruggeri17}
{Ruggeri} R.,  {Percival} W.~J.,  {Gil-Mar{\'\i}n} H.,  {Zhu} F.,  {Zhao}
  G.-B.,   {Wang} Y.,  2017, \mn@doi [\mnras] {10.1093/mnras/stw2422}, \href
  {https://ui.adsabs.harvard.edu/#abs/2017MNRAS.464.2698R} {464, 2698}

\bibitem[\protect\citeauthoryear{{Schaye} et~al.,}{{Schaye}
  et~al.}{2010}]{Schaye10}
{Schaye} J.,  et~al., 2010, \mn@doi [\mnras]
  {10.1111/j.1365-2966.2009.16029.x}, \href
  {http://adsabs.harvard.edu/abs/2010MNRAS.402.1536S} {402, 1536}

\bibitem[\protect\citeauthoryear{{Schneider} \& {Bridle}}{{Schneider} \&
  {Bridle}}{2010}]{Schneider10}
{Schneider} M.~D.,  {Bridle} S.,  2010, \mn@doi [\mnras]
  {10.1111/j.1365-2966.2009.15956.x}, \href
  {https://ui.adsabs.harvard.edu/\#abs/2010MNRAS.402.2127S} {402, 2127}

\bibitem[\protect\citeauthoryear{{Schneider} \& {Teyssier}}{{Schneider} \&
  {Teyssier}}{2015}]{Schneider&Teyssier15}
{Schneider} A.,  {Teyssier} R.,  2015, \mn@doi [\jcap]
  {10.1088/1475-7516/2015/12/049}, \href
  {http://adsabs.harvard.edu/abs/2015JCAP...12..049S} {12, 049}

\bibitem[\protect\citeauthoryear{{Seljak}}{{Seljak}}{2000}]{Seljak00}
{Seljak} U.,  2000, \mn@doi [\mnras] {10.1046/j.1365-8711.2000.03715.x}, \href
  {http://adsabs.harvard.edu/abs/2000MNRAS.318..203S} {318, 203}

\bibitem[\protect\citeauthoryear{{Semboloni}, {Hoekstra}, {Schaye}, {van
  Daalen}  \& {McCarthy}}{{Semboloni} et~al.}{2011}]{Semboloni11}
{Semboloni} E.,  {Hoekstra} H.,  {Schaye} J.,  {van Daalen} M.~P.,   {McCarthy}
  I.~G.,  2011, \mn@doi [\mnras] {10.1111/j.1365-2966.2011.19385.x}, \href
  {http://adsabs.harvard.edu/abs/2011MNRAS.417.2020S} {417, 2020}

\bibitem[\protect\citeauthoryear{{Seo} \& {Eisenstein}}{{Seo} \&
  {Eisenstein}}{2003}]{Seo03}
{Seo} H.-J.,  {Eisenstein} D.~J.,  2003, \mn@doi [\apj] {10.1086/379122}, \href
  {https://ui.adsabs.harvard.edu/\#abs/2003ApJ...598..720S} {598, 720}

\bibitem[\protect\citeauthoryear{{Seo} \& {Eisenstein}}{{Seo} \&
  {Eisenstein}}{2007}]{Seo07}
{Seo} H.-J.,  {Eisenstein} D.~J.,  2007, \mn@doi [\apj] {10.1086/519549}, \href
  {https://ui.adsabs.harvard.edu/\#abs/2007ApJ...665...14S} {665, 14}

\bibitem[\protect\citeauthoryear{{Sheth} \& {Tormen}}{{Sheth} \&
  {Tormen}}{1999}]{Sheth&Tormen99}
{Sheth} R.~K.,  {Tormen} G.,  1999, \mn@doi [\mnras]
  {10.1046/j.1365-8711.1999.02692.x}, \href
  {https://ui.adsabs.harvard.edu/\#abs/1999MNRAS.308..119S} {308, 119}

\bibitem[\protect\citeauthoryear{{Smith} et~al.,}{{Smith}
  et~al.}{2003}]{Smith03}
{Smith} R.~E.,  et~al., 2003, \mn@doi [\mnras]
  {10.1046/j.1365-8711.2003.06503.x}, \href
  {http://adsabs.harvard.edu/abs/2003MNRAS.341.1311S} {341, 1311}

\bibitem[\protect\citeauthoryear{{Sprenger}, {Archidiacono}, {Brinckmann},
  {Clesse}  \& {Lesgourgues}}{{Sprenger} et~al.}{2018}]{Sprenger18}
{Sprenger} T.,  {Archidiacono} M.,  {Brinckmann} T.,  {Clesse} S.,
  {Lesgourgues} J.,  2018, arXiv e-prints, \href
  {https://ui.adsabs.harvard.edu/\#abs/2018arXiv180108331S} {p.
  arXiv:1801.08331}

\bibitem[\protect\citeauthoryear{{Takada} \& {Jain}}{{Takada} \&
  {Jain}}{2004}]{Takada&Jain04}
{Takada} M.,  {Jain} B.,  2004, \mn@doi [\mnras]
  {10.1111/j.1365-2966.2004.07410.x}, \href
  {http://adsabs.harvard.edu/abs/2004MNRAS.348..897T} {348, 897}

\bibitem[\protect\citeauthoryear{{Takahashi}, {Sato}, {Nishimichi}, {Taruya}
  \& {Oguri}}{{Takahashi} et~al.}{2012}]{Takahashi12}
{Takahashi} R.,  {Sato} M.,  {Nishimichi} T.,  {Taruya} A.,   {Oguri} M.,
  2012, \mn@doi [\apj] {10.1088/0004-637X/761/2/152}, \href
  {https://ui.adsabs.harvard.edu/#abs/2012ApJ...761..152T} {761, 152}

\bibitem[\protect\citeauthoryear{{Taylor}, {Kitching}, {Bacon}  \&
  {Heavens}}{{Taylor} et~al.}{2007}]{Taylor07}
{Taylor} A.~N.,  {Kitching} T.~D.,  {Bacon} D.~J.,   {Heavens} A.~F.,  2007,
  \mn@doi [\mnras] {10.1111/j.1365-2966.2006.11257.x}, \href
  {http://adsabs.harvard.edu/abs/2007MNRAS.374.1377T} {374, 1377}

\bibitem[\protect\citeauthoryear{{Tegmark}, {Taylor}  \& {Heavens}}{{Tegmark}
  et~al.}{1997}]{Tegmark97}
{Tegmark} M.,  {Taylor} A.~N.,   {Heavens} A.~F.,  1997, \mn@doi [\apj]
  {10.1086/303939}, \href {http://adsabs.harvard.edu/abs/1997ApJ...480...22T}
  {480, 22}

\bibitem[\protect\citeauthoryear{{Teyssier}, {Pontzen}, {Dubois}  \&
  {Read}}{{Teyssier} et~al.}{2013}]{Teyssier13}
{Teyssier} R.,  {Pontzen} A.,  {Dubois} Y.,   {Read} J.~I.,  2013, \mn@doi
  [\mnras] {10.1093/mnras/sts563}, \href
  {http://adsabs.harvard.edu/abs/2013MNRAS.429.3068T} {429, 3068}

\bibitem[\protect\citeauthoryear{{Troxel} \& {Ishak}}{{Troxel} \&
  {Ishak}}{2012}]{Troxel12}
{Troxel} M.~A.,  {Ishak} M.,  2012, \mn@doi [\mnras]
  {10.1111/j.1365-2966.2012.20987.x}, \href
  {https://ui.adsabs.harvard.edu/\#abs/2012MNRAS.423.1663T} {423, 1663}

\bibitem[\protect\citeauthoryear{{Troxel} \& {Ishak}}{{Troxel} \&
  {Ishak}}{2015}]{Troxel15}
{Troxel} M.~A.,  {Ishak} M.,  2015, \mn@doi [\physrep]
  {10.1016/j.physrep.2014.11.001}, \href
  {https://ui.adsabs.harvard.edu/\#abs/2015PhR...558....1T} {558, 1}

\bibitem[\protect\citeauthoryear{{Vagnozzi}, {Giusarma}, {Mena}, {Freese},
  {Gerbino}, {Ho}  \& {Lattanzi}}{{Vagnozzi} et~al.}{2017}]{Vagnozzi17}
{Vagnozzi} S.,  {Giusarma} E.,  {Mena} O.,  {Freese} K.,  {Gerbino} M.,  {Ho}
  S.,   {Lattanzi} M.,  2017, \mn@doi [\prd] {10.1103/PhysRevD.96.123503},
  \href {https://ui.adsabs.harvard.edu/abs/2017PhRvD..96l3503V} {96, 123503}

\bibitem[\protect\citeauthoryear{{Vagnozzi}, {Brinckmann}, {Archidiacono},
  {Freese}, {Gerbino}, {Lesgourgues}  \& {Sprenger}}{{Vagnozzi}
  et~al.}{2018}]{Vagnozzi18}
{Vagnozzi} S.,  {Brinckmann} T.,  {Archidiacono} M.,  {Freese} K.,  {Gerbino}
  M.,  {Lesgourgues} J.,   {Sprenger} T.,  2018, \mn@doi [\jcap]
  {10.1088/1475-7516/2018/09/001}, \href
  {https://ui.adsabs.harvard.edu/abs/2018JCAP...09..001V} {2018, 001}

\bibitem[\protect\citeauthoryear{{Wang}, {Chuang}  \& {Hirata}}{{Wang}
  et~al.}{2013}]{Wang13}
{Wang} Y.,  {Chuang} C.-H.,   {Hirata} C.~M.,  2013, \mn@doi [\mnras]
  {10.1093/mnras/stt068}, \href
  {https://ui.adsabs.harvard.edu/#abs/2013MNRAS.430.2446W} {430, 2446}

\bibitem[\protect\citeauthoryear{{White}, {Song}  \& {Percival}}{{White}
  et~al.}{2009}]{White09}
{White} M.,  {Song} Y.-S.,   {Percival} W.~J.,  2009, \mn@doi [\mnras]
  {10.1111/j.1365-2966.2008.14379.x}, \href
  {https://ui.adsabs.harvard.edu/\#abs/2009MNRAS.397.1348W} {397, 1348}

\bibitem[\protect\citeauthoryear{{Yao}, {Ishak}, {Lin}  \& {Troxel}}{{Yao}
  et~al.}{2017}]{Yao17}
{Yao} J.,  {Ishak} M.,  {Lin} W.,   {Troxel} M.,  2017, \mn@doi [Journal of
  Cosmology and Astro-Particle Physics] {10.1088/1475-7516/2017/10/056}, \href
  {https://ui.adsabs.harvard.edu/\#abs/2017JCAP...10..056Y} {2017, 056}

\bibitem[\protect\citeauthoryear{{Yao}, {Ishak}  \& {Troxel}}{{Yao}
  et~al.}{2019}]{Yao19}
{Yao} J.,  {Ishak} M.,   {Troxel} M.~A.,  2019, \mn@doi [\mnras]
  {10.1093/mnras/sty3188}, \href
  {https://ui.adsabs.harvard.edu/\#abs/2019MNRAS.483..276Y} {483, 276}

\bibitem[\protect\citeauthoryear{{Y{\`e}che}, {Palanque-Delabrouille}, {Baur}
  \& {du Mas des Bourboux}}{{Y{\`e}che} et~al.}{2017}]{Yeche17}
{Y{\`e}che} C.,  {Palanque-Delabrouille} N.,  {Baur} J.,   {du Mas des
  Bourboux} H.,  2017, \mn@doi [\jcap] {10.1088/1475-7516/2017/06/047}, \href
  {http://adsabs.harvard.edu/abs/2017JCAP...06..047Y} {6, 047}

\bibitem[\protect\citeauthoryear{{Zentner}, {Semboloni}, {Dodelson}, {Eifler},
  {Krause}  \& {Hearin}}{{Zentner} et~al.}{2013}]{Zentner12}
{Zentner} A.~R.,  {Semboloni} E.,  {Dodelson} S.,  {Eifler} T.,  {Krause} E.,
  {Hearin} A.~P.,  2013, \mn@doi [\prd] {10.1103/PhysRevD.87.043509}, \href
  {http://adsabs.harvard.edu/abs/2013PhRvD..87d3509Z} {87, 043509}

\bibitem[\protect\citeauthoryear{{Zhang}}{{Zhang}}{2010}]{Zhang10b}
{Zhang} P.,  2010, \mn@doi [\apj] {10.1088/0004-637X/720/2/1090}, \href
  {https://ui.adsabs.harvard.edu/\#abs/2010ApJ...720.1090Z} {720, 1090}

\bibitem[\protect\citeauthoryear{{van Daalen}, {Schaye}, {Booth}  \& {Dalla
  Vecchia}}{{van Daalen} et~al.}{2011}]{vandaalen11}
{van Daalen} M.~P.,  {Schaye} J.,  {Booth} C.~M.,   {Dalla Vecchia} C.,  2011,
  \mn@doi [\mnras] {10.1111/j.1365-2966.2011.18981.x}, \href
  {http://adsabs.harvard.edu/abs/2011MNRAS.415.3649V} {415, 3649}

\bibitem[\protect\citeauthoryear{{van Daalen}, {Schaye}, {McCarthy}, {Booth}
  \& {Dalla Vecchia}}{{van Daalen} et~al.}{2014}]{vanDaalen14}
{van Daalen} M.~P.,  {Schaye} J.,  {McCarthy} I.~G.,  {Booth} C.~M.,   {Dalla
  Vecchia} C.,  2014, \mn@doi [\mnras] {10.1093/mnras/stu482}, \href
  {http://adsabs.harvard.edu/abs/2014MNRAS.440.2997V} {440, 2997}

\makeatother
\end{thebibliography}

\vspace{-5.mm}

\appendix 

\section{The halo model}
\label{sec:halo_model}
The halo model provides a highly useful framework for interpreting the behaviour of the matter distribution on non-linear scales \citep{Seljak00, Peacock00} (note that several fundamentals of the halo model we outline here overlap with a primer in \citet{Copeland18}). The prescription of spherically-collapsed haloes positioned randomly in the linear density field permits the matter power spectrum, which describes the two-point statistics of density perturbations, to be written as the sum of two separate contributions,
\begin{equation}
\Delta^2\left(k\right) = \Delta_{1h}^2\left(k\right) + \Delta^2_{2h}\left(k\right).
\end{equation} 
Here we have used the dimensionless form of the power spectrum, $\Delta^2\left(k\right)$. Correlations between different haloes are represented by the 2-halo term, $\Delta_{2h}^2\left(k\right)$, which is commonly approximated as the linear matter power spectrum because it is relevant on large scales. Internal halo structure is described by the 1-halo term, $\Delta_{1h}^2\left(k\right)$. This is constructed by summing the self-convolutions of haloes with mass $M$ over the full mass range,
\begin{equation}
\Delta_{1h}^2\left(k\right) = \frac{k^3}{2\pi^2}\int_0^{\infty}\mathrm{d}M\, \frac{M^2 n\left(M\right)}{\bar{\rho}^2} u^2\left(k\mathopen{|}\mathclose M\right).
\label{eq:1halo} 
\end{equation}
The integrand is weighted by the total number density of halo pairs with mass $M$. The halo mass function for the comoving number density of haloes within $\mathrm{d}M$ is provided by $n\left(M\right)$ while $\bar{\rho}$ is the background matter density. \citet{Press&Schechter74} showed that the mass function is nearly independent of cosmology if expressed in terms of the density peak height, $\nu \equiv \delta_c/\sigma\left(M\right)$, given by the spherical collapse overdensity, $\delta_c$, when the density field is smoothed at the mass scale, $M$. \citet{Sheth&Tormen99} determined a form of the `universal' mass function,
\begin{equation}
f\left(\nu\right) = \frac{M}{\bar{\rho}}n\left(M\right)\,\frac{\mathrm{d}M}{\mathrm{d}\nu} 
\label{eq:massfunc}
\end{equation} 
that fit simulations to be
\begin{equation}
f\left(\nu\right) = 0.2162\left[1+\left(0.707\nu^2\right)^{-0.3}\right]\exp\left[-\frac{0.707\nu^2}{2}\right].
\end{equation} 
This mass function depends on the square root of the variance of the matter distribution, 
\begin{equation}
\sigma^2\left(R\right) = \int_0^{\infty} \mathrm{d}\ln k\; \Delta_{\rm{lin}}^2\left(k\right) W^2\left(k R\right),
\end{equation}
at scale, $R$, defining the radius of a sphere enclosing homogeneous mass, $M$. The field is smoothed by the window function of the spherical top-hat profile. The dimensionless linear matter power spectrum is equivalent to the fractional contribution to $\sigma^2\left(R\right)$ per logarithmic interval of $k$, and therefore indirectly influences the 1-halo power. 

When normalized by its mass, the Fourier space density profile of a spherical halo is given by
\begin{equation}
u\left(k\mathopen{|}\mathclose M\right) = \frac{4\pi}{M}\int_0^{r_v}r^2\mathrm{d}r\, \frac{\sin\left(kr\right)}{kr}\rho\left(r, M\right),
\label{eq:fourierprofile}
\end{equation}
which transforms a real-space density profile, such as the common NFW case \citep{NFW97}, 
\begin{equation}
\rho\left(r,M\right) = \frac{\rho_s}{\left(\frac{r}{r_s}\right)\left[1+\left(\frac{r}{r_s}\right)\right]^2}.
\label{eq:NFW}
\end{equation}
The inner (linear) and outer (cubic) regions of the halo are separated at the scale radius, $r_s$, which also determines the normalization density, $\rho_s$. As a halo stops collapsing once it reaches a virial equilibrium of gravitationally interacting matter shells exchanging energy, the integral in equation~\eqref{eq:fourierprofile} is truncated at the associated virial radius, 
\begin{equation}
r_v=\left(\frac{3M}{4\pi\bar{\rho}\Delta_v}\right)^{\frac{1}{3}}.
\end{equation}     
This quantity and the related virial density, $\Delta_v$, which depends on the specific cosmology in which spherical collapse occurs, are useful parameters for characterizing the final, virialized halo. The relationship between the scale radius and virial radius, $r_s=r_v/c$, is determined through the \citet{Bullock01} concentration factor relation, 
\begin{equation}
c\left(M,z\right)=A_B\frac{1+z_f}{1+z},
\label{eq:conc}
\end{equation}  
which is sensitive to both cosmology and halo mass via the formation redshift, $z_f$ at which a fraction $f$ of the total matter in a density perturbation collapses. \citet{Bullock01} find $f=0.01$ produces the most accurate fits of their model to N-body simulations. The amplitude, $A_B$, is discussed in detail in \S~\ref{subsec:baryonparams}.

\section{Probes of Large Scale Structure}
\label{sec:probes}
In this work we explore the impact that marginalizing over baryons and the amplitude of the intrinsic alignment signal has on the capacity of Stage IV surveys to measure $\Sigma$ and distinguish neutrino mass hierarchies. To do so we perform Fisher analyses using weak lensing and spectroscopic galaxy clustering as signals. Here we briefly review the theory behind these probes and the Fisher formalism used to make forecasts.

\subsection{Weak gravitational lensing}
\label{subsec:wgl}
We approximate the source galaxy distribution as
\begin{equation}
n\left(z\right) \propto z^2\exp\left[-\left(\frac{z}{z_0}\right)^{\frac{3}{2}}\right],
\end{equation}  
\\
where $z_0=0.636$, which roughly matches that expected for a Euclid-like survey \citep{Laurejis11}\footnote{The weak lensing formalism presented here has also been detailed in \citet{Copeland18}, except for intrinsic alignments which were not discussed in that work.}. Within $\mathrm{d}\chi$ of comoving position, $\chi$, the number of source galaxies is $\mathrm{d}\chi n\left(\chi\right)$. This is an important ingredient in the weak lensing convergence power spectrum, which, in the Limber approximation is
\begin{equation}
C_{\ell,ij} = \frac{9}{4}\Omega_m^2\left(\frac{H_0}{c}\right)^4\int_0^{\chi_{\mathrm{max}}}\,\mathrm{d}\chi\, \frac{g_i\left(\chi\right)g_j\left(\chi\right)}{a^2\left(\chi\right)}P\left(k=\frac{\ell}{f_K\left(\chi\right)},\chi\right),
\label{eq:convergencepower}
\end{equation}
where $\left(i,j\right)$ are tomographic redshift bin labels, $f_K\left(\chi\right)$ is the comoving angular distance, and $g_i\left(\chi\right)$ is the total weighting function, determined by the separation between sources and lenses over the distribution up to the survey limit, $\chi_{\rm{\max}}$. This is given by
\begin{equation}
g_i\left(\chi\right) = \int_{\chi}^{\chi_{\mathrm{max}}}\mathrm{d}\chi^{\prime}\, n_i\left(\chi^{\prime}\right) \frac{f_K\left(\chi-\chi^{\prime}\right)}{f_K\left(\chi^{\prime}\right)}.
\end{equation}
It is necessary to address photometric redshift errors. Within each redshift bin the galaxy distribution when normalized over the survey is given by 
\begin{equation}
n_i\left(z\right) = \frac{n\left(z\right)\, \int_{z_{i,-}}^{z_{i,+}}\mathrm{d}z_{ph}\, p_{ph}\left(z_{ph}|z\right)}{\int_{z_{\rm{min}}}^{z_{\rm{max}}}\mathrm{d}z^{\prime}\,n\left(z^{\prime}\right)\, \int_{z_{i,-}}^{z_{i,+}}\mathrm{d}z_{ph}\,  p_{ph}\left(z_{ph}|z^{\prime}\right)}, 
\end{equation}
where $p_{ph}\left(z_{ph}|z\right)$ is the probability of galaxies at $z$ being measured at redshift $z_{ph}$. The integrals are equivalent to convolving the true distribution with a binning top-hat function, whilst accounting for the probability of galaxies being measured at $z_{ph}$ outside the bin. The denominator is the result of normalizing over the bin. This can be shown to be equivalent to the distribution used in \citet{Taylor07}. We use the frequently employed probability distribution \citep{Ma06, Taylor07}, 
\begin{multline}
p_{ph}\left(z_{ph}|z\right) = \frac{1}{\sqrt{2\pi}\sigma_z\left(1+z\right)}\exp\left\{-\left[\frac{z-z_{ph}}{\sqrt{2}\sigma_z\left(1+z\right)}\right]^2\right\}. 
\end{multline}
For the photometric redshift error, we use the \citet{Laurejis11} value of $\sigma_z=0.05$.

\subsubsection{Intrinsic alignments}
The tidal forces that correlate ellipticities, $\epsilon$, and shears, $\gamma$, are an intrinsic contribution to the observed signal. Accounting for this to linear order by writing the observed ellipticity as $\epsilon = \gamma + \epsilon^I$, the weak lensing power spectrum becomes
\begin{equation}
C_{\ell,ij} = C_{\ell,ij}^{\gamma\gamma} + C_{\ell,ij}^{I\gamma} + C_{\ell,ij}^{II}, 
\label{eq:iacelltot}
\end{equation}
where $C_{\ell,ij}^{\gamma\gamma}$ is given by equation~\eqref{eq:convergencepower} and the term, $C_{\ell,ij}^{I\gamma}$, encompasses correlations of foreground ellipticities with background shear and of foreground shear with background ellipticities. The latter correlations should be very small under the assumption that redshift bins do not overlap. The IA-shear term is given by 
\begin{align}
C_{\ell,ij}^{I \gamma} = \frac{3}{2}\Omega_m\left(\frac{H_0}{c}\right)^2\int_0^{\chi_{\mathrm{max}}}\,\mathrm{d}\chi\, \frac{g_i\left(\chi\right)n_j\left(\chi\right) + g_j\left(\chi\right)n_i\left(\chi\right)}{a\left(\chi\right)\chi}
 \nonumber
\\
\times \, P_{mI}\left(k=\frac{\ell}{f_K\left(\chi\right)},\chi\right),
\end{align}
and accounts for terms arising from the misordering of redshift bins as well as those from a physical signal. The IA-IA term is given by
\begin{equation}
C_{\ell,ij}^{II} = \int_0^{\chi_{\mathrm{max}}}\,\mathrm{d}\chi\, \frac{n_i\left(\chi\right)n_j\left(\chi\right)}{\chi^2}P_{II}\left(k=\frac{\ell}{f_K\left(\chi\right)},\chi\right).
\end{equation}
The matter-IA and IA-IA source power spectra in the non-linear alignment model \citep{Bridle&King07} are
\begin{equation}
P_{mI}\left(k,\chi\right) = -a_{\rm{IA}}\frac{c_{\rm{IA}}\rho_{\rm{crit}}\Omega_m}{D\left(z\left(\chi\right)\right)}P_m\left(k,\chi\right)
\end{equation}
and
\begin{equation}
P_{II}\left(k,\chi\right) = \left(a_{\rm{IA}}\frac{c_{\rm{IA}}\rho_{\rm{crit}}\Omega_m}{D\left(z\left(\chi\right)\right)}\right)^2P_m\left(k,\chi\right).
\end{equation}
$D\left(z\right)$ is the linear growth factor.
 
The amplitude, $a_{\rm{IA}}$, can be taken as a free parameter with a fiducial value of unity, while fits to SuperCOSMOS galaxy data determine the normalization, $c_{\rm{IA}}\approx 0.0134/\rho_{\rm{crit}}$, in terms of the critical density \citep{Brown02, Hirata04}. In Figure~\ref{fig:lenspower049} we show comparisons of the shear-shear, IA-shear and IA-IA auto- and cross- lensing power spectra for several combinations of redshift bins. The IA-IA contributions to the signal are most significant for low redshift auto-power spectra, while the IA-shear term has its greatest impact on cross-power spectra of widely separated bins because it describes the correlations of foreground galaxy ellipticities with background shear. 

\subsection{Spectroscopic galaxy clustering}
The clustering of galaxies provides an additional probe through the anisotropic galaxy power spectrum, $P_{gg}\left(k,\mu;z\right)$, for which the specific observable is H$\alpha$-emitting galaxies for a Euclid-like survey. We outline an approach to constructing $P_{gg}\left(k,\mu;z\right)$ that closely follows e.g., \citet{Ballinger96} and \citet{Seo03}. Sources of apparent anisotropy make it necessary to include the dependence on the cosine of the wave mode and radial vector, $\mu$. These include redshift-space distortions arising from a peculiar velocity component in observed redshifts \citep{Kaiser87}. As a result, the observed strength of galaxy clustering will be sensitive to the orientation of the pair-separation vector. To account for this, the Kaiser factor is incorporated into $P_{gg}\left(k,\mu;z\right)$. The underlying dark matter distribution is biased by the galaxies that trace it \citep[see e.g.,][]{Desjacques18}, so the galaxy power spectrum can then be constructed from the matter power spectrum according to
\begin{equation}
P_{gg}\left(k,\mu;z\right) = \left(b\sigma_8\left(z\right) + f\sigma_8\left(z\right)\mu^2\right)^2\frac{P_m\left(k;z\right)}{\sigma_8^2\left(z\right)}, 
\end{equation}
where $b\left(z\right)=\sqrt{1+z}$ is the effective bias of the sample of H$\alpha$-emitters used in \citet{Amendola18} and $f\left(z\right)$ is the logarithmic growth rate of structure, which relates the peculiar velocity to the density. The redshift-dependence of the (linear) matter power spectrum is removed by factoring out $\sigma_8\left(z\right)$, which leads to $b\sigma_8\left(z\right)$ and $f\sigma_8\left(z\right)$ becoming free parameters in galaxy clustering forecasts \citep{White09}.    

Converting measured angles and redshifts into radial and transverse distances requires assumptions about cosmology. If these are incorrect, distortions arise through the AP effect. The ratios,
\begin{equation}
\alpha_{\perp}\left(z\right) \equiv \frac{D_A\left(z\right)}{D_{A,\rm{fid}}\left(z\right)}
\end{equation}
and
\begin{equation}
\alpha_{\parallel}\left(z\right) \equiv \frac{H_{\rm{fid}}\left(z\right)}{H\left(z\right)},
\end{equation}
of the assumed and true angular diameter distance and Hubble parameter \citep{Alcock&Paczynski79} change wavenumbers and angles from their `fiducial' values to \citep{Ballinger96}
\begin{equation}
k\left(k_{\rm{fid}},\mu_{\rm{fid}};z\right) = \frac{k_{\rm{fid}}}{\alpha_{\perp}\left(z\right)}\sqrt{1+\mu_{\rm{fid}}^2\left(\frac{\alpha_{\perp}^2\left(z\right)}{\alpha_{\parallel}^2\left(z\right)} - 1\right)}
\end{equation}
and
\begin{equation}
\mu\left(\mu_{\rm{fid}};z\right) = \mu_{\rm{fid}}\frac{\alpha_{\perp}\left(z\right)}{\alpha_{\parallel}\left(z\right)}\frac{1}{\sqrt{1+\mu_{\rm{fid}}^2\left(\frac{\alpha_{\perp}^2\left(z\right)}{\alpha_{\parallel}^2\left(z\right)} - 1\right)}}.
\end{equation}
Under these projections the matter power spectrum transforms as
\begin{equation}
P_m\left(k_{\rm{fid}},\mu_{\rm{fid}};z\right) \longrightarrow \frac{1}{\alpha_{\perp}^2\alpha_{\parallel}}P_m\left(k,\mu;z\right).
\end{equation}
Hereafter, unless explicitly stated, $k$ and $\mu$ may be assumed to account for the AP effect.  

Finally, redshift uncertainties along the line-of-sight are accounted for via the factor \citep[e.g.,][]{Wang13},
\begin{equation}
F_z\left(k;z\right) = \exp\left[-k^2\mu^2\sigma_r^2\left(z\right)\right],
\end{equation}
where
\begin{equation}
\sigma_r\left(z\right) = \frac{c}{H\left(z\right)}\left(1+z\right)\sigma_{z=0},
\end{equation}
with the spectroscopic redshift error given by $\sigma_{z=0}=0.001$ \citep{Laurejis11}. The full observed linear galaxy power spectrum is then
\begin{align}
P_{gg}\left(k,\mu;z\right) = \frac{1}{\alpha_{\perp}^2\alpha_{\parallel}}\left(b\sigma_8\left(z\right) + f\sigma_8\left(z\right)\mu^2\right)^2\frac{P_m\left(k;z\right)}{\sigma_8^2\left(z\right)}F_z\left(k;z\right)  \nonumber \\
+ P_{\rm{shot}}\left(z\right).
\end{align}
The final term represents any residual shot noise that is not accounted for by Poisson sampling the underlying CDM density field with H$\alpha$-emitting galaxies. We assign a fiducial value of $P_{\rm{shot,fid}}=0$ for all $z$. Non-linear contributions, such as the Finger-of-God effect \citep{Hamilton98} or the damping of the BAO component of $P\left(k;z\right)$ \citep{Eisenstein07, Seo07}, are beyond the scope of this work and are not included.

\subsection{Fisher formalism}
The Fisher information matrix is given by the expectation of the curvature of a multi-dimensional likelihood function, $L\left(\bmath{x}\lvert\Theta\right)$, evaluated at the most likely values of a parameter set, $\Theta=\left(\theta_1,...,\theta_N\right)$, for the data set, $\bmath{x}$ \citep*[e.g.,][]{Tegmark97}. These values are taken to be the fiducial ones in our model. The inverse of this matrix provides the parameter covariances that underpin our survey forecasts. Therefore, we write the Fisher matrix as
\begin{equation}
F_{\alpha\beta} \equiv \Big\langle \frac{\partial^2 \mathcal{L}}{\partial\theta_{\alpha}\partial\theta_{\beta}} \Big\rangle,
\end{equation} 
where $\mathcal{L}=-\ln L$. Assuming a Gaussian likelihood,
\begin{equation}
L=\frac{1}{\sqrt{\left(2\pi\right)^N\det \bmath{C}}}\exp\left[-\frac{1}{2}\left(\bmath{x}-\boldsymbol{\mu}\right)\bmath{C}^{-1}\left(\bmath{x}-\boldsymbol{\mu}\right)^{T}\right],
\end{equation}
where $\bmath{C}$ is the data covariance matrix and $\boldsymbol{\mu}=\langle\bmath{x}\rangle$ is the mean data vector, the result,
\begin{equation}
F_{\alpha\beta} = \frac{1}{2} \mathrm{Tr}\left[\bmath{C}^{-1}\bmath{C}_{,\alpha}\bmath{C}^{-1}\bmath{C}_{,\beta} + \bmath{C}^{-1}M_{\alpha\beta}\right],
\end{equation}
can be derived. Here, $M_{\alpha\beta}=\boldsymbol\mu_{,\alpha}\boldsymbol\mu_{,\beta}^T+\boldsymbol\mu_{,\beta}\boldsymbol\mu_{,\alpha}^T$ is the expectation value of the second derivative of the data matrix, $\left(\bmath{x}-\boldsymbol{\mu}\right)\left(\bmath{x}-\boldsymbol{\mu}\right)^{T}$, under the Gaussian approximation. Derivatives with respect to parameters are written via the shorthand $,\alpha \equiv \partial/\partial\Theta_{\alpha}$. 

We apply the Fisher formalism for two observables: the weak lensing convergence power spectrum and the galaxy power spectrum, which are both treated as Gaussian. For the former we have \citep[][]{Tegmark97, Takada&Jain04}
\begin{equation}
F_{\alpha\beta} = \frac{1}{2}f_{\rm{sky}}\sum_{\ell}  \left(2\ell + 1\right) \sum_{\left(ij\right)}\sum_{\left(pq\right)} C^{ij}_{\ell,\alpha}C^{pq}_{\ell,\beta} \left[\rm{Cov}^{-1}\right]_{\ell,\left(ij\right),\left(pq\right)},
\label{eq:fishertrace}
\end{equation}
which sums over the spherical harmonic $\ell$ and $m$ modes. The fraction of the sky within the scope of the survey, $f_{\rm{sky}}$, is included as a limiting factor. The auto- and cross-correlations of observed power in redshift bins, $i,j,p,q=\left(1,...,N_{z,\rm{WL}}\right)$, are used to construct the covariance matrix,
\begin{equation}
{\rm{Cov}}_{\ell,\left(ij\right),\left(pq\right)} = \hat{C}^{ip}_{\ell}\hat{C}_{\ell}^{jq} + \hat{C}_{\ell}^{iq}\hat{C}_{\ell}^{jp}.
\end{equation}
If we ignore non-linear couplings between modes (see e.g., \citet{Kiessling11} for explorations of how to account for these additional correlations within the Fisher formalism), then the covariance matrix can be treated as block diagonal in $\ell$. Shape noise, averaged over the number of galaxies, is added to the auto-correlations of power within each bin such that 
\begin{equation}
\hat{C}_{\ell,ij} = {C}_{\ell,ij} + \frac{\sigma_e^2}{n_i}\delta_{ij},
\end{equation}
where $n_i$ is the number density of galaxies in redshift bin, $i$, and $\sigma_e=0.3$.  

The Fisher matrix for galaxy clustering \citep{Seo03},
\begin{flalign}
F_{\alpha\beta} = \frac{1}{8\pi^2} & \sum_{i=1}^{N_{z,\rm{GC}}}\bigintsss_{-1}^{1}\, \mathrm{d}\mu \bigintsss_{k_{\rm{min}}}^{k_{\rm{max}}}\, k^2\,\mathrm{d}k\,  \nonumber \\
&\times\, \frac{\partial\ln P_{gg}\left(k,\mu;z_i\right)}{\partial \theta_{\alpha}} V_{\rm{eff}}\left(k,\mu;z_i\right)  \frac{\partial\ln P_{gg}\left(k,\mu;z_i\right)}{\partial \theta_{\beta}},
\end{flalign}
is computed by integrating over the angular and radial contributions of the logarithmic galaxy power derivatives across $N_{z,\rm{GC}}$ redshift bins. An upper bound of $k_{\rm{max}}=0.2\, h^{-1} \mathrm{Mpc}$ is chosen, as we restrict our GC forecasts to linear scales. The choice of lower bound has very limited bearing on results so we use the common value, $k_{\rm{min}}=0.001\, h^{-1} \mathrm{Mpc}$. 
Cross-correlations between redshift bins will have limited impact so they are not included. The effective volume, 
\begin{equation}
V_{\rm{eff}}\left(k,\mu;z_i\right) = V_{\rm{survey}}\left[\frac{1}{P_{gg}\left(k,\mu;z_i\right) + \frac{1}{n\left(z\right)}}\right]^2.
\end{equation}
encodes the covariances and is calculated from the survey volume, $V_s$, and the number density, $n\left(z\right)$, of H$\alpha$-emitters, for which we use Euclid survey values provided by the reference case in Table 3 of \citet{Amendola18}. 

Note that all scales and angles refer to the `fiducial' case, for which we have dropped the label for convenience. Free parameters governing the AP effect, growth, bias and residual shot noise within each redshift bin independently, $\left\{D_A\left(z_i\right), H\left(z_i\right), f\sigma_8\left(z_i\right), b\sigma_8\left(z_i\right),P_{\rm{shot}}\left(z_i\right)\vert i=1,...,N_{z,\rm{GC}}\right\}$, can then be propagated in a Fisher analysis alongside redshift-independent parameters, $\left\{\omega_m,\omega_b,\omega_{\nu},h,n_s\right\}$, controlling the shape of $P\left(k\right)$. The physical densities are related to our original density parameters via $\left\{\omega_i = \Omega_i\, h^2 \vert i=m,b,\nu\right\}$. We marginalize over the nuisance parameters, $\left\{b\sigma_8\left(z_i\right),P_{\rm{shot}}\left(z_i\right)\vert i=1,...,N_{z,\rm{GC}}\right\}$, by inverting the Fisher matrix, removing rows and columns corresponding to these parameters, and then inverting the remaining sub-block \citep{Seo03}. The resulting Fisher matrix, $\tilde{F}$, treats $\left\{D_A\left(z_i\right), H\left(z_i\right), f\sigma_8\left(z_i\right)\vert i=1,...,N_{z,\rm{GC}}\right\}$ as free parameters contributing information alongside $\left\{\omega_m,\omega_b,\omega_{\nu},h,n_s\right\}$. Collecting all these parameters into the set, $\tilde{\Theta}$, we derive a final Fisher matrix for the $w$CDM parameter set, $\Theta=\left\{\Omega_m,\Omega_b,\Omega_{\nu},h,n_s,\sigma_8,w\right\}$, by performing the transformation,
\begin{equation}
F = J^T \tilde{F} J,
\end{equation}
where $J=\partial \tilde{\Theta} / \partial \Theta$ is the Jacobian matrix of parameter derivatives. Note that because $\tilde{\Theta}$ is evaluated in the $w$CDM model it can be completely determined by $\Theta$.  

We evaluate the Fisher information available to a Euclid-like spectroscopic galaxy survey in the redshift range $0.75 < z \leq 2.05$, with bin width $\Delta z = 0.1$. For lower redshifts we refer to BOSS as it provides superior information at these $z$ than Euclid-like surveys are designed to achieve. This entails using covariance information provided by \citet{BOSS17} on $\left\{D_A\left(z_i\right),H\left(z_i\right),f\sigma_8\left(z_i\right)\right\}$, where the $i$ label runs over three partially overlapping redshift bins, spanning the total range $0.2 < z \leq 0.75$.  We perform a matrix inversion to derive the corresponding Fisher information matrix. The BOSS bins overlap with the full $z$ range available to the Euclid-like survey, so to ensure that covariance between the surveys is limited we only use redshift bins for the latter that lie beyond the range of BOSS. We then evaluate our final galaxy clustering Fisher matrix from the combination of Euclid-like high $z$ forecasts and BOSS low $z$ data. 

When evaluating derivatives, step sizes are determined for each parameter as 1\% of the fiducial value for that parameter. We have established that modest changes in the step size, up to doubling it, has no significant impact, suggesting our derivatives are stable. Similarly, the results are insensitive to whether we use 2-point or 5-point derivatives. For weak lensing and galaxy clustering, confidence regions for pairs of parameters are determined by inverting the corresponding Fisher matrix to marginalize over the remaining parameters, and extracting the resulting errors, $\sigma_{\alpha}=\sqrt{\left[F^{-1}\right]_{\alpha\alpha}}$, from the elements. Specifically, we compute the 1-$\sigma$ errors.

\section{Complete Fisher Forecasts}
\label{appendix:completeforecasts}
\begin{figure*}
\includegraphics[width=\textwidth]{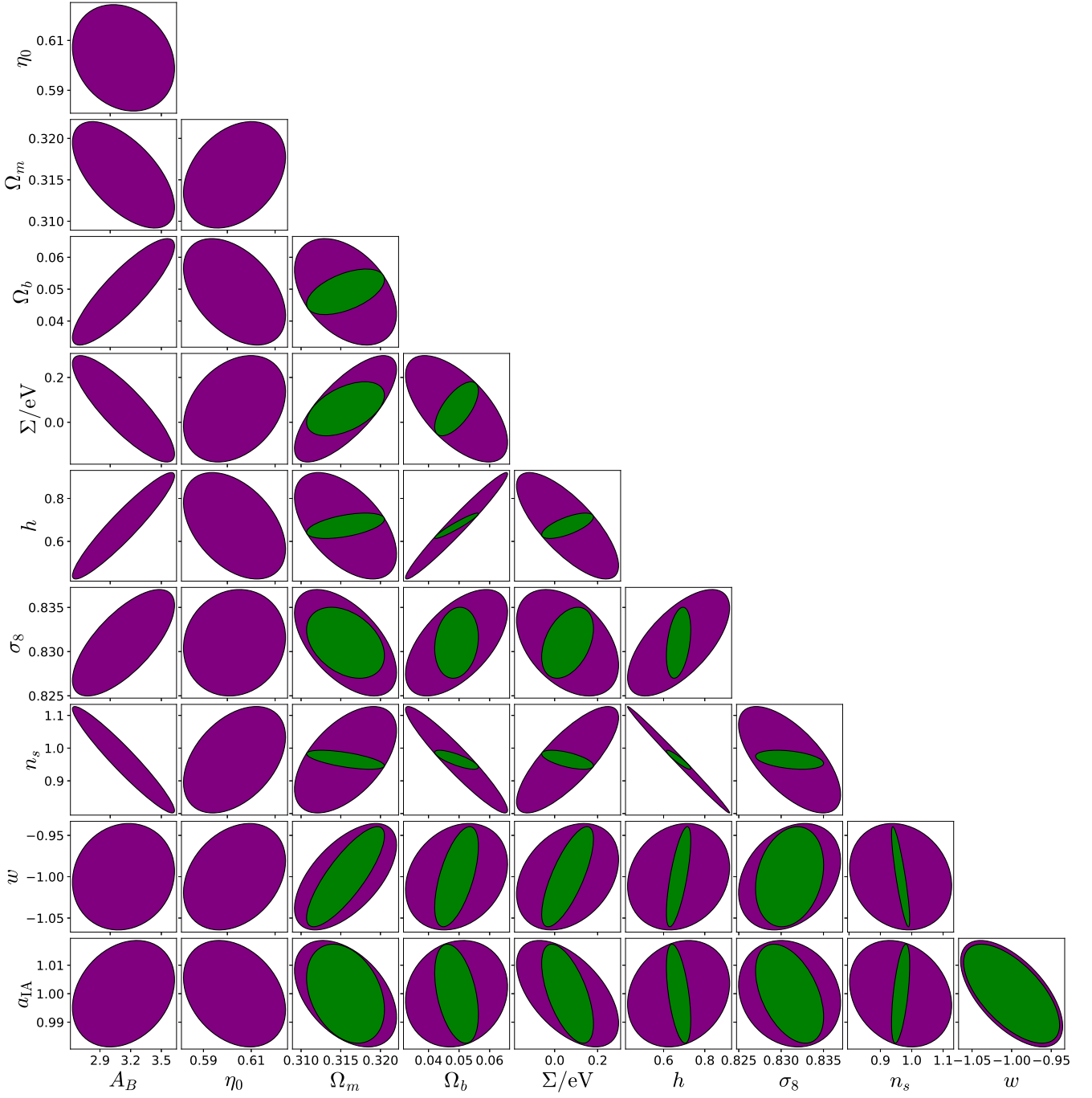}
\caption{1-$\sigma$ 2-parameter confidence ellipses for the normal hierarchy with all parameters in $\Theta=\left(A_B,\eta_0,\Omega_m,\Omega_b,\Sigma_{\rm{NH}}, h,\sigma_8,n_s,w,a_{\rm{IA}}\right)$ marginalized over (purple); and with the baryon parameters, $A_B$ and $\eta_0$, fixed to their fiducial values (dark green).}
\label{fig:fisherlensbar_nh} 
\end{figure*}
\par{
In Figure~\ref{fig:fisherlensbar_nh} we present the results of the full weak lensing Fisher analysis for every parameter combination in $\Theta=\left(A_B,\eta_0,\Omega_m,\Omega_b, \Sigma, h,\sigma_8,n_s,w,a_{\rm{IA}}\right)$ for the NH. These include plots showing comparisons of cases where baryon parameters are fixed to their fiducial values to those where they are marginalized over.}

\section{Intrinsic alignment power spectra responses}
\label{appendix:09ia}
\begin{figure*}
\centering
\includegraphics[width=\textwidth]{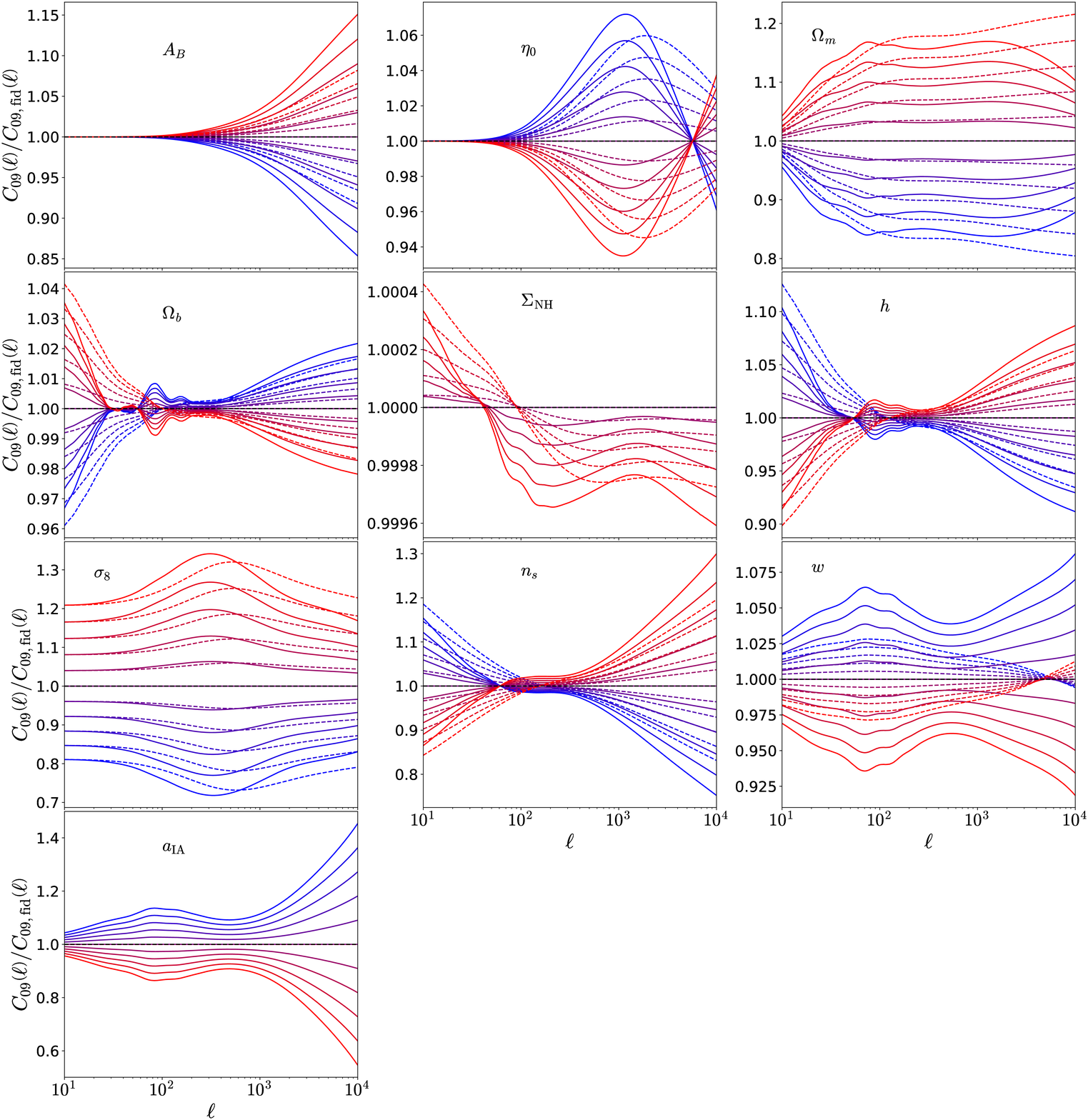}
\caption{Lensing power spectrum responses using NH in the $ij=09$ redshift bin. Blue (red) lines correspond to the lowest (highest) parameter values for $\Theta=\left(A_B,\eta_0,\Omega_m,\Omega_b,\Sigma_{\rm{NH}},h,\sigma_8,n_s,w,a_{\rm{IA}}\right)$ in the range $0.9\Theta_{\rm{fid}} \leq \Theta \leq 1.1\Theta_{\rm{fid}}$, except in the case of the neutrino mass parameter,  which varies between $\Sigma_{\rm{NH,min}} \leq \Sigma_{\rm{NH}} \leq 1.1\,\Sigma_{\rm{NH,min}}$ with purple curves representing the minimal mass case, $\Sigma_{\rm{NH,min}}=0.06\, \rm{eV}$. Solid lines correspond to the power spectrum including IA and dashed lines correspond to the $\gamma\gamma$ power spectrum where IA are not included.}
\label{fig:09ia}
\end{figure*}

In Figure~\ref{fig:09ia} we provide a comparison of the $C_{\ell}$ responses to varying parameters, both with and without intrinsic alignments. To display this clearly, we show results for a redshift bin combination, $ij=09$, at which the differences between the two cases are most pronounced.



\bsp 
\label{lastpage}
\end{document}